\documentclass[twocolumn]{aastex7}

\usepackage{appendix}
\usepackage[english]{babel} 
\usepackage{blindtext}
\usepackage{amsmath}

\newcommand\msol{M$_{\odot}$}

\newcommand\mstar{$M_{\star}$}

\newcommand\sml{$\Upsilon_{\star}$}

\newcommand\atlas{ATLAS$^{3\mathrm{D}}$}
\newcommand\nd{ ~$\cdots$~ }
\newcommand\err{$\pm$~}
\mathchardef\mh="2D

\newcommand{\uvic}{Department of Physics and Astronomy, University of Victoria, Victoria, BC V8P 1A1, Canada}
\newcommand{\nrchaa}{National Research Council of Canada, Herzberg Astronomy \& Astrophysics Research Centre, Victoria, BC, V9E 2E7, Canada}
\newcommand{\csa}{Canadian Space Agency, 6767 Route de l’A\'eroport, Saint-Hubert, QC J3Y 8Y9, Canada}
\newcommand{\noirlab}{NSF’s NOIRLab, 950 N. Cherry Avenue, Tucson, AZ 85719, USA}

\newcommand{\psec}{\(\stackrel{\:''}{\textstyle.}\)}

\newcommand{\pdeg}{\(\stackrel{\:\circ}{\textstyle.\rule{0pt}{0.65ex}}\)}

\begin{document}

\title{The Next Generation Virgo Cluster Survey (NGVS). II. A Catalog of Galaxies in the Virgo Cluster}

\correspondingauthor{Laura Ferrarese}

\author[0000-0002-0363-4266]{Laura Ferrarese}
\affiliation{National Research Council of Canada, Herzberg Astronomy \& Astrophysics Research Centre, 5071 West Saanich Road, Victoria, BC, V9E 2E7, Canada}
\email{Laura.Ferrarese@nrc-cnrc.gc.ca}

\author[0000-0003-1184-8114]{Patrick C\^ot\'e}
\affiliation{National Research Council of Canada, Herzberg Astronomy \& Astrophysics Research Centre, 5071 West Saanich Road, Victoria, BC, V9E 2E7, Canada}
\email{patrick.cote@nrc-cnrc.gc.ca}

\author[0009-0003-5548-6773]{Lauren A. MacArthur}
\affiliation{Princeton University, Department of Astrophysical Sciences, Princeton, NJ 08544, USA}
\email{lauren@astro.princeton.edu}

\author[0000-0002-0363-4266]{Joel C. Roediger}  
\affiliation{\csa}
\email{joel.roediger@gmail.com}

\author[0000-0002-5213-3548]{John P. Blakeslee}
\affiliation{\noirlab}
\email{john.blakeslee@noirlab.edu}

\author[0000-0003-2072-384X]{Michele Cantiello}
\affiliation{INAF, Osservatorio Astronomico di Teramo, Via Maggini, 1-64100 Teramo, Italy}
\email{michele.cantiello@inaf.it}

\author[0000-0002-3263-8645]{Jean-Charles Cuillandre} 
\affiliation{AIM, CEA, CNRS, Universit\'e Paris-Saclay, Universit\'e de Paris, F-91191 Gif-sur-Yvette, France}
\email{jc.cuillandre@cea.fr}

\author[0000-0001-8867-4234]{Puragra Guhathakurta}
\affiliation{UCO/Lick Observatory, Department of Astronomy and Astrophysics, University of California Santa Cruz, Santa Cruz, CA, 95064, USA}
\email{raja@ucolick.org}

\author[0000-0001-8221-8406]{Stephen Gwyn} 
\affiliation{National Research Council of Canada, Herzberg Astronomy \& Astrophysics Research Centre, 5071 West Saanich Road, Victoria, BC, V9E 2E7, Canada}
\email{Stephen.Gwyn@nrc-cnrc.gc.ca}

\author[0000-0001-5208-4697]{Max M. Kurzner}
\affiliation{\uvic}
\email{mkurzner@uvic.ca}

\author[0000-0002-2073-2781]{Eric W. Peng}
\affiliation{\noirlab}
\email{eric.peng@noirlab.edu}

\author[0009-0006-1653-3882]{Matthew Santos}
\affiliation{National Research Council of Canada, Herzberg Astronomy \& Astrophysics Research Centre, 5071 West Saanich Road, Victoria, BC, V9E 2E7, Canada}
\email{matthewsantos@ieee.org}

\author[0009-0008-9732-2407]{Eleanore B. Todd}
\affiliation{\uvic}
\email{blonsbrough@uvic.ca}

\author[0000-0001-6443-5570]{Elisa Toloba}
\affiliation{Department of Physics and Astronomy, University of the Pacific, Stockton, CA, USA}
\email{etoloba@pacific.edu}

\author[0000-0003-3343-6284]{Pierre-Alain Duc}
\affiliation{Universit\'e de Strasbourg, CNRS, Observatoire astronomique de Strasbourg, UMR 7550, F-67000 Strasbourg, France}
\email{pierre-alain.duc@astro.unistra.fr}

\author[0000-0001-9427-3373]{Patrick R. Durrell}
\affiliation{Department of Physics, Astronomy, GIS, Geology and Environmental Sciences, Youngstown State University, One Tressel Way, Youngstown, OH 44555 USA}
\email{prdurrell@ysu.edu}

\author[0000-0003-3816-3254]{Nicholas Fantin}
\affiliation{\uvic}
\email{Nicholas.Fantin@gov.bc.ca}

\author[]{Yuting Feng}
\affiliation{Department of Astronomy and Astrophysics, University of California Santa Cruz, 1156 High Street, Santa Cruz, CA 95064, USA}
\email{yfeng47@ucsc.edu}

\author[0000-0002-7214-8296]{Ariane Lan\c{c}on}
\affiliation{Universit\'e de Strasbourg, CNRS, Observatoire astronomique de Strasbourg, UMR 7550, F-67000 Strasbourg, France}
\email{ariane.lancon@astro.unistra.fr}

\author[0000-0002-5049-4390]{Sungsoon Lim}
\affiliation{Division of Science Education, Kangwon National University, Chuncheon, Republic of Korea}
\email{sslim00@gmail.com}

\author[0000-0002-4718-3428]{Chengze Liu}
\affiliation{State Key Laboratory of Dark Matter Physics, Shanghai Key Laboratory for Particle Physics and Cosmology, School of Physics and Astronomy \& Tsung-Dao Lee Institute, Shanghai Jiao Tong University, Shanghai 200240, China}
\email{czliu@sjtu.edu.cn}

\author[0000-0002-2406-7344]{Deborah Lokhorst} 
\affiliation{\nrchaa}
\email{Deborah.Lokhorst@nrc-cnrc.gc.ca}

\author[0000-0001-5569-6584]{Alessia Longobardi} 
\affiliation{Dipartimento di Fisica G. Occhialini, Universit\'a degli Studi di Milano-Bicocca, Piazza della Scienza 3, I-20126 Milano, Italy; INAF - Osservatorio Astronomico di Brera, via Bianchi 46, I-23087 Merate (LC), Italy}
\email{alessia.longobardi@unimib.it}

\author[0000-0002-2849-559X]{Simona Mei}
\affiliation{Universit\'e Paris Cit\'e, CNRS/IN2P3, Astroparticule et Cosmologie, 75013, Paris, France ; Jet Propulsion Laboratory and Cahill Center for Astronomy \& Astrophysics, California Institute of Technology, 4800 Oak Grove Drive, Pasadena, CA, 91011, USA}
\email{simona.mei@obspm.fr}

\author[0000-0002-7089-8616]{J. Christopher Mihos}
\affiliation{Department of Astronomy, Case Western Reserve University, Cleveland, OH 44106, USA}
\email{mihos@case.edu}

\author[0000-0003-2922-6866]{Sanjaya Paudel} 
\affiliation{Department of Astronomy \& Center for Galaxy Evolution Research, Yonsei University, Seoul 03722, Republic of Korea}
\email{sanjpaudel@gmail.com}

\author[0000-0003-0350-7061]{Thomas H. Puzia}
\affiliation{Departamento de Astronom\'{\i}a y Astrof\'{\i}sica, Pontificia Universidad Cat\'olica de Chile, 7820436 Macul, Santiago, Chile} 
\email{tpuzia@astro.puc.cl}

\author[0000-0001-5999-7923]{Anand Raichoor}
\affiliation{Universit\'e Paris Cit\'e, CNRS, Astroparticule et Cosmologie, F-75013 Paris, France; Lawrence Berkeley National Laboratory, 1 Cyclotron Road, Berkeley, CA 94720, USA}
\email{ARaichoor@lbl.gov}

\author[0000-0003-4945-0056]{Rub\'en S\'anchez-Janssen}
\affiliation{Isaac Newton Group of Telescopes, Apartado 321, E-38700 Santa Cruz de la Palma, Tenerife, Spain}
\email{ruben.sanchez-janssen@stfc.ac.uk}

\author[0000-0003-4672-8497]{Kaixiang Wang} 
\affiliation{Department of Astronomy, Peking University, Beijing, China Kavli Institute for Astronomy and Astrophysics, Peking University, Beijing, China}
\email{kkwang.astro@gmail.com}

\author[0000-0003-1632-2541]{Hongxin Zhang}
\affiliation{Key Laboratory for Research in Galaxies and Cosmology, Department of Astronomy, University of Science and Technology of China, Hefei, Anhui 230026, People's Republic of China; School of Astronomy and Space Science, University of Science and Technology of China, Hefei, Anhui 230026, People's Republic of China}
\email{hzhang18@ustc.edu.cn}

\author[0000-0002-9795-6433]{Alessandro Boselli}
\affiliation{Aix Marseille Universit\'e, CNRS, CNES, LAM, Marseille, France}
\email{alessandro.boselli@lam.fr}

\author[0000-0003-4849-9536]{Michael L. Balogh}
\affiliation{Department of Physics and Astronomy, University of Waterloo, Waterloo, Ontario N2L 3G1, Canada}
\email{mbalogh@uwaterloo.ca}

\author[0000-0002-9091-2366]{Samuel Boissier}
\affiliation{Aix Marseille Universit\'e, CNRS, CNES, LAM, Marseille, France}
\email{samuel.boissier@lam.fr}

\author[0000-0002-6155-7166]{Eric Emsellem}
\affiliation{European Southern Observatory, Karl-Schwarzschild-Stra{\ss}e 2, 85748 Garching, Germany; Universit\'e Lyon1, ENS de Lyon, CNRS, Centre de Recherche Astrophysique de Lyon UMR5574, F-69230 Saint-Genis-Laval France}
\email{eric.emsellem@eso.org}

\author[0000-0001-7032-5255]{J.J. Kavelaars}
\affiliation{National Research Council of Canada, Herzberg Astronomy \& Astrophysics Research Centre, 5071 West Saanich Road, Victoria, BC, V9E 2E7, Canada}
\email{JJ.Kavelaars@nrc-cnrc.gc.ca}

\author[0000-0003-4666-6564]{Alan W. McConnachie}
\affiliation{National Research Council of Canada, Herzberg Astronomy \& Astrophysics Research Centre, 5071 West Saanich Road, Victoria, BC, V9E 2E7, Canada}
\email{alan.mcconnachie@nrc-cnrc.gc.ca}

\author[0009-0009-2522-3685]{Cameron R. Morgan}
\affiliation{Waterloo Centre for Astrophysics, University of Waterloo, Waterloo, ON N2L 3G1, Canada; Department of Physics and Astronomy, University of Waterloo, Waterloo, ON N2L 3G1, Canada}
\email{crmorgan@uwaterloo.ca}

\author[0000-0002-1685-4284]{Chelsea Spengler}
\affiliation{\uvic; Departamento de Astronom\'{\i}a y Astrof\'{\i}sica, Pontificia Universidad Cat\'olica de Chile, 7820436 Macul, Santiago, Chile}
\email{g.spengleri@gmail.com}

\author[0000-0002-6639-4183]{James E. Taylor}
\affiliation{Department of Physics and Astronomy, University of Waterloo, Waterloo, Ontario N2L 3G1, Canada, Waterloo Centre for Astrophysics, University of Waterloo, Waterloo, ON N2L 3G1, Canada}
\email{taylor@uwaterloo.ca}

\author[0000-0003-3009-4928]{Matthew A. Taylor}
\affiliation{University of Calgary, 2500 University Drive NW, Calgary Alberta T2N 1N4, Canada}
\email{matthew.taylor2@ucalgary.ca}

\author[0000-0003-1845-0934]{Toby Brown}
\affiliation{\nrchaa; \uvic}
\email{tobias.brown@nrc-cnrc.gc.ca}

\author[0000-0003-2239-7988]{S\'ebastien Fabbro}
\affiliation{\nrchaa}
\email{Sebastien.Fabbro@nrc-cnrc.gc.ca} 

\author[0000-0002-5540-6935]{Raphael Gavazzi}
\affiliation{Aix-Marseille Universit\'e, CNRS, CNES, LAM, Marseille, France; Institut d’Astrophysique de Paris, UMR 7095, CNRS, and Sorbonne Universit\'e, 98 bis boulevard Arago, 75014 Paris, France}
\email{raphael.gavazzi@lam.fr}

\author[0000-0002-9814-3338]{Hendrik Hildebrandt}
\affiliation{Ruhr University Bochum, Faculty of Physics and Astronomy, Astronomical Institute (AIRUB), German Centre for Cosmological Lensing, 44780, Bochum, Germany}
\email{hendrik@astro.ruhr-uni-bochum.de}

\author[0000-0001-5339-9017]{Rashi Jain}
\affil{School of Physics and Astronomy, University of Southampton, Southampton SO17 1BJ, UK}
\email{rashi.jain@soton.ac.uk}

\author[0000-0002-5389-3944]{Andr\'es Jord\'an}
\affil{Facultad de Ingenier\'ia y Ciencias, Universidad Adolfo Ib\'{a}\~{n}ez, Av. Diagonal las Torres 2640, 7941169 Pe\~{n}alol\'{e}n, Santiago, Chile;Departamento de Astronom\'ia, Universidad de Chile, Casilla 36-D, Santiago, Chile}
\email{andres.jordan@uai.cl}

\author[0000-0001-5303-6830]{Rory Smith}
\affiliation{Departamento de Fisica, Universidad Tecnica Federico Santa Maria, Avenida Espa\~na 1680, Valparaíso, Chile; Millennium Nucleus for Galaxies (MINGAL), Valpara\'iso, Chile}
\email{rorysmith274@gmail.com}

\author[0009-0006-7485-7463]{Solveig Thompson}
\affiliation{University of Calgary, 2500 University Drive NW, Calgary Alberta T2N 1N4, Canada}
\email{solveig.thompson@ucalgary.ca}

\author[0009-0006-9760-9315]{Haruka Yoshino}
\affiliation{\uvic}
\email{harukayoshino@uvic.ca}

\author[0000-0003-4770-9829]{Wim van Driel}
\affiliation{LUX, Observatoire de Paris, Universit\'e PSL, Sorbonne Universit\'e, CNRS, 5 place Jules Janssen, 92190 Meudon, France}
\email{wim.vandriel@obspm.fr}

\author[0000-0002-2637-8728]{Ludovic van Waerbeke} 
\affiliation{Department of Physics and Astronomy, University of British Columbia, 6224 Agricultural Road, Vancouver, BC V6T 1Z1, Canada}
\email{waerbeke@phas.ubc.ca}

\author[0000-0003-1428-5775]{Tyrone E. Woods}
\affiliation{Department of Physics and Astronomy, Allen Building, 30A Sifton Road, University of Manitoba, Winnipeg MB R3T 2N2, Canada}
\email{Tyrone.Woods@umanitoba.ca}

\author[0009-0002-6928-6353]{Jade Yeung}
\affiliation{Department of Physics and Astronomy, Allen Building, 30A Sifton Road, University of Manitoba, Winnipeg MB R3T 2N2, Canada}
\email{yeungj2@myumanitoba.ca}

\begin{abstract}
The Next Generation Virgo Cluster Survey (NGVS) is a deep, high resolution imaging campaign that used the 1 deg$^2$ MegaCam instrument on the Canada-France-Hawaii Telescope to carry out a comprehensive optical survey of the Virgo cluster, from its core to its virial radius. The NGVS covers a contiguous area of 104 deg$^2$ (8.63 Mpc$^2$ at the 16.5 Mpc distance of Virgo) in the $u^*$-,$g$-,$i$-, and $z$-band, with additional limited coverage in $r$. In this paper, we present the final catalog of Virgo galaxies across the entire NGVS area. The catalog includes 3680 galaxies considered to be {\it bona fide} members of the cluster, spanning a factor of 2.5 million in luminosity, from $g = 8.42$ mag to $g = 24.41$ mag ($M_g = -22.67$ mag to $M_g = -6.68$ mag). With 2100 previously uncataloged galaxies, the NGVS catalog augments the number of known Virgo members by a factor 2.3. The catalog is complete down to $g = 18.6$ mag ($M_g=-12.5$ mag, corresponding to a stellar mass \mstar $\sim 1.6\times10^7$ M$_{\odot}$ for an old stellar population) and 50\% complete at $g = 22.0$ mag ($M_g=-9.1$ mag, \mstar $\sim 6.2\times10^5$ M$_{\odot}$), three magnitudes deeper than the venerable Virgo Cluster Catalog (VCC), which for over 40 years has served as the reference standard for Virgo. Photometric and structural parameters are derived for all NGVS galaxies and presented in a series of tables, alongside nuclear and morphological classification, as well as stellar masses and, when available, radial velocities. 
\end{abstract}

\keywords{Galaxies (573) --- Catalogs (205) --- Virgo Cluster (1772) --- Stellar populations (1622) --- Galaxy photometry(611) --- Dwarf galaxies(416) --- Giant elliptical galaxies(651) --- Spiral galaxies(1560)}

\section{Introduction} \label{sec:intro}

As the most massive concentration of galaxies in the local universe, the Virgo Cluster serves as the primary laboratory for studying galaxy evolution within dense environments. An essential bridge between detailed ``microscopic" studies of the Local Group and ``macroscopic" statistical surveys of the distant universe, Virgo offers an unparalleled opportunity to resolve stellar populations, identify and characterize rare objects such as the most compact or extended galaxies, map the intricate web of tidal features that reveal the ongoing assembly of cosmic structures, and test cosmological models on small scales. It is therefore hardly surprising that a large number of targeted and blind surveys have turned to Virgo to address questions that range from probing the low-mass galaxy regime, to quantifying the role of the environment on galaxy evolution, to studying the incidence of Active Galactic Nuclei (AGNs). 

As a demonstration of the importance of Virgo in tackling a range of scientific questions, we highlight a specific survey: the HST ACS Virgo Cluster Survey \citep[ACSVCS,][]{Cote+04}. The ACSVCS imaged 100 early-type Virgo galaxies with the Advanced Camera for Survey (ACS) in the $g$- and $z$-band; among its most significant findings are the discovery of a link between supermassive black holes (SMBHs) and Nuclear Star Clusters (NSCs) \citep{Ferrarese+06b}; the demonstration that the structural properties of early-type galaxies vary smoothly and systematically from the smallest dwarfs to the largest giants \citep{Ferrarese+06a}; the characterization of compact stellar nuclei \citep{Cote+06}, Ultra Compact Dwarfs \citep[UCDs,][]{Hasegan+05}, and Globular Clusters \citep[GCs,][]{Jordan+05,Jordan+07,Peng+08}; and precision distance mapping using the Surface Brightness Fluctuation (SBF) technique \citep{Mei+07}. Representing a modest (100-orbit) investment of HST time, the ACSVCS' impact (as measured by the number of citations per paper) is comparable to that of much larger HST Treasury programs, and second only to pioneering deep field programs such as the HDF or GOODS --- evidence for Virgo's uniqueness in addressing a diversity of important and consequential astrophysical questions. 

Beyond the optical regime, targeted surveys of Virgo include AMUSE-Virgo \citep{Gallo+08}, a Chandra X-ray survey of the ACSVCS galaxies; Virgo Redux \citep{Spengler+17}, a UV and IR follow-up of the ACSVCS sample; the Herschel Reference Survey \citep[HRS,][]{Boselli+10}, a near-infrared targeted survey including 65 galaxies in Virgo; the VLA Imaging of Virgo in Atomic Gas \citep[VIVA,][]{Chung+09}, a high-resolution HI survey of 53 late-type galaxies; the Multiphase Astrophysics to Unveil the Virgo Environment (MAUVE) survey \citep{Cortese+26}, a VLT/MUSE, ALMA and HST followup of 40 VIVA galaxies; the Virgo Environment Traced in CO Survey \citep[VERTICO,][]{Brown+12}, an ALMA program mapping molecular gas in 51 galaxies; and the Atomic gas in Virgo Interacting Dwarf galaxies ({\tt AVID}) survey, designed to obtain high-resolution VLA HI observations of 14 post-merger dwarfs  \citep{Sun+20a,Sun+20b}. 

Blind surveys of the cluster include the Herschel Virgo Cluster Survey \citep[HeViCS,][]{Davies+10}, which mapped 64 deg$^2$ in Virgo to detect cold dust in the Interstellar Medium (ISM) and Intracluster Medium (ICM); the GALEX Ultraviolet Virgo Cluster Survey \citep[GUViCS,][]{Boselli+11}, a program covering 120 deg$^2$ to map recent star formation across the cluster; the Virgo Environmental Survey Tracing Ionised Gas Emission \citep[VESTIGE,][]{Boselli+18}, a blind narrow-band H$\alpha$ survey using CFHT/MegaCam to map ionized gas tails and star-forming regions; and the Virgo Cluster multi-Telescope Observations in Radio of Interacting galaxies and AGN \citep[ViCTORIA,][]{deGasperin+25}, a low- and high-band LOFAR and MeerKAT radio survey covering 132 deg$^2$. Even surveys mapping large extents of the sky have, for the most part, been designed to include Virgo in their footprint; this is the case, for example, for the Arecibo Legacy Fast ALFA Survey \citep[ALFALFA,][]{Giovanelli+05}, a blind HI survey that detected hundreds of gas-bearing sources across the Virgo region, and WALLABY \citep{Koribalski+20}, an HI survey covering 14,000 deg$^2$, and extending to the southern part of Virgo. Finally, it is worth mentioning that despite the northern location of the cluster, Virgo was the most publicized of the early-release observations from the  Vera C. Rubin Observatory, and the Virgo region has been explicitly added to the footprint of Rubin's Legacy Survey of Space and Time (LSST) Wide Fast Deep (WFD) survey \citep{Ivezic+19}.

Most of the above surveys have one thing in common: whether analyzing X-ray emission from the ICM, neutral hydrogen, or dust heating maps, they rely on optical catalogs to provide the necessary context -- stellar mass, morphology, and membership -- as well as targets in the case of targeted surveys. Producing a comprehensive catalog of galaxies within the Virgo cluster is therefore not merely an exercise in bookkeeping; it is a fundamental requirement for addressing several of the most pressing questions in modern astrophysics. 

For nearly four decades, the Virgo Cluster Catalog \citep[VCC,][]{Binggeli+85} has served as the benchmark optical galaxy catalog for Virgo. Based on photographic plates from the Du Pont telescope at Las Campanas Observatory, the VCC lists 2,096 galaxies, of which 1851 are deemed {\it bona fide}, i.e. certain or probable, Virgo members, within an area of roughly 140 deg$^2$. The VCC provided the first systematic morphological classification for cluster members and established ``standard" membership criteria based on galaxy luminosity, surface brightness, and ``resolution into knots", the basic tenant of which have been followed essentially ever since.

We will show later in this paper that the VCC is complete to a $g$-band magnitude of $\sim 16.5$, 50\% complete at $g \sim 18.8$ mag, and fully incomplete beyond $g \sim 20$ mag. Since the VCC, a few notable studies have extended the Virgo galaxy luminosity function further into the regime of dwarf and low surface brightness galaxies, at the expense of smaller footprints: \citet{Impey+88} specifically targeted low surface brightness galaxies selected from UK Schmidt plates covering 7.7 deg$^2$; \citet{Trentham+02} and \citet{Sabatini+03} used data from the INT Wide Field Survey covering two strips totaling 25 deg$^2$ in the $B$-band \citep[and $I$ in the case of][]{Sabatini+03}, to reach $M_B \sim -10$ mag. 

The most significant advancement since the VCC, however, did not occur until 2014, with the publication of the Extended Virgo Cluster Catalog (EVCC) by \citet{Kim+14}. The EVCC used radial velocities from (mostly) the Sloan Digital Sky Survey (SDSS) Data Release 7 \citep[DR7,][]{Abazajian+09} to establish spectroscopic membership for 1,589 galaxies, including 676 not found in the original VCC, within 725 deg$^2$, an area over five times larger than the VCC. Due to the limited depth of the SDSS, the EVCC did not improve on the magnitude limit of the VCC, although it added color information useful for stellar population studies. However, by reaching out to 3.5 times Virgo's virial radius, the EVCC is unique in enabling studies of the ``infall" populations of galaxies that are being drawn into the cluster for the first time. 
    
This paper represents the next chapter in the effort to produce a reliable, unbiased, deep galaxy catalog for the Virgo cluster. The Next Generation Virgo Cluster Survey \citep[NGVS,][hereafter NGVS-I]{Ferrarese+12} was designed to provide a deep, wide-field imaging census of baryonic structures in the cluster --- galaxies, globular and nuclear star clusters, tidal features, UCDs, etc. Utilizing the MegaCam instrument on the Canada-France-Hawaii Telescope (CFHT), the survey covers 104 deg$^2$ out to the cluster's virial radius in the $u^*$-,$g$-,$i$- and $z$-band. 

The NGVS has already boosted our census of the low-mass regime, uncovering a previously unknown population of dwarf galaxies and globular clusters. A galaxy catalog for the core of the cluster was published in \citet{Ferrarese+20} (hereafter NGVS-XIV). It contains 404 {\it bona fide} Virgo galaxies within a 3.71 deg$^2$ (0.3 Mpc$^2$) region centered on M87, Virgo's dominant galaxy, and spanning the range $17.8$ mag $< g < 23.7$ mag ($-13.4$ mag $< M_g < -7.4$ mag at the 16.5 Mpc distance of Virgo). Of these galaxies, 154, or almost 40\%,  were previously uncataloged. \citet{Ferrarese+20} shows that the NGVS is complete down to $g \sim 18.6$ mag ($M_g\sim -12.5$ mag, corresponding to a stellar mass \mstar\ $\sim 1.6 \times 10^7$ M$_{\odot}$ for an
old stellar population) and 50\% complete at $g \sim 22.0$ mag ($M_g \sim -9.1$ mag, or \mstar $\sim 6.2 \times 10^5$ M$_{\odot}$), a full two magnitudes deeper than the magnitude beyond which no VCC galaxies are known. 

This paper presents the final galaxy catalog over the full 104 deg$^2$ of the NGVS. Representing the culmination of over a decade of data acquisition, processing, and analysis, this catalog is intended to serve as the foundational dataset for the next generation of studies regarding environment-driven evolution in the local universe.

The paper is organized as follows. In \S2, we summarize the data acquisition and data reduction strategies, although the reader is referred to NGVS-I for a more extensive exposition. \S3 discusses the methodology and algorithm used to detect potential Virgo galaxies in the NGVS images and assess their cluster membership (again, the reader is referred to NGVS-I for a more extensive discussion). \S3 also describes the final catalog and presents a comparison with the VCC and the EVCC.  \S4 discusses the completeness and purity of the NGVS galaxy sample, while the derivation of the photometric and structural parameters for the nearly 3700 {\it bona fide} Virgo galaxies that comprise the final catalog is summarized in \S5, which also discusses the identification and characterization of NSCs and the morphological classification of Virgo's galaxies. \S6 describes in some detail the procedure to derive stellar masses from the NGVS photometry. Figures and Tables are interspersed throughout the text but, for the reader's convenience, \S7 provides concise explanations for all parameters included in the several Tables that comprise the final catalog. Finally, \S8 presents an overview of the science results that have been obtained so far from the NGVS. Conclusions are summarized in \S9.  

Throughout this paper, we assume a distance modulus to the Virgo cluster of 31.09 mag, corresponding to a distance of 16.5 Mpc \citep[][see also \citealt{Casertano+25}]{Mei+07,Blakeslee+09,Cantiello+24}, and we consider as spectroscopic members of the cluster all galaxies with radial velocity $v < 3,500$ km s$^{-1}$, a cut that includes all objects belonging to the different cluster substructures and their surrounding regions  \citep[e.g.][]{Binggeli+93,Boselli+14}.

\section{The NGVS CFHT Imaging Data: Observing Procedures and Data Reductions} \label{sec:data}

NGVS-I provides a detailed description of the survey and its science goals. We also refer to that paper for an in-depth exposition of field placements, observing strategy, data-reduction pipeline, and data quality and depth; only the most salient points will be summarized here. 

\begin{figure}
    \includegraphics[width=\linewidth]{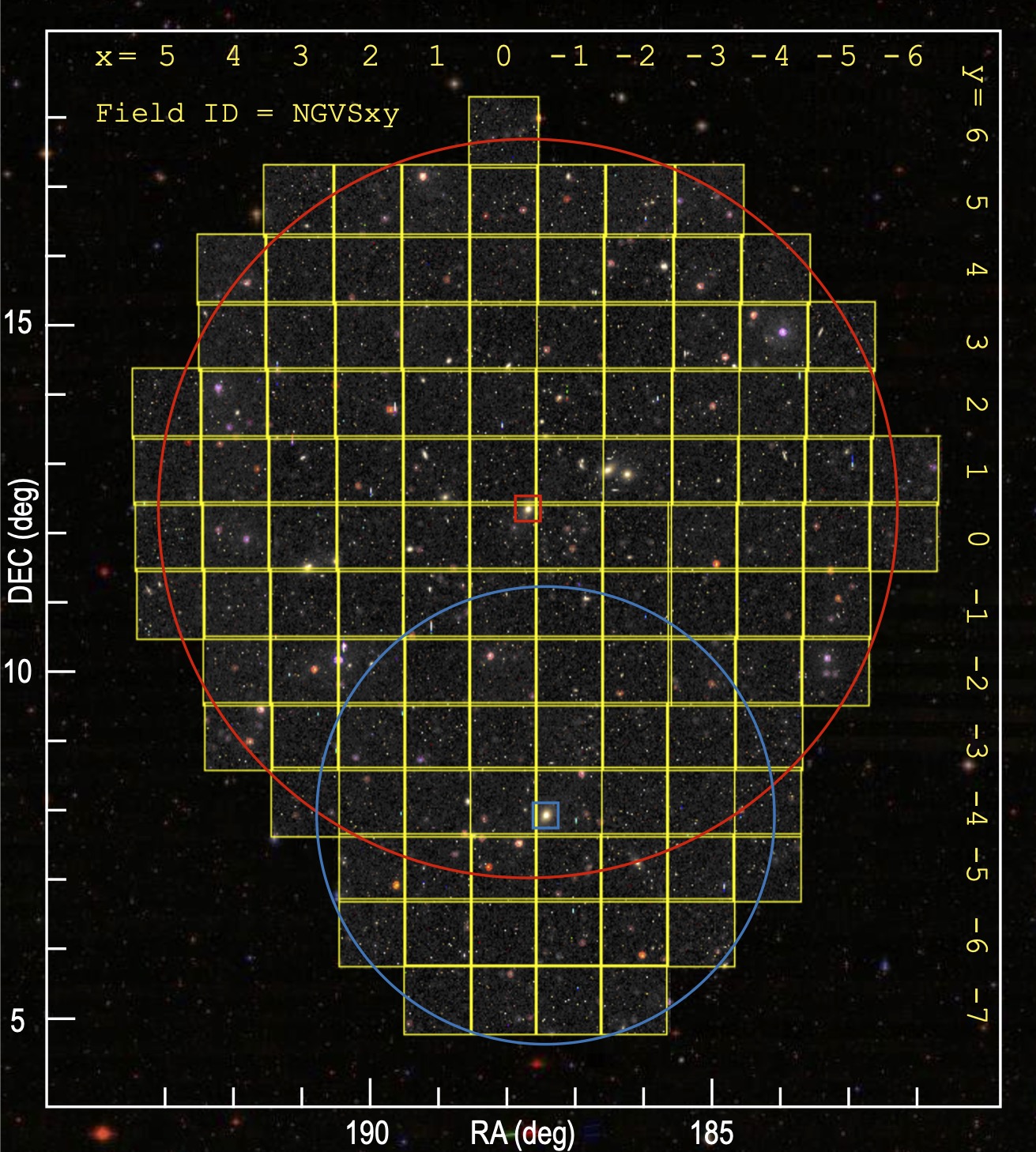}
    \caption{The location of the 117 NGVS fields in the plane of the sky. Each field is identified as NGVSxy, where x and y are the number of fields separating the field in question from the one containing M87, identified as  NGVS+0+0 (see text footnote). All fields have $u^*$-,$g$-,$i$-, and $z$-band long exposures, and $g$- and $z$-band short exposures, but short exposures coverage in the $u^*$- and $i$-band is not complete (see Figure \ref{fig:fig2} and Figure \ref{fig:fig3}). The background image within the NGVS footprint is a color mosaic of NGVS images, while the background outside the NGVS footprint is from the SDSS. The small red and blue squares mark the locations of M87 and M49, respectively, while the large red and blue circles represent the viral radii of Virgo's A and B subclusters, respectively.}
    \label{fig:fig1}
\end{figure}

The NGVS' main science goal -- the study of baryonic structures in a cluster environment, including galaxies, UCDs, NSCs, and GCs --- dictates strict requirements on the areal and wavelength coverage of the data, as well as on its depth and image quality. To enable the study of environmental effects (albeit still within a cluster setting), the survey covers the 104 deg$^2$ region (see NGVS-I Figures 1 and 4, as well as Figure \ref{fig:fig1} in this paper) defined by the union of the two partially overlapping circles mapping Virgo's two main subclusters from their center out to their virial radius \citep[][NGVS-I]{McLaughlin99}. Subcluster A is centered on M87 (VCC1316, RA = 12h30m49s; DEC = $+12^\circ23\arcmin28\arcsec$) and has a virial radius of 5\pdeg383 (1.55 Mpc), while the smaller subcluster B is centered on M49 (VCC1226, RA = 12h29m47s; DEC = $+08^\circ00\arcmin02\arcsec$) and has a virial radius of 3\pdeg334 (0.96 Mpc). 

The entire survey area was imaged with the MegaCam wide field camera on the 3.6m Canada France Hawaii Telescope (CFHT). After stacking multiple dithered exposures (see NGVS-I, \S4.2), each MegaCam field\footnote{By convention, each field is uniquely identified as {\sf NGVSXY} (Figure \ref{fig:fig1}), where {\sf X} and {\sf Y} denote the separation, in degrees, between the field in question and the one including M87 ({\sf NGVS+0+0}): {\sf X} is measured along the Right Ascension (with positive numbers to the East and negative to the West of {\sf NGVS+0+0}) and {\sf Y} is measured along the Declination (positive numbers to the North and negative to the South of {\sf NGVS+0+0}).} covers an area of $1^\circ\times1^\circ$ and overlaps adjacent fields by $3\arcsec$, requiring a total of 117 pointings to cover the 104 deg$^2$ NGVS footprint. 

This footprint was imaged with the $u^*$-,$g$-,$i$-, and $z$-band (see NGVS-I, \S3.3 and Figure 6), a combination that allows the study of stellar populations and aids in the separation of Virgo members from foreground/background contaminants. Although the original plan also called for full $r$-band coverage, losses due to weather and technical problems meant that full depth $r$-band exposures were only acquired for the 4 deg$^2$ surrounding M87 (fields {\sf NGVS+0+0, NGVS+0+1, NGVS-1+0}, and {\sf NGVS-1+1}; collectively, we refer below to these fields as the ``core region", see Figure \ref{fig:fig1}). Shallower depth $r$-band data were obtained for additional fields; for completeness, we list these data in Appendix \ref{app:short} although we do not make use of them in the remainder of this paper\footnote{We note that VESTIGE \citep{Boselli+18} covered the entire NGVS area in the $r$-band to constrain the continuum in the $H\alpha$ images, although at a shallower depth than the original NGVS images (reaching a $g$-band magnitude of 24.5 at 5$\sigma$)} 

The point- and extended-source limits of the data are summarized in Tables \ref{tab:Tab1} and \ref{tab:Tab2} while more detailed information, including the average image quality, are given in  Tables \ref{tab:Tab3} and \ref{tab:Tab4}. The main dataset consists of images (we will refer to these as ``long" exposures) with total integration times ranging from 2055s in the $i$-band to 6402s in the $u^*$-band, and reaching, in the $g$-band, a point- and extended-source limit of 25.9 mag (10$\sigma$) and 29 mag arcsec$^{-2}$ (2$\sigma$), respectively. At this depth, the NGVS can detect dwarf galaxies in Virgo to a completeness limit comparable to what is achieved for classical dwarf spheroidal galaxies in the Local Group, and can characterize low surface brightness features resulting from the interaction of galaxies with the surrounding environment \citep{Duc+15}. For compact and unresolved sources, the NGVS samples the brightest 90\% of the luminosity function of GCs in Virgo at S/N $\geq$ 10. All images were acquired in sub-arcsecond seeing conditions (see Tables \ref{tab:Tab2} and \ref{tab:Tab4}). The best seeing, with a mean across all exposures of 0.58\err 0.03 arcsec, was achieved in the $i$-band, and is sufficient to marginally resolve the brightest GCs, UCDs and NSCs. 

For these long exposures, the total exposure in each field and filter was built from a series of individual, dithered exposures following the ``step-dither'' pattern described in NGVS-I \S3.5\footnote{This pattern involves taking multiple sequences of single exposures of several fields with small dithers in between each sequence. Typically, 6 to 7 not necessarily adjacent fields are observed in a single, uninterrupted sequence, and the sequence is then repeated between 5 and 11 times, depending on the filter, each time at a slightly different position, shifted from the first by $8\arcsec$ to $18\arcsec$ in right ascension and $12\arcsec$ to $30\arcsec$ in declination --- see NGVS-I Figures 9 and 10.}. Combined with the ``{\sf Elixir-LSB}" reduction procedure developed by the NGVS team (see NGVS-I, \S4.2), this strategy allows us to greatly reduce the systematics associated with scattered light as well as the spatially varying component of the sky: typical residuals in the {\sf Elixir-LSB}-processed, scattered-light-subtracted images are 0.2\% of the sky background in all filters, corresponding to a surface brightness of $\sim29$ mag arcsec$^{-2}$ (NGVS-I, Table 2 and Figure 11). Nevertheless, some systematics are still present in the processed data, most notably, the intra-chip regions ($13\arcsec$ wide between columns and $80\arcsec$ wide between rows) are not imaged at full depth and therefore have lower S/N. Additionally, the halos of bright, saturated foreground stars, as well as scattered light from Galactic cirrus (both of which can reach surface brightness $\leq 27$ mag arcsec$^{-2}$ in the $g$-band) are not removed by the reduction procedure and can compromise the detection of low surface brightness features in the affected areas.  

\begin{figure}
    \centering
    \includegraphics[width=\linewidth]{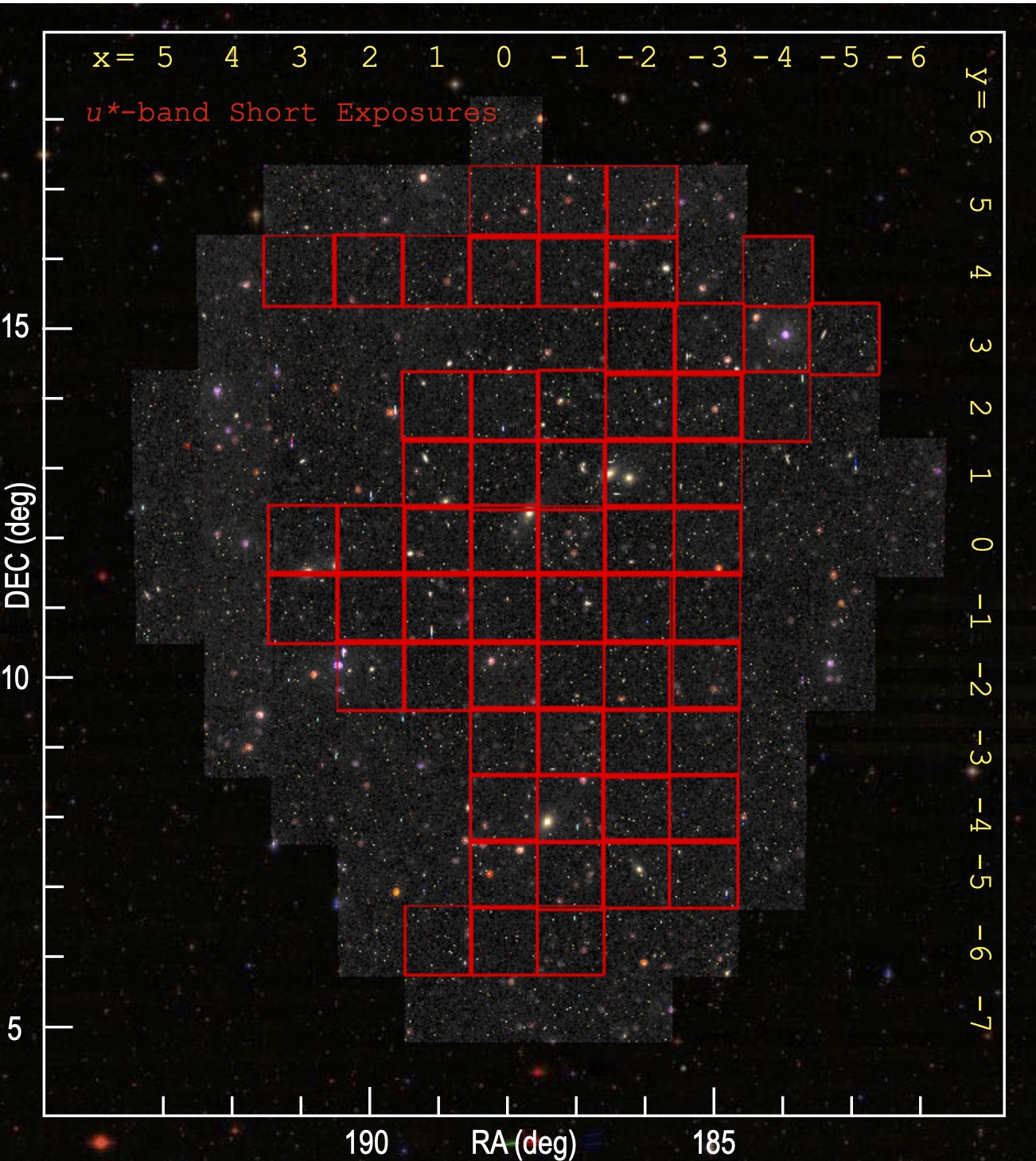}
    \caption{NGVS fields with $u^*$-band short exposures are shown in red. Note that short exposures were obtained for all fields containing galaxies whose cores  saturated in the $u^*$-band long exposures.}
    \label{fig:fig2}
\end{figure}

\begin{figure}
    \centering
    \includegraphics[width=\linewidth]{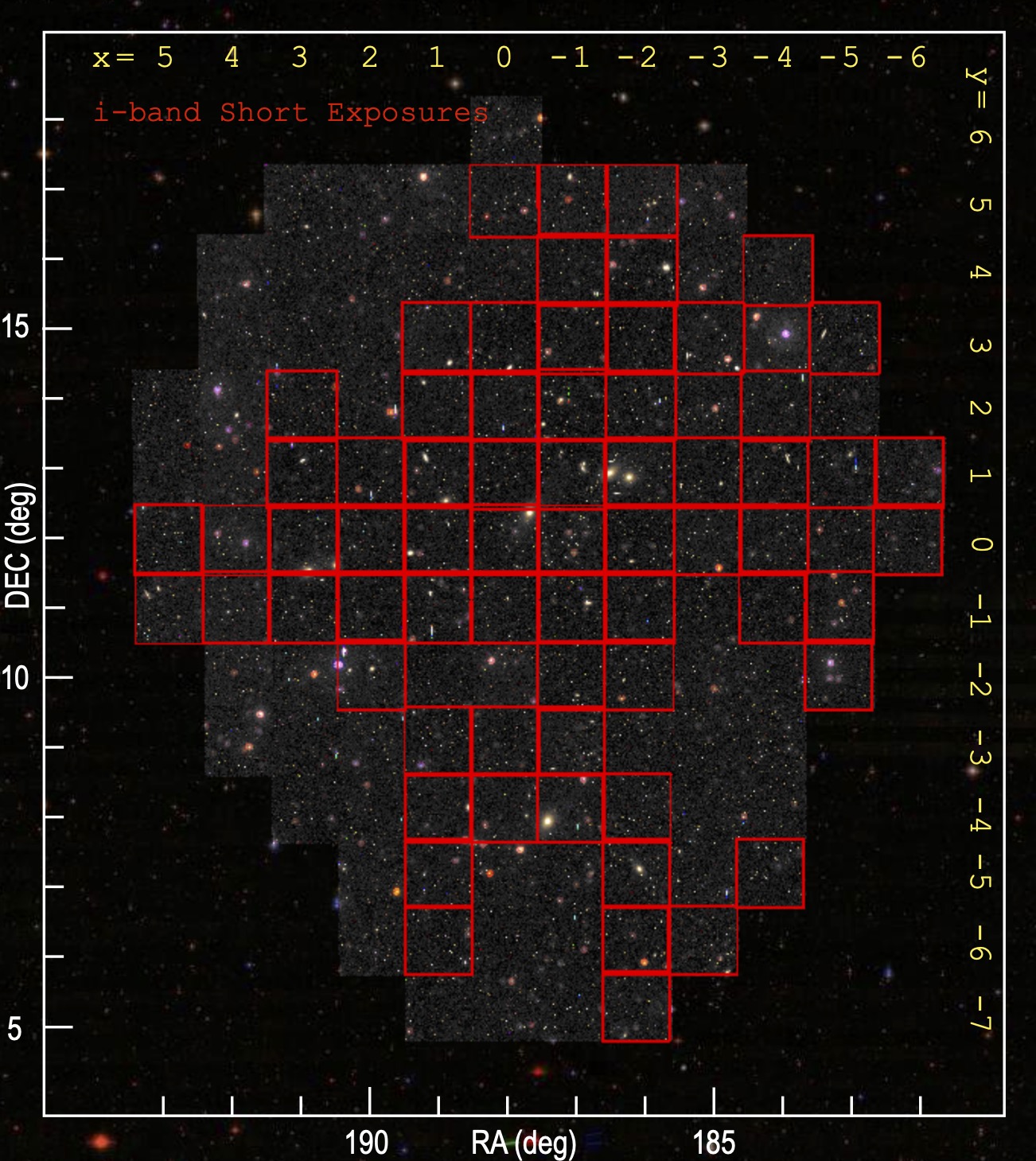}
    \caption{NGVS fields with $i^*$-band short exposures are shown in red. Note that short exposures were obtained for all fields containing galaxies whose cores  saturated in the $i^*$-band long exposures.}
    \label{fig:fig3}
\end{figure}

There is one notable exception to the procedure outlined above. As detailed in NGVS-XIV, \S2, the four fields in the core region surrounding M87 were observed between 2008 March 1 and 2008 March 12 prior to the start of the main survey, using a standard observing strategy that is not conducive to the {\sf Elixir-LSB} processing. To control systematics in the background, the same fields were observed again between 2010 January 13 and 2010 June 10 using the step-dither observing strategy, but with lower image quality (average seeing $\sim 1\farcs3$) than required for the main survey. The analysis of galaxies in this 3.71 deg$^2$ region uses both sets of images --- the lower resolution {\sf Elixir-LSB-}processed images for the study of the extended halos and low surface brightness features, and the higher resolution standard images for the study of the galaxies' inner regions and compact objects, including NSCs, UCDs, and GCs.

In addition to the long exposures, short, dithered exposures were acquired to recover the centers of galaxies that saturate in the long exposures (NGVS-I, \S3.4 and Table 2, as well as Tables \ref{tab:Tab1} and \ref{tab:Tab3} in this paper). These exposures, which did not use the ``step-dither'' pattern adopted for the longer exposures described above, have also proven useful for recovering (albeit at low S/N) areas affected by the halos and diffraction spikes of foreground stars. Although short exposures in the $g$- and $z$-band were acquired for every NGVS field, due to weather/technical losses, only a subset of fields (but all those including galaxies that saturated in the corresponding longer exposures) were observed in the $u^*$- and $i$-band (Figures \ref{fig:fig2} and \ref{fig:fig3}).

Finally, four “background” fields, located three virial radii ($\sim16\deg$) from M87 at Galactic latitudes corresponding to the lower and upper boundaries of the NGVS footprint, were observed with the identical observing strategy used for the long exposures of the NGVS Virgo fields (see NGVS-I, \S3.2). These background fields, which are not expected to contain any Virgo galaxies, are helpful for assessing the reliability of the detection algorithm used to identify Virgo members and to provide a statistical assessment of the purity and contamination of the final Virgo galaxy catalog (see \S\ref{subsec:VCC}).

\section{Catalog of Virgo Cluster Galaxies} \label{sec:cat}

\subsection{Summary of the Detection Procedure: {\sf VCands}} \label{sec:vcands}
In order to identify Virgo galaxies, and specifically faint, low surface brightness dwarfs, the NGVS team developed a dedicated algorithm, {\sf VCands}, described in detail in NGVS-XIV, \S 3. {\sf VCands} was extensively tested and optimized using a training set of 393 galaxies, independently identified by three NGVS team members in the core region. This sample includes 253 already cataloged Virgo members, as well as 140 previously unidentified galaxies. Given the extreme range in structural and photometric parameters spanned by Virgo galaxies, {\sf VCands} was optimized to detect galaxies fainter than $g \sim 16$ mag; presumably all galaxies brighter than this limit are already cataloged (and, with few exceptions, spectroscopically confirmed) Virgo members.

While we refer to NGVS-XIV for the details, the salient features of {\sf VCands} are as follows. {\sf VCands} is composed of several modules (see NGVS-XIV, Figure 2). The first module is based on an optimized {\sf SExtractor} run \citep{Bertin+Arnouts96} performed on NGVS images, to which a mask and then a circular ring filter have been applied. This pre-processing is crucial not only to reduce (by over an order of magnitude) the number of spurious detections but also, and most importantly, to detect diffuse, low surface brightness galaxies that would elude a standard {\sf SExtractor} run on unprocessed images (see NGVS-XIV \S 3.2 and Figures 3 and 4). This initial step produces nearly two million detections across the entire 104 deg$^2$ NGVS footprint. For reference, at the NGVS depth, and based on reasonable assumptions for the faint-end slope of the luminosity function, the number of Virgo galaxies is expected to be in the range of a few to several thousand: it is therefore clear that, at this stage of the procedure, the overwhelming majority of the detected objects are foreground stars or background galaxies.

The second {\sf VCands} module is therefore designed to make a first attempt at reducing the number of contaminants, and is based on the diagnostic plots discussed in NGVS-XIV \S 3.2 and Figures 5 to 7. These plots employ a combination of several of the parameters measured during the {\sf SExtractor} run carried out by the first module, namely: the area subtended by each detection (to be specific, the area within the isophote corresponding to 2.4$\sigma$ of the background RMS), and the fluxes at the peak, at the 2.4$\sigma$ isophote mentioned above, and averaged between the two. In these diagnostic plots, extended objects segregate into different areas according to their concentration (NGVS-XIV \S 3.2). Conservative cuts are then applied to reduce the number of contaminants while ensuring that none of the Virgo galaxies in the training set is rejected; these cuts reduce the number of contaminants by slightly more than 40\%.

The remaining $\sim1.1$ million objects are then further processed using the third module of {\sf VCand}. This involves performing, for each detection,  2-D parametric fits using a combination of {\sf GalFit} \citep{Peng+02} and GALAPAGOS \citep{Barden+12} ---  a very CPU-intensive process carried out on the cloud-based Canadian Advanced Network for Astronomical Research (CANFAR). {\sf GalFit} allows us to measure, albeit crudely, structural and photometric parameters for all objects, and these, in turn, can be used to optimally define a ``membership probability", $\wp$, as a combination of photometric and structural parameters, photometric redshifts (measured using {\sf LePhare}; \citealt{Arnouts+99, Ilbert+06}), and parameters designed to assess the likelihood that a detection might be compromised by the presence of nearby bright stars (see Equation 2 of NGVS-XIV). $\wp$, which ranges between 0 and 1, is not a probability in the strict sense. It is  empirically defined so that objects in our training set have high $\wp$ (generally above 0.5), while background/foreground contaminants have low $\wp$ (generally below 0.5). We found that eliminating all objects with $\wp < 0.22$ reduces the number of detections by $\sim 93\%$, without excising known Virgo member galaxies. 

This leads to a final sample of $\sim 11,000$ objects. These, plus several hundred additional objects for which the {\sf GalFit} fit failed and therefore $\wp$ could not be computed,  were then inspected by eye to assess membership -- a task that, in our experience, cannot be automated. For every detection, three NGVS team members independently inspected FITS images as well as color images, without any prior knowledge about the object (e.g., existing classifications, spectroscopic redshifts, etc.). Each galaxy was assigned a ``membership class" of 0, 1, 2, or 4, according to whether it was judged to be a certain Virgo cluster member (0), a likely member (1), a possible member (2), or a background object (4). There was excellent agreement among the classification performed by each individual, and a consensus was reached for the objects for which the classification differed.  

Based on the steps described above, 2847 galaxies (about a quarter of the $\sim 11,000$ galaxies that survived the last automated step in {\sf VCands}) passed visual inspection and were classified as {\it bona fide} Virgo member candidates. When visually inspecting the images of {\sf VCands} objects with $\wp > 0.22$, the cutouts were kept purposely significantly larger than the galaxy being targeted, so that the surrounding environment could also be inspected. This led to the identification of several hundred additional galaxies that were deemed to be members of Virgo:
\begin{itemize}
\item 493 galaxies (13\% of the final sample) brighter than $g \sim 16$ mag that were not, by design, analyzed by {\sf VCands}. We stress that these galaxies were identified during the final visual inspection, i.e., they were not simply added to the final catalog based on an already existing identification. Indeed, for instance, 13 of these galaxies do not have a Virgo Cluster Catalog \citep[VCC,][]{Binggeli+85} classification;
\item 
349 galaxies (9\% of the final sample) fainter than $g \sim 16$ mag, again identified while visually inspecting the {\sf VCands} candidates. Further investigation showed that all of these galaxies were detected by {\sf VCands} but had $\wp < 0.22$ and were therefore not directly subjected to visual inspection. 
\end{itemize}
At this stage, therefore, the NGVS catalog comprises 2847+493+349=3689 galaxies, of which 1691 (46\%) classified as certain members (Membership Class 0), 808 (22\%) as likely members (Membership Class 1), 1190 (32\%) as possible members (Membership Class 2). This classification was slightly changed following the sanity checks described in \S\ref{subsec:spec} and \S\ref{subsec:EVCC}, and a final count will be presented in \S \ref{subsec:cat}.

\subsection{Are You Out? The Spectroscopic Catalog} \label{subsec:spec}

\begin{figure*}
    \centering
    \includegraphics[width=1\textwidth]{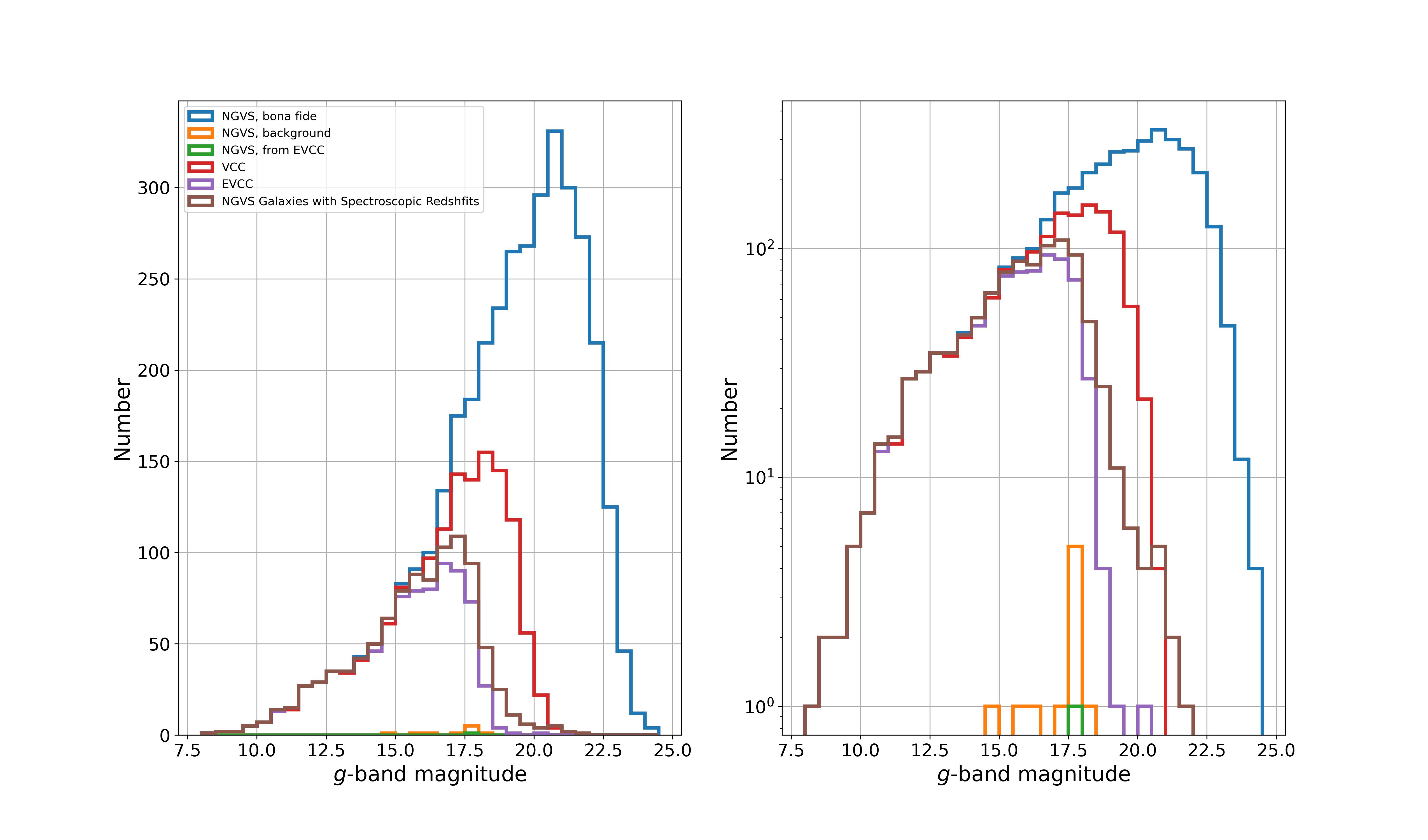}
    \caption{Histograms showing the $g$-band magnitude distributions of all NGVS galaxies. The blue and orange histograms show, respectively, {\it bona fide} members and galaxies found, after the fact, to have velocities that place them beyond the $v = 3,500$ km s$^{-1}$ boundary adopted for Virgo. The green histogram shows the one galaxy, EVCC1010, identified as a Virgo member in the EVCC, but not originally included in the NGVS catalog. The red and violet histograms show NGVS galaxies identified in the VCC and in the EVCC, respectively. Finally, the brown histogram show NGVS galaxies with a measured recessional velocity. The panel to the right is plotted on a logarithmic y-axis to better highlight small samples.}
    \label{fig:fig4}
\end{figure*}

After the NGVS catalog was assembled, a search for radial velocity measurements for all galaxies in the catalog was performed. Heliocentric radial velocity measurements were retrieved primarily through a search of the NASA/IPAC Extragalactic Database (NED). Insofar as our targets are concerned, most of the entries in NED are from the Third Reference catalog of Bright Galaxies (RC3), the Sloan Digital Sky Survey (SDSS, mostly Data Release 13, DR13; \citealt{Albareti+17}), and the Arecibo Legacy Fast Alfa Survey \citep{Kent+08}, although other sources are occasionally available. All objects within $5\arcsec$ of each galaxy's NGVS coordinates were retrieved, and from these all objects found to be within 0\farcs01 of an object in the GAIA catalog (and thus presumed to be a star), as well as all duplicate entries were removed. Finally, for objects with multiple measurements differing by more than 500 km s$^{-1}$, the literature references were reviewed to resolve the discrepancy. 

The resulting velocity catalog was supplemented with radial velocity measurements from our own Keck, MMT, AAT, Hectospec, and HST programs, carried out over several observing seasons, to produce the final catalog of radial velocities. When multiple velocity measurements exist for the same galaxy, we computed their average, with error equal to the standard deviation of the mean --- given the inhomogeneity of the sample, we opted not to compute weighted averages. 

\begin{figure*}
    \centering
    \includegraphics[width=1\textwidth]{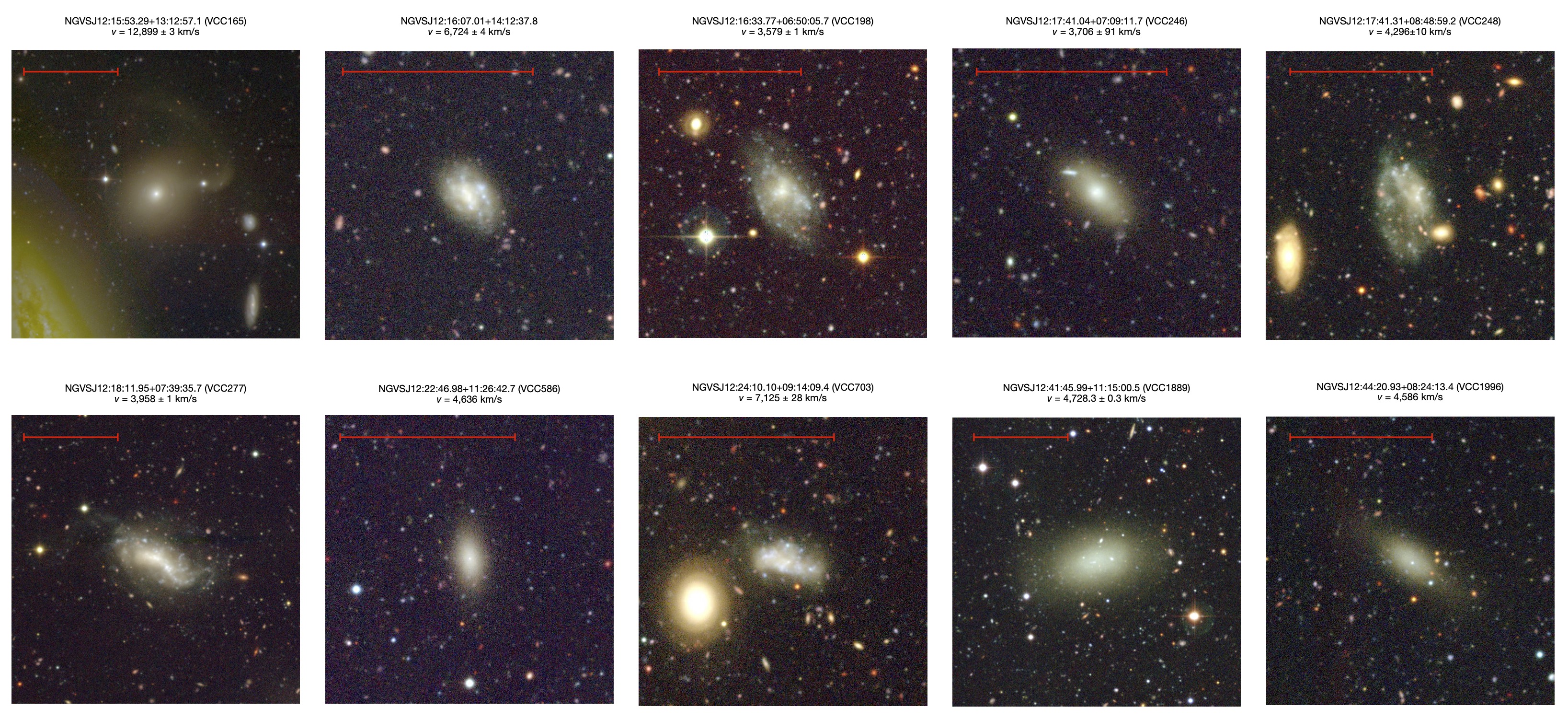}
    \caption{$u^*gi$ color images for the 10 galaxies selected as \textit{bona fide} Virgo member candidates based on the NGVS criteria, but subsequently found to have radial velocity exceeding the 3,500 km s$^{-1}$ boundary adopted for Virgo. These galaxies are retained in the catalog but labeled as background sources (Membership Class 4). The red bar in each image corresponds to 1 arcminute (see \S \ref{subsec:spec} for further details)}
    \label{fig:fig5}
\end{figure*}

Of the 3689 NGVS galaxies, 996 (27\% of the sample) have a measured heliocentric radial velocity. Figure \ref{fig:fig4} shows the magnitude distribution of these galaxies compared to the entire NGVS sample ($g$-band magnitudes are derived as described in \S\ref{subsec:typhon} and \S\ref{subsec:galfit}). For ten galaxies (0.2\% of the sample), the measured radial velocity places the galaxy beyond our adopted upper limit of $v=3,500$ km s$^{-1}$ for Virgo membership; for completeness, these galaxies are retained in the NGVS catalog, but labeled as ``background" sources (Membership Class 4) even though their true nature is unclear in some cases, as we discuss below. Finding charts for all 10 galaxies can be found in Figure \ref{fig:fig5}. In more detail:

\begin{itemize}
\item NGVS J121553.29+131257.1 (NGVS183, VCC165), originally classified as a ``certain" member in both the NGVS and the VCC, is a large early-type galaxy with a clear system of shells and a stellar stream. SDSS DR13 reports a radial velocity of $v=12,899 \pm 3$ km s$^{-1}$, although values of 255 km s$^{-1}$ are attributed to various sources in NED. We were unable to track down the original source for the lower value, and therefore adopt the DR13 measurement, even if the galaxy appearance is difficult to reconcile with a distance of almost 200 Mpc. We also note that the galaxy appears to be an interacting member of a group containing NGVS185 (VCC167), NGVS218 (VCC187), and NGVS228 (VCC200), all of which are spectroscopically confirmed Virgo members. 

\item NGVS J121607.01+141237.8 (NGVS197), originally classified as a ``possible" member in the NGVS, is a compact, star-forming irregular galaxy. The velocity of $v=6,724 \pm 4$ km s$^{-1}$ is from SDSS DR13. 

\item NGVS J121633.77+065005.7 (NGVS229, VCC198), originally classified as a ``certain" member in the NGVS (and a ``probable" member in the VCC) is an irregular, star-forming galaxy with a hint of spiral structure. The measured radial velocity $v=3,579 \pm 2$ km s$^{-1}$ is from HI data \citep{Haynes+18} and is confirmed by SDSS DR13.  

\item NGVS J121741.04+070911.7 (NGVS294, VCC246), is a regular early-type galaxy, originally classified as a ``certain" member in the NGVS, and as a ``probable" member in the VCC. The SDSS DR13 radial velocity of $v=135,508 \pm 82$ km s$^{-1}$ must be in error, and indeed SDSS DR4 and DR5, as well as the EVCC, report the velocity as 3,706 $\pm$ 91 km s$^{-1}$, which we adopt here. 

\item NGVS J121741.31+084859.2 (NGVS295, VCC248) originally classified as a ``likely" member in the NGVS and as a ``probable" member in the VCC, is a star-forming, irregular galaxy. It has a single velocity measurement, $v=4,296 \pm 10$ km s$^{-1}$, from the RC3 (reportedly from HI observations, although the exact source is unclear).

\item NGVS J121811.95+073935.7 (NGVS327, VCC277) is morphologically very similar to NGVS229 (VCC198) and, like NGVS229, was classified as a ``certain" member in the NGVS and as a ``probable" member in the VCC. The radial velocity of $v=3,958 \pm 1$ km s$^{-1}$ is from SDSS DR13.  

\item NGVS J122246.98+112642.7 (NGVS742, VCC586) is a regular, compact, nucleated low surface brightness galaxy, classified as a ``certain" member in both the NGVS and the VCC. The 49,417 $\pm$ 41 km s$^{-1}$ radial velocity from SDSS DR13 must be in error; other sources, including SDSS DR4 and the EVCC, place the galaxy at 4,636 km s$^{-1}$.

\item NGVS J122410.10+091409.4 (NGVS954, VCC703), originally classified as a ``likely" member in the NGVS and as a ``member" in the VCC, is a star-forming, irregular galaxy. The radial velocity of $v=7,125 \pm 28$ km s$^{-1}$ is from HI data \citep{Haynes+18}.

\item NGVS J124145.99+111500.5 (NGVS3277, VCC1889), originally classified as a ``certain" member in both the NGVS and the VCC, is a large early-type galaxy with a hint of shells and/or spiral structure. The RC3 \citep{devaucouleurs+91} reports a radial velocity of $v=4,728.3 \pm 0.3$ km s$^{-1}$, although the original source is unclear. 

\item NGVS J124420.93+082413.4 (NGVS3432, VCC1996) is a smooth, very low surface brightness galaxy, originally classified as a ``possible" member in the NGVS, and  as a ``members" in the VCC. The 172,894 km s$^{-1}$ radial velocity from SDSS DR13 must be in error. Other sources, including the EVCC, place the galaxy at 4,586 km s$^{-1}$.  

\end{itemize}

\begin{figure*}
    \centering
    \includegraphics[width=1.0\textwidth]{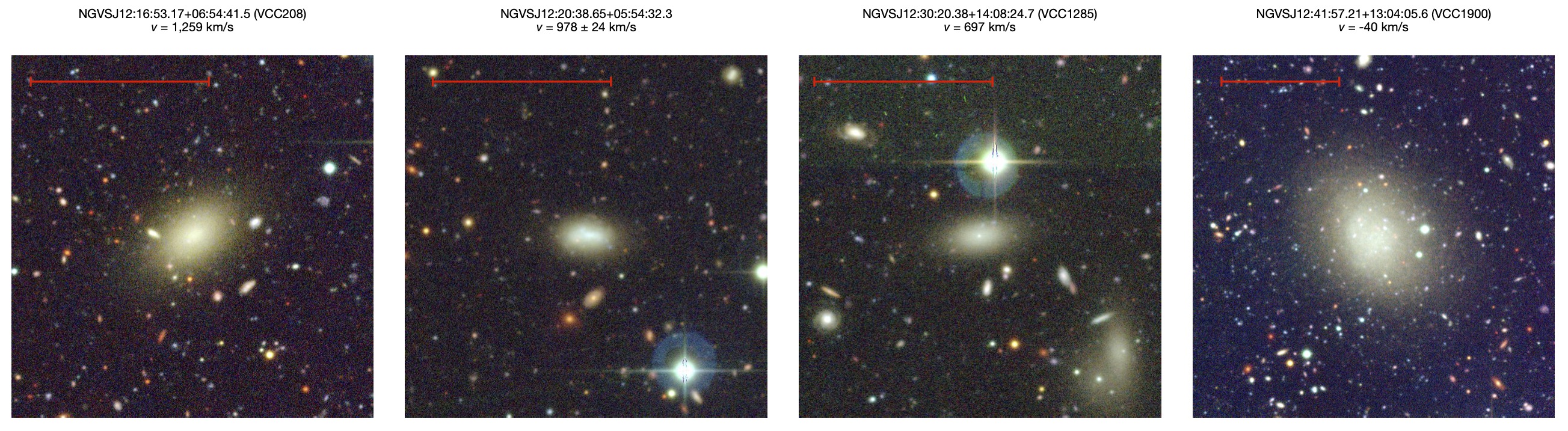}
    \caption{$u^*gi$ color images for the four galaxies selected as \textit{bona fide} Virgo member candidates based on the NGVS criteria, and retained as such despite the fact that for all, at least one measured velocity is above the 3,500 km s$^{-1}$ boundary adopted for Virgo. The red bar in each image corresponds to 1 arcminute (see \S \ref{subsec:spec} for further details)}
    \label{fig:fig6}
\end{figure*}

\begin{figure}
    \centering
    \includegraphics[width=1.0\linewidth]{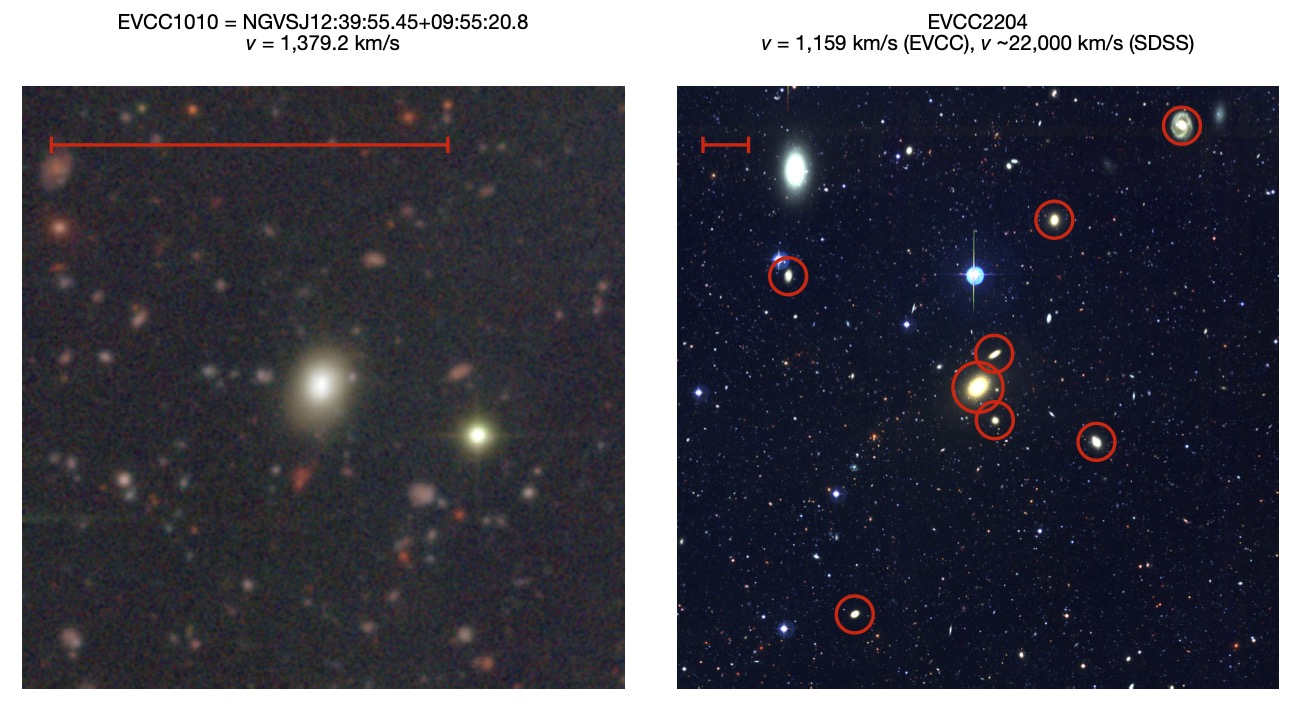}
    \caption{$u^*gi$ color images for two galaxies for which the EVCC reports a velocity within 3,000 km s$^{-1}$, but not originally included in the NGVS catalog. The red bar in each image corresponds to 1 arcminute. The red circles in the panel to the right identify galaxies for which  SDSS DR13 reports a velocity of $\sim 22,000$ km s$^{-1}$. We conclude that EVCC2204 is part of a background cluster and therefore do not include it in the final NGVS catalog. See \S \ref{subsec:EVCC} for further details}
    \label{fig:fig7}
\end{figure}

We note that four of these galaxies (NGVS197, NGVS229, NGVS295, and NGVS954) show signs of star formation activity in the form of clumpy, blue associations. With the exception of NGVS229, all are imaged by VESTIGE, although not always at full depth, yet no H$\alpha$ emission was detected. The measured radial velocities for these galaxies (4,000 to 7,100 km s$^{-1}$) would shift the H$\alpha$ emission beyond the 3,500 km s$^{-1}$ upper cutoff of the transmission curve of the narrow-band filter employed by VESTIGE, supporting the conclusion that these are indeed background objects (although only marginally). 

Finally, the four galaxies shown in Figure \ref{fig:fig6} have conflicting velocity measurements. Only two (NGVS515 and NGVS3289) show some low level of star formation, and although neither was detected in H$\alpha$ by VESTIGE, NGVS515 is in a region of very limited H$\alpha$ depth, and NGVS3289 does not show signs of the very recent \citep[$< 10$ Myr, e.g.,][]{Kennicutt+98, Boselli+09} star formation required to produce H$\alpha$ emission. For all four of these galaxies, we therefore assign more credence to the velocity measurement that is consistent with Virgo membership, and therefore they are retained as Virgo members in the NGVS catalog. In more detail:

\begin{itemize}

\item NGVS J121653.17+065441.5 (NGVS252, VCC208), is a regular, quiescent early-type galaxy, classified as a ``certain" member in the NGVS, and as a ``probable" member in the VCC. SDSS DR4 and DR5 report a velocity of $\sim 7,700$ km s$^{-1}$, however SDSS DR13 reports a velocity of 1,259 km s$^{-1}$. We adopt this latest value, but note the discrepancy.

\item NGVS J122038.65+055432.3 (NGVS515), classified as a ``likely" member, is a relatively low surface brightness, irregular galaxy with low level star-forming regions and a hint of shells and filaments. The 214,690 $\pm$ 7 km s$^{-1}$ radial velocity from SDSS DR13 must be in error. Other sources place the galaxy between 900 and 1,000 km s$^{-1}$. We adopt a radial velocity of 978 $\pm$ 24 km s$^{-1}$ from HI observations \citep{Haynes+18}.

\item NGVS J123020.38+140824.7 (NGVS2034, VCC1285), classified as a ``likely" member in the NGVS, is a regular, compact, quiescent, nucleated low surface brightness galaxy, classified as a ``member" in the VCC. The 94,193 $\pm$ 47 km s$^{-1}$ radial velocity from SDSS DR13 must be in error. Several other sources \citep{Rines+08,Castignani+22}, place the galaxy at 697 km s$^{-1}$, which we adopt here. 

\item NGVS J124157.21+130405.6 (NGVS3289, VCC1900) is classified as a ``likely" member in the NGVS and as a ``member" in the VCC. SDSS DR13 places the galaxy at 10,289 $\pm$ 9 km s$^{-1}$. The galaxy is a relatively large, relatively low surface brightness galaxy with a disturbed morphology and low level of star formation activity, and the DR13 velocity appears too large given the galaxy appearance. NED reports a velocity of $-$40 km s$^{-1}$ from \citet{Rines+08} which we adopt, although the original source is unclear. 
\end{itemize}

Finally, we note that of the NGVS {\it bona fide} Virgo members, 133 show signs of active star formation (see \S\ref{subsec:morph}) and are within the VESTIGE footprint. Of these, 126 are detected in H$\alpha$, strongly supporting their cluster membership. Of the nine galaxies not detected by VESTIGE, three (NGVS1319 == VCC124, NGVS1663, and NGVS137 == VCC124) show presence of low level star formation in the form of blue associations, while the others show diffuse blue colors but no evidence of the very recent ($< 10$ Myr) star formation expected to produce strong emission lines, suggesting a recent quenching of their activity \citep[e.g.,][]{Kennicutt+98, Boselli+09}. Of the three galaxies listed above, the first two do not have a measured velocity, while NGVS137 has a velocity of 2,085 km s$^{-1}$ \citep[from the RC3,][]{devaucouleurs+91}. In assessing membership for these three galaxies, the lack of an H$\alpha$ detection is not conclusive, as it could be due either to the fact that star formation has ceased more than a few tens of Myrs ago, or that the radial velocity is beyond 3,500 km s$^{-1}$. In either case, these observations imply an upper limit of  $< 2$\% in the number of active NGVS galaxies that might have been misclassified as Virgo members.

\subsection{Are You In? The EVCC}
\label{subsec:EVCC}

The spectroscopic catalog discussed in \S\ref{subsec:spec} can tell us whether a galaxy deemed to be a Virgo member based on the NGVS analysis is in reality in the background of the cluster. It cannot, however, tell us whether any spectroscopically confirmed Virgo members were {\it not} included in the NGVS catalog in the first place. For this, we turn to the Extended Virgo Cluster Catalog \citep[EVCC;][]{Kim+14} which includes all objects that are spectroscopically confirmed Virgo members (defined as having $v < 3,000$ km s$^{-1}$), based mostly, but not exclusively, on radial velocities from SDSS DR7. Of the 1589 galaxies listed in the EVCC, 850 are within the NGVS footprint (see Figure \ref{fig:fig4}), and all but two are classified independently as members in the NGVS catalog. The two exceptions are EVCC1010 and EVCC2204 (see Figure \ref{fig:fig7}). EVCC1010 is a very compact, high surface brightness, isolated early-type galaxy. Its velocity is reported as 1,379.2 km s$^{-1}$ from SDSS DR7. 

EVCC2204 is a high surface brightness early-type galaxy, located at the center of a small group of several galaxies that are clearly in the background of Virgo. A velocity of 1,159 km s$^{-1}$ is available from \citet{Rines+08}. However, several other independent sources, including 2MASS and SDSS DR13, report a velocity of 20,000 to 21,000 km~s$^{-1}$; additionally, several galaxies within 15 arcmin of EVCC2204 have a velocity from SDSS DR13 of $\sim$ 22,000 km~s$^{-1}$ (see Figure \ref{fig:fig7}). We take this as evidence that EVCC2204 is part of a small background group, and not a {\it bona fide} Virgo member. 

Based on this, we include EVCC1010, but not EVCC2204, in the NGVS catalog\footnote{As will be mentioned in the next section, each galaxy is given a ``Nickname" equal to ``NGVS" followed by a sequential integer ordered by increasing Right Ascention. Since the Nickname had already been established and used in NGVS publications prior to the addition of EVCC1010 (NGVS J123955.45+095520.8) to the catalog, the galaxy nickname, NGVS3690, is out of sequence.}.

Finally, we note that VESTIGE detected five objects that are not included in the NGVS catalog \citep[Boselli, private comm.,][]{Bellazzini+18}. In the broad-band NGVS images these objects are, with no exception, bright blue point sources surrounded by very small (typical total extent 2\arcsec~ to 3\arcsec), faint, irregular envelopes, and are morphologically indistinguishable from several hundreds galaxies within the NGVS footprint that are classified as being in the background of Virgo (and are not detected by VESTIGE). Work is underway to establish whether these five VESTIGE detections are indeed HII regions at the distance of Virgo, or background emission-line galaxies for which a random emission line happens to fall within the passband of VESTIGE's H$\alpha$ filter. Given these uncertainties, these objects are not included in the present work.”

\subsection{The Final Catalog of Virgo Galaxies}
\label{subsec:cat}

The final galaxy catalog, therefore, comprises 3680 {\it bona fide} Virgo members (3679 plus EVCC1010) and 10 galaxies originally classified as members but found to be in the background based on the spectroscopic information presented in \S \ref{subsec:spec}. Of the {\it bona fide} members,  1690 (46\%) are classified as certain members (Membership Class 0), 802 (22\%) as likely members (Membership Class 1), and 1188 (32\%) as possible members (Membership Class 2). 

\begin{figure*}
    \centering
    \includegraphics[width=0.9\textwidth]{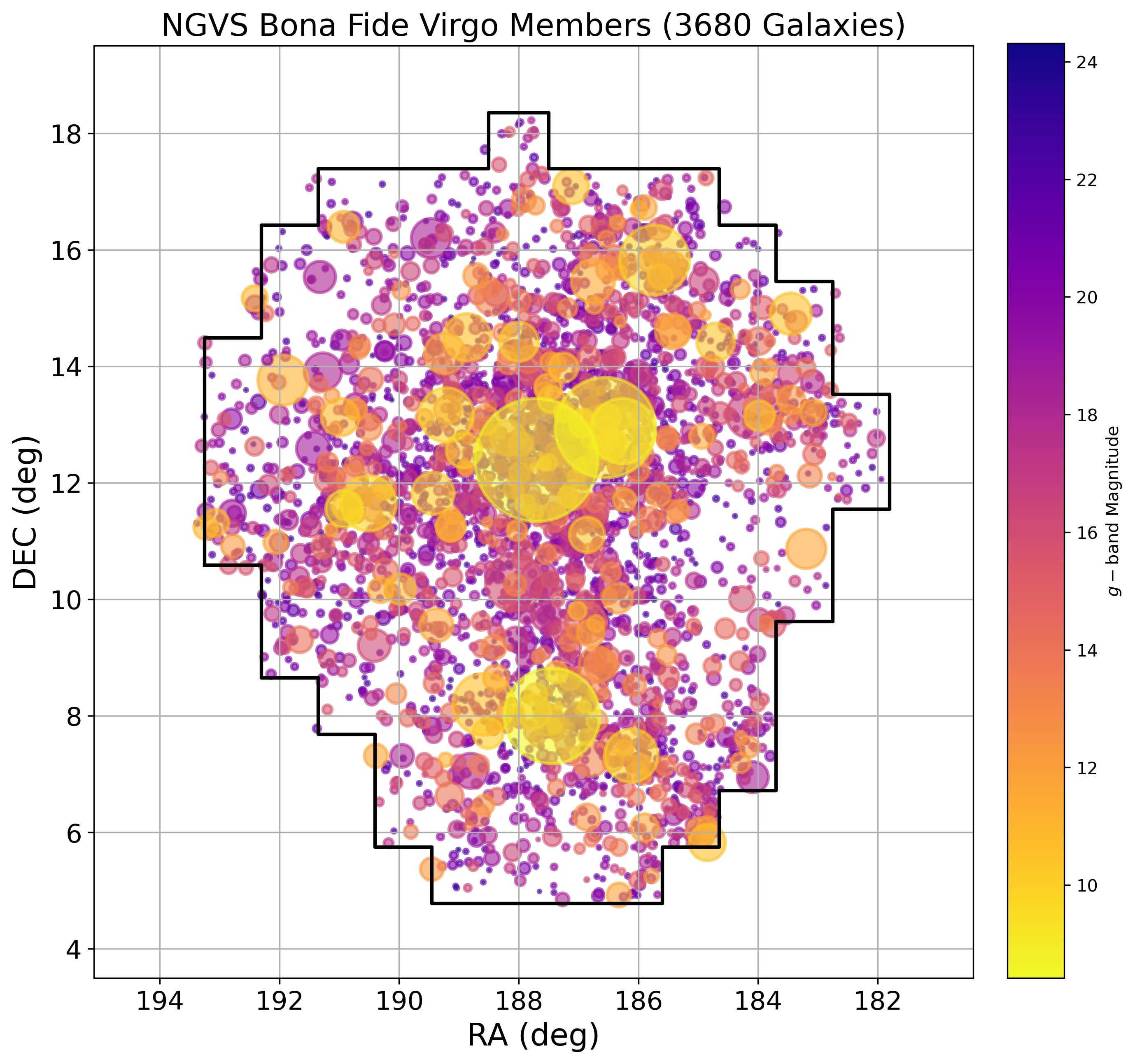}
    \caption{The location of the 3680 NGVS galaxies classified as certain, likely, or possible members of the Virgo cluster based on the NGVS analysis (the 10 galaxies found, after the fact, to be beyond the 3,500 km s$^{-1}$ boundary for Virgo are not included). Galaxies are plotted with sizes proportional to their geometric effective radius (elevated to the 0.8 power for ease of displaying) and are color-coded according to their $g$-band magnitude as shown in the bar to the right (see Table \ref{tab:Tab8}). The black outline shows the NGVS footprint.}
    \label{fig:fig8}
\end{figure*}

\begin{figure*}
    \centering
    \includegraphics[width=0.9\textwidth]{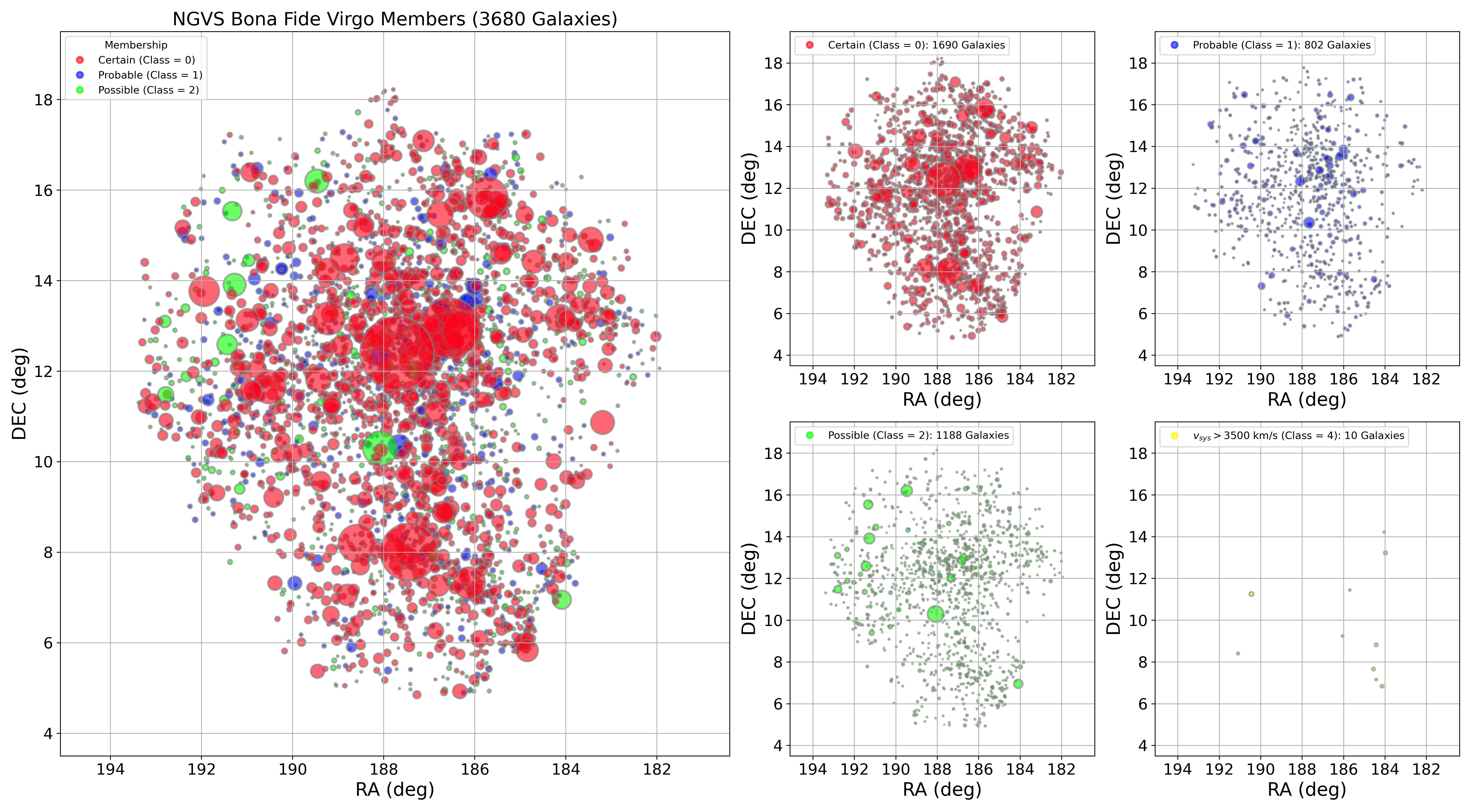}
    \caption{As Figure~\ref{fig:fig8}, but with galaxies color-coded according to their Membership Class (as listed in Table~\ref{tab:Tab5}). Certain (Membership Class = 0), likely (Membership Class = 1), and possible (Membership Class = 2) members are shown separately in the three smaller panels to the right. The fourth small panel on the bottom right shows the 10 galaxies classified as Virgo members based on our photometric/structural criteria, but subsequently shown to have velocities above 3,500 km~s$^{-2}$.}
    \label{fig:fig9}
\end{figure*}

\begin{figure}
    \centering
    \includegraphics[width=\linewidth]{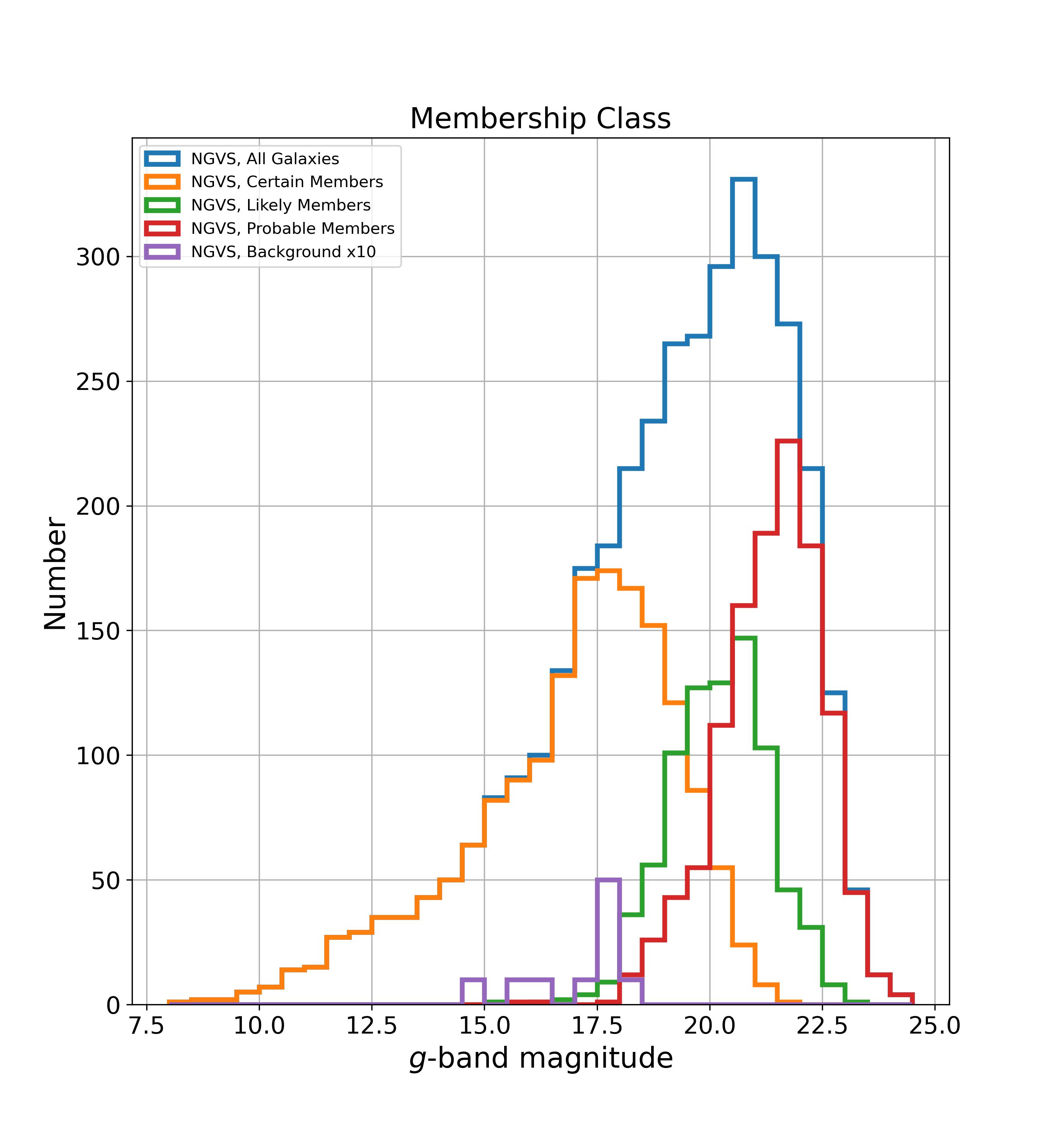}
    \caption{Histogram showing the $g$-band magnitude distribution of NGVS galaxies classified as certain (Membership Class 0), likely (Membership Class 1) and possible (Membership Class 2) members, as well as galaxies found, after the fact, to have velocities that place them beyond the $v = 3,500$ km s$^{-1}$ boundary adopted for Virgo. Note that the latter histogram has been multiplied by a factor 10 to make it easier to see it.}
    \label{fig:fig10}
\end{figure}

\begin{figure*}
    \centering
    \includegraphics[width=0.45\textwidth]{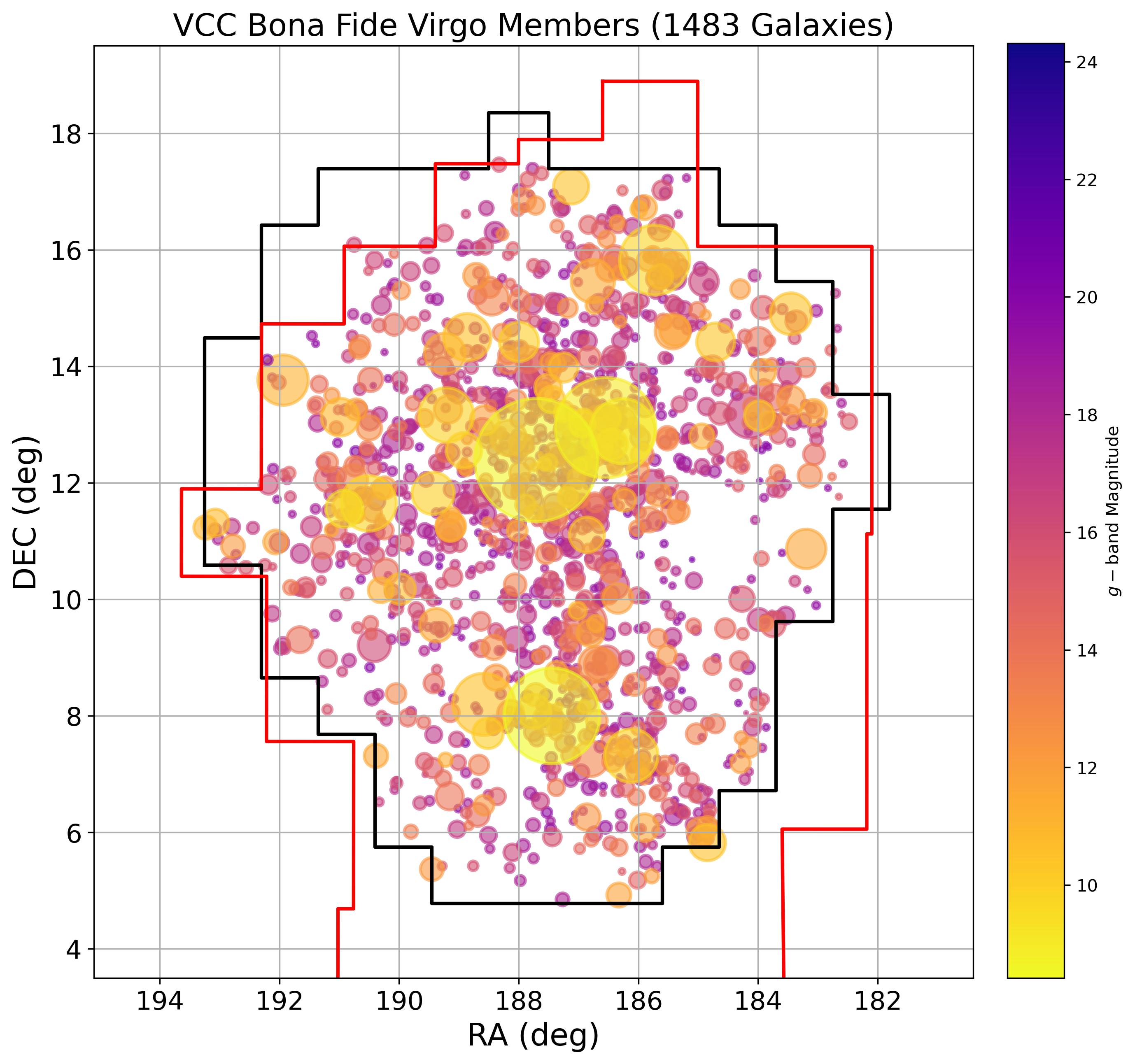}
    \includegraphics[width=0.45\textwidth]{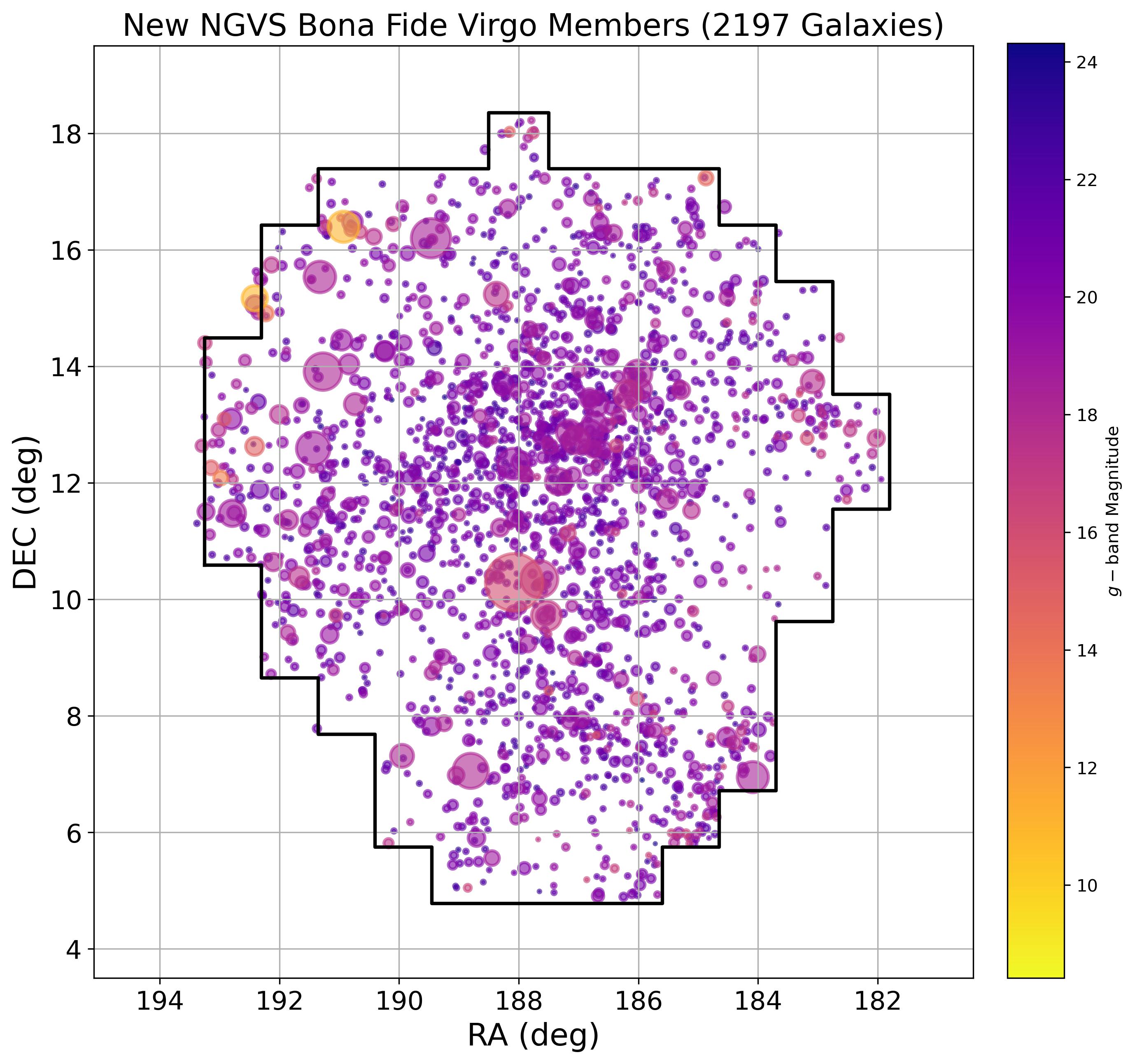}
    \caption{{\it Left panel}: as Figure \ref{fig:fig8}, but showing only the 1483 {\it bona fide} Virgo galaxies in the NGVS footprint (black outline) also cataloged in the VCC (see Table \ref{tab:Tab5}), the footprint of which is shown by the red outline. {\it Right panel}: as Figure \ref{fig:fig4}, but showing only the 2199 galaxies identified in the NGVS and not in the VCC.}
    \label{fig:fig11}
\end{figure*}

\begin{figure*}
    \centering
    \includegraphics[width=1.0\textwidth]{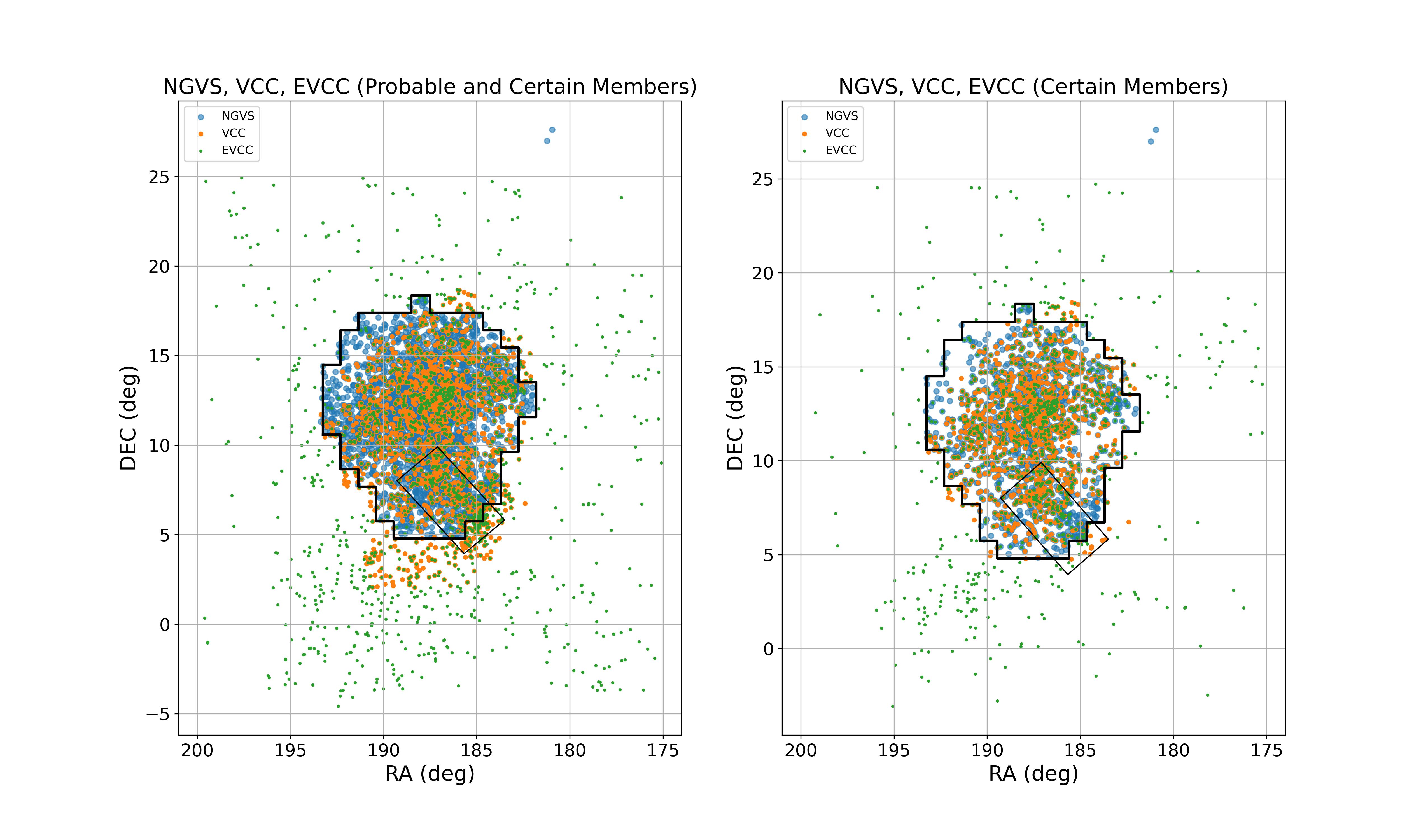}
    \caption{A wider view of Virgo, showing all galaxies identified in the VCC (orange dots), EVCC (green dots), and NGVS (blue dots). The left panel shows all galaxies, while the right panel shows galaxies whose membership is considered certain. The larger black outline shows the NGVS footprint, while the smaller rectangular outline shows the footprint of the Vera C. Rubin Observatory's early-release observations that targeted Virgo.}
    \label{fig:fig12}
\end{figure*}

The location of the 3680 galaxies deemed to be {\it bona fide} Virgo cluster member candidates (membership classes 0, 1, and 2) is shown in Figure \ref{fig:fig8}. Galaxies are color-coded according to their $g$-band magnitude and plotted with size proportional to their effective radius (both derived as described in \S \ref{subsec:typhon} and \S \ref{subsec:galfit}). Figure~\ref{fig:fig9} shows the spatial distribution of NGVS members, this time color-coded according to their membership class, while a histogram showing the distribution of the different membership classes as a function of the galaxies' $g$-band magnitude is given in Figure \ref{fig:fig10}. Figure \ref{fig:fig11} compares the spatial distribution of NGVS galaxies cataloged in the VCC (left panel) to the distribution of NGVS galaxies not identified in the VCC (right panel). As can be seen, the VCC is essentially complete to a $g$-band magnitude of $\sim 16.5$, 50\% complete at $g \sim 18.8$ mag, and essentially fully incomplete beyond $g \sim 20$ mag. Figure \ref{fig:fig12} shows a wider view of the part of the sky occupied by the Virgo cluster, showing the distribution of all NGVS, VCC and EVCC galaxies. 

General properties for all galaxies are tabulated in Tables~\ref{tab:Tab5} and \ref{tab:Tab6}, and brief notes are reported in Appendix \ref{app:notes}. Table~\ref{tab:Tab5} lists the galaxy Nickname (which will be used in all subsequent Tables for convenience) and Official Name, the Right Ascension and Declination, the Membership Class, the VCC number (when available), the spectroscopic information (updated as of July 2025, see \S\ref{subsec:spec}), and the SBF distance modulus from \citet[][]{Cantiello+24}. Table~\ref{tab:Tab6} lists the Membership Class, the morphological and nuclear classification for all NGVS galaxies (see \S \ref{subsec:nuclei} and \S \ref{subsec:morph}), the parametric model used in fitting the light distribution (see \S\ref{subsec:typhon} and \S\ref{subsec:galfit}), the FWHM of the PSF at the location of the galaxy's center (see \S 4, step 4, of NGVS-XIV), and two estimates of the galaxy surface density, the first defined as the number of galaxies per square degree within a circular area centered on the galaxy in question and with radius equal to the distance to the 10th nearest neighbor, and the second including only galaxies with an absolute $g$-band magnitude brighter than $M_g = -13.5$ mag when counting neighboring galaxies. Notes are provided in the footnotes to the Tables, and additional information on the entries in this and subsequent Tables is collected in \S \ref{sec:fin}.

\section{Completeness and Purity of the NGVS sample} \label{sec:bias}

\subsection{Artificial Galaxy Simulations} \label{subsec:art}

A rigorous test of the completeness of the NGVS sample was carried out using an extensive set of simulations, described in detail in NGVS-XIV, \S4. In total, 182,500 galaxies (36,500 in each filter) were injected at random locations within the core region. The surface brightness profile was assumed to follow a S\'ersic law with parameters (magnitudes, effective radii, and S\'ersic indexes) randomly generated to span a range encompassing that expected for Virgo members (see NGVS-XIV, \S4 for further details). 

{\sf VCands}'s recovery rate for simulated galaxies plateaus at $\sim$90\% for galaxies brighter than $g \sim 19$ mag, and declines gradually for fainter galaxies (however, we will argue in \S\ref{subsec:biasconc} that the actual recovery rate is  higher than the artificial galaxy test seems to indicate). The decline is more abrupt in surface brightness than total magnitude; in the $g$-band, essentially all galaxies become undetectable once the average surface brightness within one effective radius becomes fainter than 29 mag arcsec$^{-2}$. We refer the reader to NGVS-XIV for details and an in-depth discussion of the NGVS completeness as a function of the galaxy structural and photometric properties.

\subsection{Circumstantial Evidence as to the Reliability of the NGVS Classification}
\label{subsec:VCC}

As mentioned, NGVS imaging includes four background fields located  three virial radii ($\sim16\deg$) from M87 at Galactic latitudes matching those of the lower and upper boundaries of the NGVS footprint. These background fields were included in the {\sf VCands} analysis; the team members performing the visual inspection of the {\sf VCands} candidates with high $\wp$ had no knowledge of a galaxy's location within the cluster, and therefore candidate in the background field were inspected alongside candidates in the main footprint. Surprisingly (and, at first, a bit disappointingly), the final catalog included two {\it bona fide} Virgo members located in one of the background fields. Upon further investigation, both galaxies were found to have a measured radial velocity ($v = 549$ km s$^{-1}$ and $v = 3,500$ km s$^{-1}$, respectively) that is consistent with Virgo membership; moreover, these are the only two galaxies in all background fields to have a measured radial velocity consistent with Virgo membership. The {\sf VCands} identification and further visual confirmation for these two galaxies provides a validation of the NGVS procedure for assigning Virgo membership. 

Also of interest is a comparison between the NGVS catalog and the VCC, which still serves as a  commonly used reference for galaxies in Virgo. In the VCC, galaxies are classified as M, P or B, for ``Members", ``Probable members", and ``Background objects", respectively. In what follows, when referring to ``VCC galaxies" we will include only M and P members. 

There are a total of 1558 VCC galaxies within the NGVS footprint, of which 1492 are identified in the NGVS catalog. This leaves 66 ``orphan" VCC galaxies. One of these, VCC1570, is a duplicate entry in the VCC --- the duplicate, VCC1571, is detected, identified as a ``certain" member, and included in the NGVS catalog as NGVS J123432.18+160157.5 (NGVS2656). Further investigation reveals that only one of the remaining 65 galaxies, VCC835, was not detected by {\sf VCands}, or identified visually in the NGVS. VCC835 is listed in the VCC as a probable member, with unknown morphology and a $B$-band magnitude of 20.0, i.e., at the very limit of the VCC survey. There are no galaxies in the NGVS images at the presumed location of VCC835 so we conclude that VCC835 is a spurious detection likely due to an artifact in the VCC photographic plates. This leaves 64 VCC galaxies that were detected by {\sf VCands} but then excluded from the final catalog as background sources. A search in NED reveals that 63 of these galaxies have a measured recessional velocity, and in all cases this is above 3,500 km s$^{-1}$, validating the NGVS classification. The last galaxy, VCC1424, does not have a measured recessional velocity. In the VCC, VCC1424 is listed as a possible member but is at the very limit of the survey: its morphology is unknown and its $B$-band magnitude is 19.6. In the NGVS imaging, VCC1424 has a compact core surrounded by an extended and somewhat irregular envelope, and an overall morphology consistent with what is expected for a background object. We stand by this classification and therefore exclude the galaxy from the NGVS catalog. 

\subsection{Takeaway Points}
\label{subsec:biasconc}

From the comparisons with the spectroscopic catalog (\S \ref{subsec:spec}), the EVCC (\S \ref{subsec:EVCC}), and the VCC (\S \ref{subsec:VCC}), we can draw the following conclusions. First, within the NGVS survey area, the EVCC contains 848 galaxies as faint as $g = 20.56$ mag, and only one spectroscopically-confirmed EVCC galaxy is missed by the NGVS (see \S\ref{subsec:EVCC}) --- i.e. 0.03\% of the sample. This argues that the NGVS is highly complete for galaxies brighter than $g\sim 20$ mag, more so than the 90\% completeness rate derived from the artificial galaxy simulations (see \S \ref{subsec:art}) would indicate. These results are not necessarily inconsistent with each other; most bright galaxies are not detected and/or identified as having $\wp > 0.22$ by {\sf VCands} because they happen to fall in problematic regions, for instance behind the halo or across the diffraction spike of a bright foreground star, or near the core of a much brighter galaxy. These galaxies, however, can be (and are) identified during visual inspection, as the comparison with the EVCC confirms. We also point out that many fainter galaxies were visually detected and therefore even at magnitudes fainter than $g \sim 20$ mag, the completeness is likely higher than the artificial galaxy simulations would indicate. The results of the artificial galaxy simulations, therefore, should be seen as conservative --- at least within the range of parameters spanned by the simulated galaxies (see NGVS-XIV, Figure 20).  

We also believe the purity of the sample to be high. Of the 996 galaxies classified as Virgo members by the NGVS, and with a measured velocity (\S \ref{subsec:spec}), only 10 (1\%) are beyond our adopted Virgo boundary, and for several of these (and certainly for the most discrepant ones, e.g. NGVS J121553.29+131257.1 = NGVS183), there are reasons to doubt the reported velocities. Of these ten, six were originally classified as certain members, two as likely members, and two as possible members. All are consistent with having early-type,  early-type/irregular, or early-type/spiral morphologies. None of the outliers are brighter than $g = 14.5$ mag, and only three are brighter than $g\sim$16 (a range that contains 497 NGVS galaxies with measured velocity), while seven have $g$-band magnitudes between 17.4 and 18.3, a range that contains 156 galaxies with measured velocities. While it is quite likely that the percentage of outliers increases at fainter magnitude, where we have essentially no spectroscopic information, these results argue that the NGVS sample has high purity, at least for galaxies brighter than $g\sim18$. An extensive spectroscopic program targeting fainter galaxies would be needed to say anything conclusively in the low magnitude range. 

\section{Photometric and Structural Parameters, and Properties for NGVS Galaxies} \label{sec:par}

Photometric and structural parameters for all NGVS galaxies were derived using two separate techniques, depending on the galaxy magnitude. Bright galaxies tend to have a complex morphological structure, often displaying multiple structural components, isophotal twists, radial changes in ellipticity, star formation activity, etc. They are also larger than fainter galaxies, and therefore generally suffer from more extensive contamination from background and foreground sources, often making the analysis challenging. 

For galaxies brighter than $g \sim 16$ mag (and the occasional fainter galaxy of interest), we therefore opted to carry out a full isophotal analysis, followed by parametric fits to the surface brightness profiles, using a dedicated code, {\sf Typhon}, designed by the NGVS team specifically for this purpose (see \S\ref{subsec:typhon}).  Galaxies fainter than $g \sim 16$ mag are generally smaller and less morphologically complex than brighter galaxies, and therefore do not require the full flexibility offered by {\sf Typhon}. For these galaxies, we performed less CPU-intensive 2D parametric fits using {\sf GalFit} \citep[][see \S\ref{subsec:galfit}]{Peng+02}. To ensure that no biases are introduced by using two separate methodologies, 631 galaxies were analyzed with both {\sf Typhon} and {\sf GalFit}, including all galaxies in the Virgo core region. This leads to a total of 897 galaxies analyzed with {\sf Typhon}, and 3423 galaxies with {\sf GalFit}, of which 631 are in common.  

In both the {\sf Typhon} and {\sf GalFit} analysis, we made the decision to use at most two components to describe the galaxy light distribution --- the assumption being that one component (referred to as ``outer") describes the main body of the galaxy, and a second optional component (referred to as ``inner") describes a NSC, when present. For the outer component, we used a S\'ersic law or, for some of the brightest galaxies with depleted cores, a core-S\'ersic law \citep[][see also Appendix \ref{app:CS}]{Graham+03}. For the inner component, in the {\sf Typhon} analysis we used a S\'ersic law, a choice justified by the fact that the nuclei in the bright galaxies analyzed with {\sf Typhon} are bright and often partially resolved \footnote{To constrain the fit, it was often necessary to fix the S\'ersic index $n$ of the inner component to be equal to 2, a functional form that has been shown to work well for Galactic GCs \citep[][]{munoz+18}}. In the case of {\sf GalFit}, using two S\'ersic laws often produces unphysical results as the fits are not well constrained; additionally, fainter galaxies host fainter and smaller nuclei that as not spatially resolved. We therefore opted to model the inner (NSC) component as a point spread function (PSF) instead of a S\'ersic.

\subsection{{\sf Typhon}} \label{subsec:typhon}

{\sf Typhon} is described extensively in NGVS-XIV, \S 5.1, and we refer the reader to that paper for full details, including examples of the output for specific galaxies (NGVS-XIV, \S 5.1.6). In one sentence, {\sf Typhon} is a complete, and largely automated package that, for each galaxy and in each filter: extracts cutouts from the original images; identifies all contaminants (foreground stars, background galaxies, as well as GCs and artifacts) and creates a mask (NGVS-XIV, \S 5.1.1); performs a full isophotal analysis (NGVS-XIV, \S 5.1.2); constructs a curve of growth and estimates the background (NGVS-XIV, \S 5.1.3); and, finally, performs a parametric fit to the surface brightness profile using one of three user-selectable functional forms: a core-S\'ersic, a S\'ersic, or a double S\'ersic profile (NGVS-XIV, \S 5.1.4). 

Although {\sf Typhon} can run in a fully automated fashion, visual inspection is necessary to assess the reliability of the results. The masks, residual images, fits and residuals to the surface brightness profiles were inspected, in each filter, and any steps deemed less than satisfactory repeated with some degree of manual intervention. Common issues include contaminants not properly masked, in which case the problematic areas were masked by hand, and the steps that followed repeated; unphysical radial changes in the galaxy center, ellipticity and position angle as a function of radius, often in very low surface brightness galaxies with flat surface brightness profiles, or galaxies with very complex morphology, in which case it was often necessary to repeat the isophotal fits by holding some of the parameters fixed; and incorrect fits to the surface brightness profiles, often in the case of morphologically complex galaxies, in which case some of the parameters, or their initial value, were tweaked until a more satisfactory fit was obtained. 

As mentioned above, for each galaxy {\sf Typhon} is run independently in each filter. This is necessary to characterize wavelength-dependent changes in structural parameters (due, for instance, to radial changes in the stellar population) but is not ideal for measuring a galaxy's integrated colors or color profiles. For this purpose, one must use the same masks and apertures in each filter, or risk introducing biases in the measurement. For the exclusive aim of measuring colors, therefore, {\sf Typhon} was run again on all ($u^*$-,$g$-,$i$-, and $z$-band) images using a master mask that includes all pixels masked in any band, and fixing the isophotal solution (i.e. position of the center, ellipticity, and position angle) to the one determined in the $g$-band. Integrated colors were then measured directly by integrating the surface brightness profiles from the center out to one effective radius, the latter measured in the $g$-band.

For each galaxy and each filter, the {\sf Typhon} outputs include: 

\begin{itemize}
\item a cutout centered on the galaxy itself; a corresponding mask; a noiseless image of the model best fitting the galaxy's isophotes and a residual image created by subtracting the model from the actual image;
\item the non-parametric curve of growth magnitude, effective radius, central surface brightness, surface brightness at, and average within, the effective radius, and concentration (Table \ref{tab:Tab7});
\item the galaxy magnitude, effective radius, S\'ersic index, central surface brightness, surface brightness at, and averaged within, the effective radius, derived from the parametric fits to the surface brightness profile (Tables \ref{tab:Tab8}, \ref{tab:Tab9}, and \ref{tab:Tab10});
\item integrated colors and color profiles measured self-consistently using a common mask and aperture in each filter (Table \ref{tab:Tab11});
\item for nucleated galaxies best fitted by a double S\'ersic profile, the magnitude, effective radius, surface brightness at, and averaged within, an effective radius of the inner component (Table \ref{tab:Tab9});
\item the position of the center of the galaxy, the surface brightness,  ellipticity, major axis position angle, and Fourier coefficients that measure deviations of the galaxy isophotes from a pure ellipse \citep[see][]{Jedrzejewski87} measured as a function of radius along the galaxy major axis, and averaged across the entire galaxy (see Table \ref{tab:Tab12} for the latter).
\end{itemize}

Photometry and structural parameters derived from the curve of growth and parametric analyses are compared in Appendix \ref{app:consistency}. We note that some of the brightest galaxies fitted by {\sf Typhon} often have very complex morphological structures (e.g. spiral arms, bars, disks, etc.). For 107 galaxies in this category, the two S\'ersic components allow for a more accurate fit to the whole profile than a single S\'ersic, but neither component necessarily represents a physical structure. In particular, the inner S\'ersic does not represent an NSC (even when present). These cases are identified in  Table \ref{tab:Tab6} and described in more detail in item 8 of \S\ref{sec:fin}.     

\subsection{{\sf GalFit}} \label{subsec:galfit}

Galaxies fainter than $g \sim 16$ mag (and the occasional brighter galaxy) were analyzed using two-dimensional {\sf GalFit} fits (see NGVS-XIV, \S 5.2). The first module of {\sf Typhon} was used to construct a mask in each filter, and then a master mask by masking any pixel flagged in any of the individual masks. The {\sf GalFit} fits employed this master mask, and used a S\'ersic model to represent the main body of the galaxy, and a PSF (constrained to be centered within 5 pixels, or 0\farcs93, of the center of the galaxy) to represent a possible NSC. By construction, the fitted position of the center, ellipticity and major axis position angle are radially invariant. Finally, if the fitted $g$-band magnitude of the NSC was fainter than 26 mag (the 10$\sigma$ limit for point-source detection, see Table \ref{tab:Tab2}), the fit was discarded and the procedure repeated using a single S\'ersic model. Just as in the case of the brighter galaxies analyzed by {\sf Typhon}, the $g$-band solution was applied to the other bands for the purpose of measuring colors.

Outputs from {\sf GalFit} are, for each galaxy and each filter:

\begin{itemize}
\item a noiseless image of the model best fitting the surface brightness profile, and a residual image created by subtracting the model from the actual image;
\item the galaxy magnitude, effective radius, S\'ersic index, central surface brightness, surface brightness at, and averaged within, the effective radius, and an estimate of the background, under the assumption that the galaxy is well represented by a S\'ersic surface brightness profile  (Tables \ref{tab:Tab8} and \ref{tab:Tab10});
\item integrated colors (Table \ref{tab:Tab11});
\item for nucleated galaxies, an NSC magnitude if the fit using a S\'ersic plus central PSF parametrization returned a PSF magnitude brighter than $g = 26$ (Table \ref{tab:Tab9});
\item the position of the center, ellipticity, and major axis position angle best representing the galaxy as a whole (i.e. not measured as a function of radius, Table \ref{tab:Tab12}).
\end{itemize}

Photometric and structural parameters derived from the {\sf Typhone} and {\sf GalFit} analyses are compared in Appendix \ref{app:consistency}.

\subsection{Compilation of Structural and Photometric Parameters} \label{subsec:comp}

Global photometric and structural parameters as measured by {\sf Typhon} and {\sf GalFit}  are listed in Tables \ref{tab:Tab7} to \ref{tab:Tab12}, with additional parameters for the galaxies fitted with a core-S\'ersic law given in Appendix \ref{app:CS}\footnote{Surface brightness profiles, as well as structural parameters measured as a function of radius, are produced by {\sf Typhon}; they are not included in this paper but work is underway to make them available on a dedicated website.}. Detailed notes are provided in the footnotes of all Tables and in \S \ref{sec:fin}, and only an overview of the Tables' content is provided here.  Table \ref{tab:Tab7} lists parameters derived from the curve-of-growth analysis for galaxies analyzed with {\sf Typhon}: i.e., magnitudes, geometric effective radius ($R_e$), surface brightness at the photocenter ($\mu_0$), at the effective radius ($\mu_{R_e}$), and averaged within an effective radius ($\langle\mu_{R_e}\rangle$), as well as concentration ($C_{80/20}$). Table \ref{tab:Tab8} lists total magnitudes, $R_e$, S\'ersic index, $n$, $\mu_{R_e}$, and $\langle\mu_{R_e}\rangle$, the latter four measured in the $g$-band, derived from the parametric fits to the surface brightness profile for {\sf Typhon}, and to the 2D image for {\sf GalFit}. When a galaxy is fitted with multiple components (i.e. a double S\'ersic fit for {\sf Typhon}, or a S\'ersic plus PSF for {\sf GalFit}), the parameters reported in Table \ref{tab:Tab8} refer to the combination of the two\footnote{Note however that the S\'ersic index, $n$, listed in Table \ref{tab:Tab8} refers to the outer (main galaxy) component.}. With one exception\footnote{NGVS J124848.54+140633.5 (NGVS3613, VCC2079) is almost completely obscured by a bright star, making it impossible to carry out a photometric or structural analysis based on the NGVS data.}, these parameters are available for all galaxies, although not always in all filters, since they cannot always be measured reliably in the case of faint and/or low surface brightness systems. Note that for galaxies with both a {\sf Typhon} and {\sf GalFit} analysis, only the former are reported. Tables \ref{tab:Tab9} and \ref{tab:Tab10} are analogue to Table \ref{tab:Tab8} but list parameters for the  inner (NSC) and outer components, respectively, for nucleated galaxies (Nuclear Code equals 1, 1,off, 2 or 2,off, see \S\ref{subsec:nuclei}) fitted with a double S\'ersic ({\sf Typhon}) or a S\'ersic plus PSF ({\sf GalFit}). Table \ref{tab:Tab11} lists the foreground reddening $E(B-V)$ from \citet{Schlegel+98}, the ($u^*-g$), ($g-i$), and ($g-z$) colors, integrated between 1$\arcsec$ and one geometric effective radius (defined as the effective radius measured along the semi-major axis, multiplied by $\sqrt{(1-\epsilon)}$, where $\epsilon$ is the ellipticity averaged between 1$\arcsec$ and one effective radius), and corrected for extinction, and stellar masses, derived as described in \S \ref{sec:mass}. Finally, Table \ref{tab:Tab12} lists average ellipticities, major axis position angles, and (for the galaxies analyzed with {\sf Typhon}) deviations of the isophotes from true ellipses.

\subsection{Nuclear Star Clusters (NSCs)}\label{subsec:nuclei}

The presence of a stellar NSC is assessed in a two-step process. First, in the {\sf Typhon} analysis, both single and double S\'ersic fits are performed, and the galaxy is tentatively reported to be nucleated if a double S\'ersic is statistically preferred to a single S\'ersic fit based on an simple F-test. For {\sf GalFit}, both S\'ersic and S\'ersic+PSF fits are performed, and the galaxy is tentatively reported as nucleated if the fitted PSF $g$-band magnitude is brighter than 26 mag. Second, all galaxies are visually inspected to confirm the automated nuclear classification provided in the first step --- this includes inspection of the images, the surface brightness profiles and/or residuals images, color images, etc. For the most part, the automated and visual classification are in agreement, but there are exceptions. In some cases, a NSC might be visible, but it might either be too faint for a reliable fit, or might be found in a morphologically complex galaxy that cannot be accurately fitted with the simple two component models adopted. At the opposite end, a galaxy fitted with a two-component model might not be judged to be nucleated, either because the central component is measuring an interloper, or because it measures a larger-scale component that is not the NSC. In other words, not all galaxies with a {\sf Typhon} double-S\'ersic fit are classified as nucleated; likewise there are nucleated galaxies fitted with a single S\'ersic model (see item 8 in \S \ref{sec:fin} for further details). 

Based on the visual inspection, galaxies are classified in 4 categories (see Table \ref{tab:Tab6}): Nuclear Code 0 indicates a non-nucleated galaxy, Nuclear Code 1 a galaxy for which the presence of a NSC appears certain, and Nuclear Code 2 a galaxy for which there is some evidence of the presence of a NSC, but the evidence is not conclusive. Nuclear Code 3 is reserved for galaxies for which the presence of a NSC cannot be assessed because of the presence, in the central region, of dust or other contaminants. Additionally, NSCs that appear to be offset from the galaxy photocenter have been labelled by appending ``off" to the Nuclear Code.

\begin{figure*}
    \centering
    \includegraphics[width=0.9\textwidth]{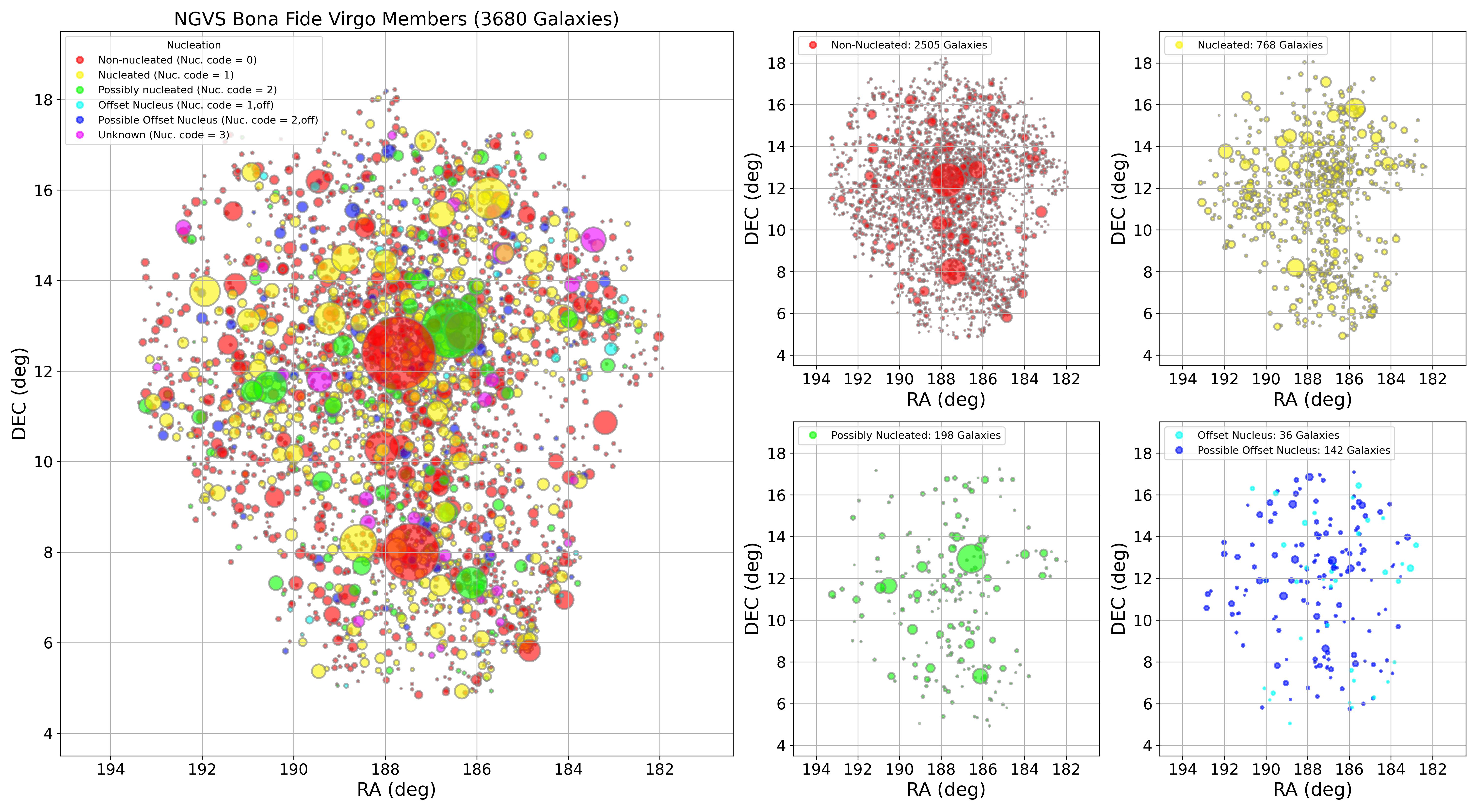}
    \caption{As Figure \ref{fig:fig8}, but with galaxies color-coded according to whether a NSC is present (as listed in Table \ref{tab:Tab6}). Galaxies classified as non-nucleated (Nuclear Code = 0), certainly nucleated (Nuclear Code = 1), possibly nucleated (Nuclear Code = 2), or having an offset NSC (certain or possible) are shown separately in the four smaller panels to the right.}
    \label{fig:fig21}
\end{figure*}

\begin{figure}
    \centering
    \includegraphics[width=\linewidth]{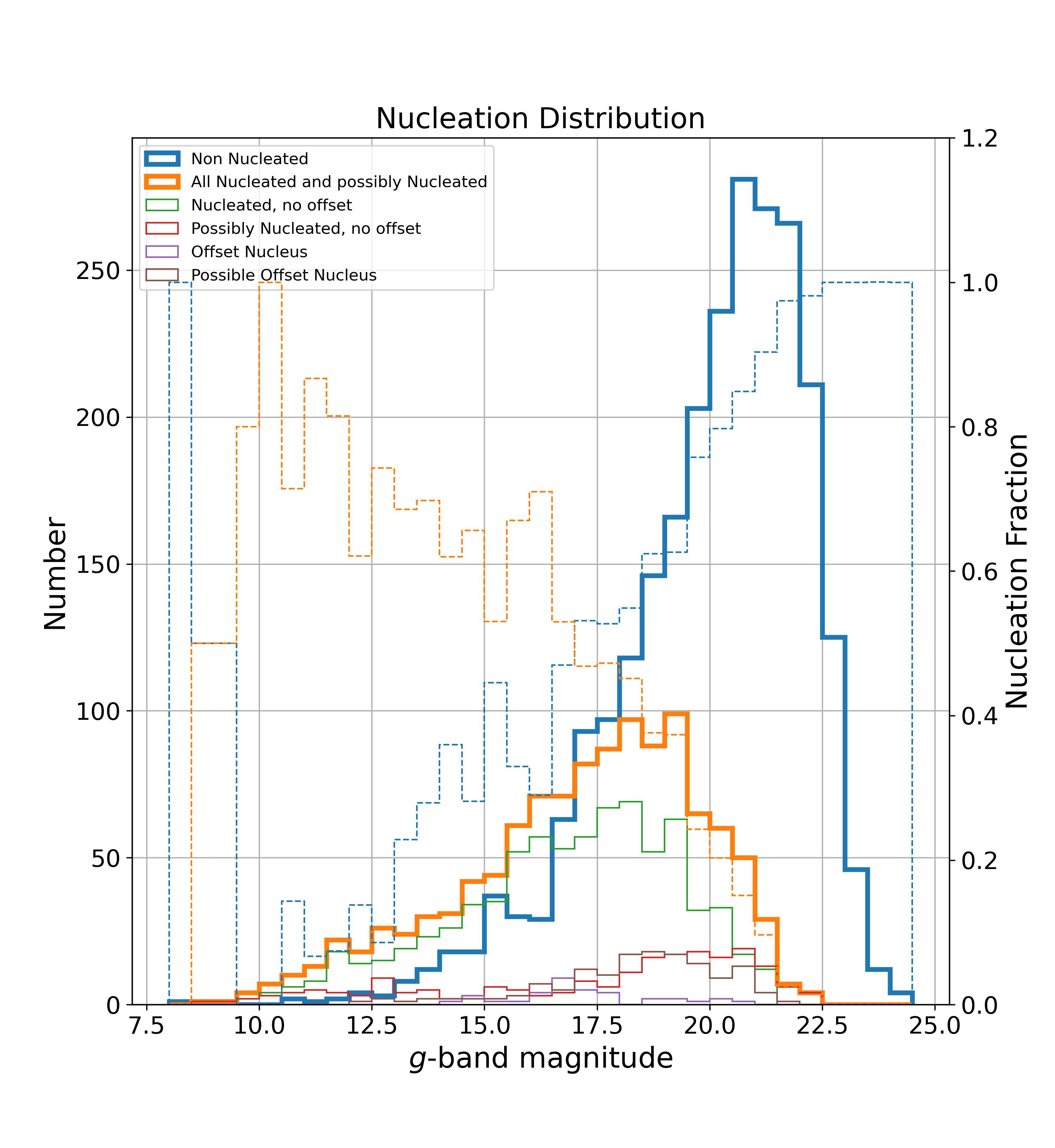}
    \caption{The number of non-nucleated (i.e., Nuclear Code 0; heavy solid blue histogram) and (possibly) nucleated (i.e., Nuclear Codes 1 and 2; heavy solid orange histogram) galaxies as a function of total galaxy $g$-band magnitude (left y-axis). The thin solid histograms show the number of galaxies that are certainly nucleated (green), possibly nucleated (red), or have a certain or possible offset NSC (magenta and brown, respectively). The dashed histograms show the nucleation fraction, i.e. the number of non-nucleated (blue) and (possibly) nucleated (orange) galaxies divided by the total number of galaxies in each bin (y-axis to the right).}
    \label{fig:fig22}
\end{figure}

Figure \ref{fig:fig21} shows the spatial distribution of NGVS galaxies, color-coded according to their nuclear classification, while the histogram in Figure \ref{fig:fig22} shows the magnitude distribution of galaxies in the various nucleation classes.

\begin{figure}
    \centering
    \includegraphics[width=\linewidth]{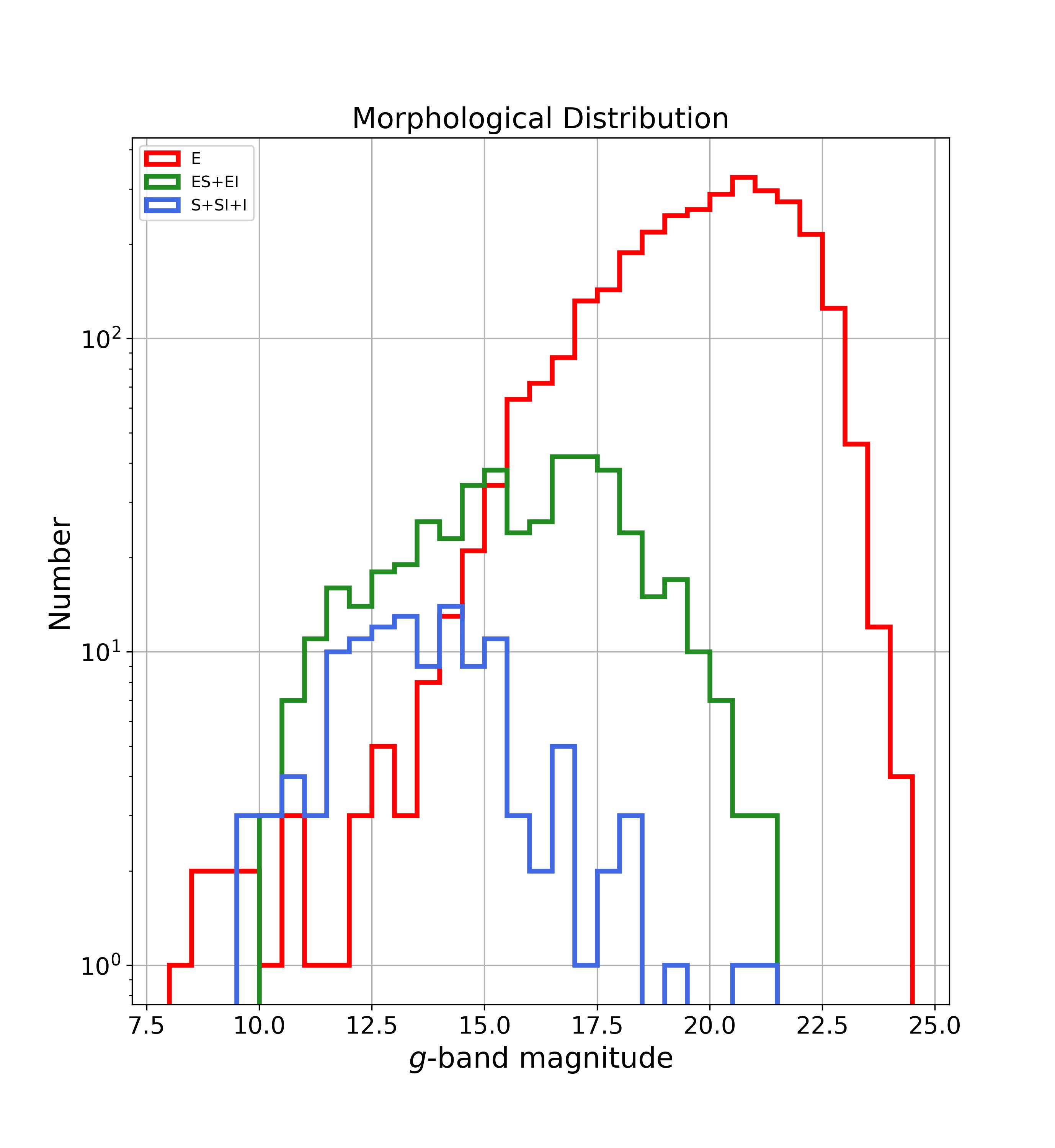}
    \caption{Histogram of the distribution of galaxy types by magnitude. Galaxies are divided in three main groups: Early-Type (E), Early-Type/Spiral (ES) and Early-Type/Irregular (EI) transition objects, and Spiral (S), Spiral/Irregular (SI) transition objects and Irregular (I).}
    \label{fig:Morphologies}
\end{figure}

\subsection{Morphological Classification}\label{subsec:morph}

Although subjective, morphological classifications offer insights into the structure and stellar content of galaxies that can be valuable for certain applications. A visual classification scheme was developed \citep[see][]{Kurzner+25} that uses two parameters to broadly describe each galaxy's global structure and current level of star formation activity. For the first parameter, galaxies are divided into six broad classes, with the following defining characteristics: 

\begin{enumerate}
\item[{\tt E}:] Smooth and regular in appearance, with {\tt elliptical} isophotes.
\item[{\tt S}:] A {\tt spiral} galaxy morphology, with features such as bars, spiral arms or flattened disks.\item[{\tt I}:] An {\tt irregular}, patchy, or filamentary appearance lacking an obvious spiral or ellipsoidal structure.
\item[{\tt ES}:] Intermediate in properties between {\tt E} and {\tt S}. These galaxies often exhibit a largely spheroidal outer light distribution characteristic of elliptical galaxies but contain inner features such as weak or embedded disks, bars, or bulge-like structures. Some edge-on systems are also placed in this category when flattened light profiles or dust lanes suggest hidden disk components. These galaxies often lie near the boundary between classic ellipticals and lenticular (S0) galaxies. 
\item[{\tt EI}:] Intermediate in properties between {\tt E} and {\tt I}. These galaxies have broadly elliptical envelopes with significant irregularities in their inner structure, including off-centered, star-forming knots, clumps, or mild asymmetries. These features suggest either a disturbed morphology or residual star formation that disrupts the appearance of a smooth spheroid.
\item[{\tt SI}:] Intermediate in properties between {\tt S} and {\tt I}. These galaxies often show a global disk-like structure but are more irregular, with patchy, asymmetric, or clumpy features replacing coherent spiral arms. Many show active star formation in irregular patterns or dust lanes that distort the underlying morphology. 
\end{enumerate}

For the second parameter in this classification scheme, galaxies are divided into three categories on the basis of their apparent state of star formation:

\begin{enumerate}
\item[{\tt Q}:] {\tt Quiescent}. Galaxies with negligible or no recent star formation, characterized by red optical colors with little or no $u^{*}$ band excess.
\item[{\tt I}:] {\tt Intermediate}. Galaxies exhibiting sporadic or low-level star formation, either localized (e.g., in specific knots or regions) or fading globally, often associated with weak $u^*$-band excess and intermediate optical colors.
\item[{\tt A}:] {\tt Active}. Galaxies with vigorous star formation across significant portions of the disk, often visible in blue optical colors and exhibiting structured features like star-forming clumps or filaments.
\end{enumerate}

Each parameter includes a variety of subcodes to denote interesting features, as summarized in \S\ref{sec:fin}. For more information on the scheme, including an overview of the morphological properties of galaxies in the cluster, we refer the reader to \citet{Kurzner+25}. The distribution of catalog galaxies by structural and star formation code is shown in Figure~\ref{fig:Morphologies}.

\section{Stellar Masses}
\label{sec:mass}

In what follows, we briefly describe the procedure used to model the integrated spectral energy distributions (SEDs) and derive stellar masses, \mstar, for the cataloged galaxies, using the multi-band NGVS imaging data. \mstar~ values are presented in Table \ref{tab:Tab11}.

\subsection{Photometric Data and Analysis}\label{subsec:SED_data}
When modeling the stellar populations of a galaxy with SED fitting, the fluxes must be measured using the same aperture across all filters so that the resultant colors reflect the same spatial region. For the galaxies analyzed with {\sf Typhon}, this was achieved by taking the radius within which the $g$-band CoG magnitude was measured, and then extracting the magnitudes in the $u^*$-, $i$-, and $z$-band (and, when available, $r$-band) directly from the corresponding CoG within the same radius \footnote{Note that this procedure is slightly different from the one detailed in \S\ref{subsec:typhon}, and was adopted out of concerns that the colors measured within one effective radius with {\sf Typhon} might not be consistent with those measure across the entire galaxy by {\sf GalFit}. In fact, the results are entirely consistent.}. For galaxies without {\sf Typhon} measurements, {\sf GalFit} total magnitudes, which are already derived using a consistent aperture in all filters, were used directly for all bands.  Photometric errors on the magnitudes were estimated based on the CCD equation  \citep{Howell+06} that considers the essential sources of error in brightness measurements (read noise, sky background, and counting statistics). We note that, for $\sim$85\% of the 631 objects in the survey that have both {\sf Typhon} and {\sf GalFit} photometry available, our modeling returns mass-to-light ratios based on both sets of measurements that agree to better than $\pm$0.2 dex.

\subsection{SED Fitting}

The software package {\tt Prospector} \citep{Johnson+21}\ was used to derive mass-to-light ratios, \sml, and \mstar\ values. {\tt Prospector} provides convenient access to a full suite of models needed to realistically reproduce the light emission from galaxies (e.g., stellar populations, H{\footnotesize II} regions, active galactic nuclei, dust) as well as statistical software to quantify the match between data and models (e.g., $\chi^2$ minimization). The backbones of {\tt Prospector} is the Flexible Stellar Population Synthesis \citep[FSPS;][]{Conroy+09} package that allows users to generate photometry and spectroscopy for a wide variety of conceivable stellar populations. We adopt a  \cite{Chabrier03} initial mass function (IMF) and make use of predictions based on the Padova isochrones \citep{Marigo+Girardi07, Marigo+08} and the MILES spectral library \citep{Sanchez-Blazquez+06}.

For our baseline fits, we assume a delayed-exponential star formation history (SFH; SFR($t$) = ($t$-$t_{\star,0}$)/$\tau_{\star}^2\times$exp\{-0.5(($t$-$t_{\star,0}$)/$\tau_{\star}$)$^2$\}), where $t_{\star, 0}$ and $\tau_{\star}$ are the start time and timescale of star formation, respectively. This SFH is commonly used for stellar population analyses in the literature since it is conceptually simple yet admits a variety of shapes; in the case of the NGVS SEDs in particular, which comprise five data points at most (and more commonly four), using more complex formulations would risk overfitting the data. In addition, \cite{Carnall+19} studied the impact of various SFH parameterizations on stellar population model inferences and found stellar masses to be robust at the 0.1 dex level, while \cite{Leja+19} analyzed the differences between parametric and non-parametric SFHs and found no reason to favor one family over the other when trying to reproduce photometry with a minimalist but flexible model. 

We combine the delayed-exponential SFH with a non-evolving metallicity. This results in {\it dustless} models with four fitted parameters, all of which are allowed to vary: the total mass of stars formed (\mstar), $t_{\star, 0}$, $\tau_{\star}$, and the stellar metallicity ($Z_{\star}$). In a second set of fits (the {\it dusty} models), we allow for a contribution of dust reddening to our galaxies' SEDs by invoking the default prescription for dust attenuation in {\tt Prospector} (via FSPS), which mimics the two-component model of \cite{Charlot+Fall00}.

We follow a Bayesian approach and use {\tt emcee} \citep{Foreman-Mackey+13} to construct the marginalized Probability Distribution Function (PDF) for all parameters via Markov Chain Monte Carlo (MCMC) sampling. For SEDs sampled with four or five points, we find the best fits using the Levenberg-Marquardt minimization method, while if only three points are available we used the Powell line search method. We adopt uninformative (i.e., flat) priors for all parameters over their respective ranges of allowed values.

Finally, we assume a distance of 16.5 Mpc \citep{Mei+07, Blakeslee+09} to all galaxies\footnote{The 1$\sigma$ scatter in this value \citep[0.6 Mpc;][]{Mei+07} imparts a systematic uncertainty of only $\sim$0.03 dex in our masses, although there is also an unquantified random error due to the back-to-front depth of the cluster.}, correct our photometry for foreground reddening by adopting the E(B-V) maps from \cite{Schlegel+98} and the extinction coefficients quoted in \cite{Ferrarese+16} (see also footnotes to Table \ref{tab:Tab11}), and ignore possible AGN contributions to the SEDs. We also determine the marginalized PDF of the surviving stellar mass by calculating the age-dependent mass loss associated with each posterior sample.

\begin{figure}[hbt!]
\epsscale{1.00}
\plotone{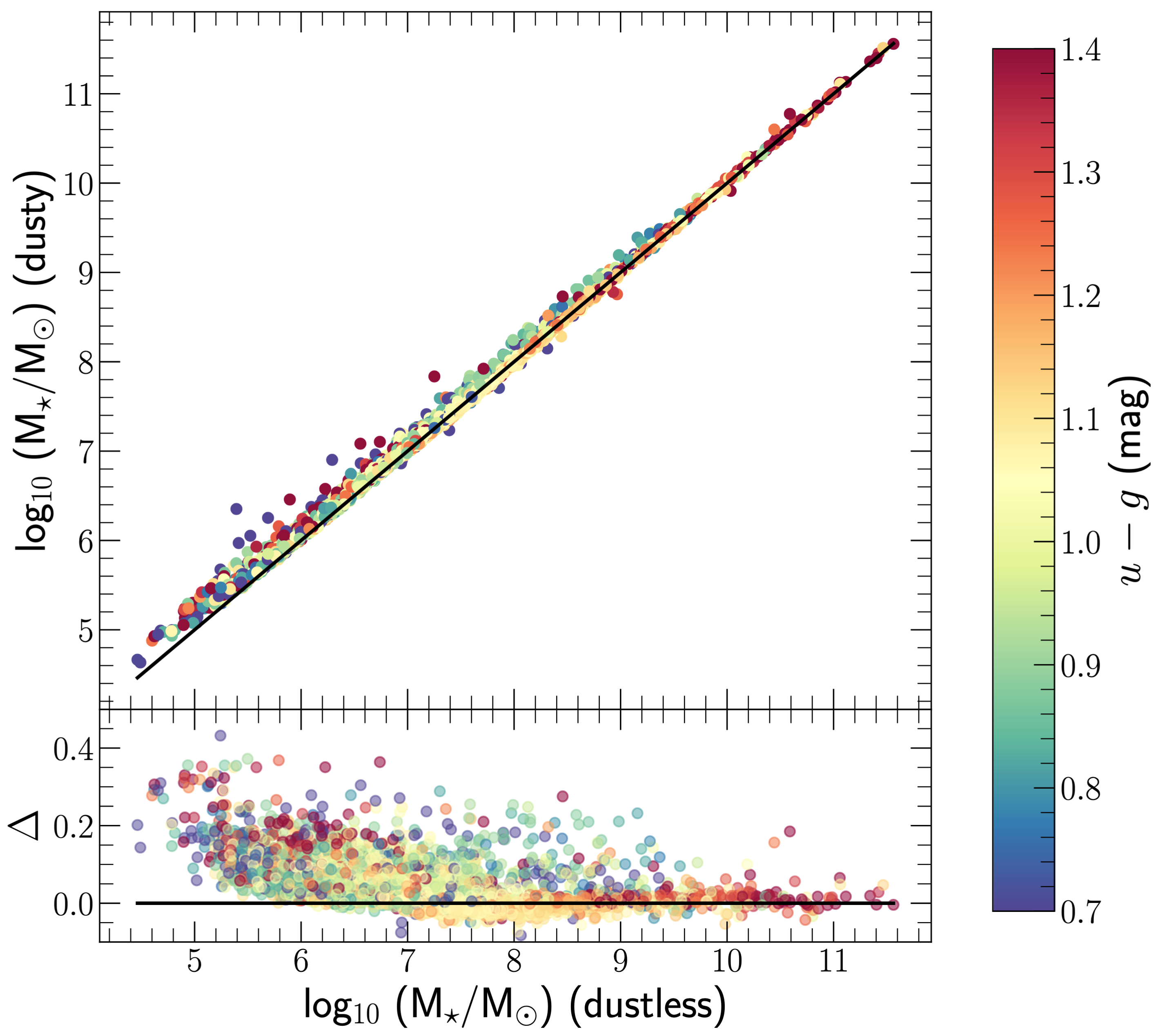}
\caption{({\it Top panel}) Dusty versus dustless masses for our full sample with points colored-code by ($u^*-g$) color. The black line represents the 1:1 relation. ({\it Bottom panel}) Difference (in dex) between the dusty and dustless \mstar~ estimates. 
\label{fig:dustlessVsDusty_mass}}
\end{figure}

\subsection{Treatment of Dust}

The models described in the preceding section differ only in their treatment of dust --- one allows for a dust contribution to the SED (the ``dusty'' model) and the other does not (the ``dustless'' model). In general, the dusty model provides a more realistic representation of a galaxy's optical emission and will typically result in a higher maximum likelihood simply by virtue of having more parameters than a dustless model. On the other hand, with only (at most) five data points, adding dust attenuation risks over-fitting the data. 

In the top panel of Figure \ref{fig:dustlessVsDusty_mass}, we plot the mass estimates from our two models against each other, and in the bottom panel we show the residuals about the 1:1 line as a function of the dustless \mstar. For galaxies with \mstar\ $\gtrsim$ 10$^{9.5}$ \msol , there is no significant difference between the two sets of mass estimates. Below this threshold, however, there is a tendency for the dusty models to produce systematically higher masses than the dustless models, and increasingly so as we move to lower masses, although the difference is rarely more than $\sim 0.2$ dex. 
The difference in \mstar\ between the models does not show a significant correlation with the observed $(u^*-g)$ color, which is interesting as it refutes the expectation that the results for redder galaxies would be more sensitive to the inclusion of dust in their modeling (i.e. galaxies with more dust should be redder in color). Further investigation shows that the above trend is due to the fact that in SEDs with higher photometric errors, as is the case for fainter galaxies, the inclusion of dust attenuation in the dusty models is compensated by a lower age and/or metallicity at constant color, according to the age-metallicity-dust degeneracy \citep{Worthey94}. This, in turn, translates to higher \sml~ and therefore \mstar. 

We note that star formation, as traced by H$\alpha$ emission, has been detected in $\sim$10\% of the NGVS galaxy sample \citep{Boselli+23} which suggests that dust may be important in a relatively small number of systems. This view is consistent with the results of \citet{Kurzner+25} who found just 17.5\% of Virgo galaxies to show clear signs of active or intermediate-level star formation, and reported the unambiguous presence of dust in only $\sim$6\% of NGVS member galaxies. To further explore this issue, we have used the Bayesian Information Criterion (BIC) statistic \citep{Schwarz78} 
to determine, for each galaxy in the survey, which model is statistically preferred over the other. 
This exercise showed evidence in favor of the dustless model for practically the entire sample ($\sim$99\%), with the dusty model preferred for only 50 galaxies. Of these, 23 lie within the footprint of the HeViCS survey \citep{Davies+10}, and for 15 the presence of dust was definitely established in one or more of the five FIR bands studied \citep{Auld+13}; upper limits could only be established for the remaining eight galaxies. While the frequency of such dusty galaxies is low, this should not be interpreted as there being an absence of dust in the rest of the sample. Rather, it should be concluded that we can explain most of the variance in our data with just a variable SFH and metallicity, and including dust attenuation very often leads to over-fitting the data. We expect that adding more measurements to the SEDs would result in the dusty model being preferred for a larger fraction of our sample. For the time being, though, we adopt the dustless model as the fiducial for our stellar mass measurements.

\begin{figure*}
\centering
\includegraphics[width=0.9\textwidth]{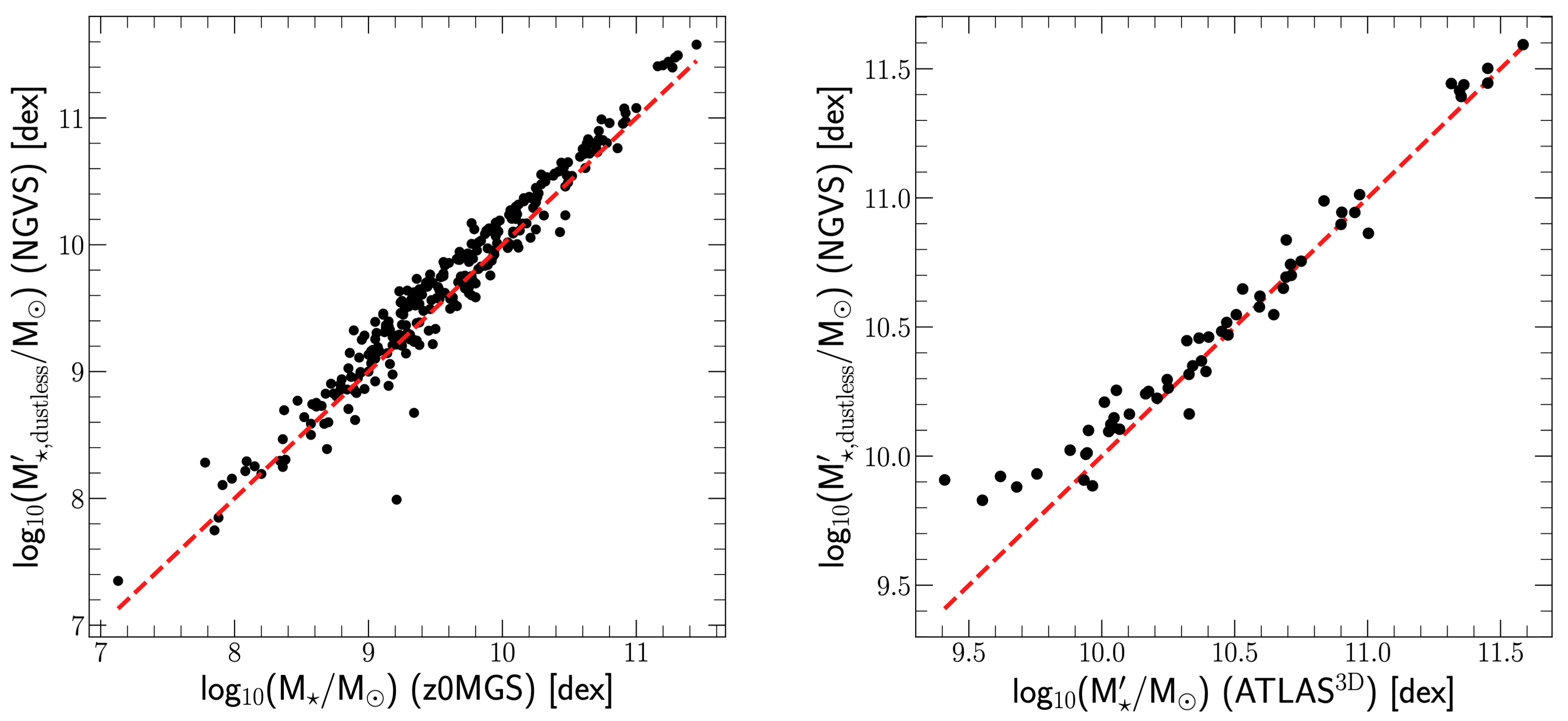}
\caption{({\it Left panel}) Comparison of masses from our dustless models versus those from the {\tt z0MGS} catalog. The dashed line represents the 1:1 relation. ({\it Right panel}) As in previous panel but now plotting our dustless masses against masses from the {\tt \atlas} survey. A constant (--0.3 dex) has been applied to correct the {\tt \atlas} masses to the IMF \citep{Chabrier03} used in our analysis. 
\label{fig:litComp}}
\end{figure*}

\subsection{Literature Comparison}

Stellar masses for 385 mainly late-type Virgo galaxies were presented in \citet{Boselli+23} using the CIGALE code \citep{Boquien+19} but similar IMF and star formation history adopted in this paper. In addition to the optical data from the NGVS, \citet{Boselli+23} supplemented their SEDs with FUV and far-IR data, allowing them to include a full treatment of dust in their analysis. They found excellent agreement between their mass estimates and the ones presented in this paper, with a mean ratio of 0.05 dex and a dispersion of $\sim$0.13 dex.

An additional validation of our stellar masses can be carried out by drawing on two catalogs. 

The first is the $z$ = 0 Multiwavelength Galaxy Synthesis catalog \citep[{\tt z0MGS};][]{Leroy+19}.
The authors obtained masses for their sample using a calibration of \sml\ in the WISE1 band (3.4 $\mu$m) based on the \mstar\ and specific star formation rate (sSFR) estimates from the GALEX-SDSS-WISE Legacy Catalog 2 \citep[GSWLC-2;][]{Salim+18}. 

The second catalog is the {\tt \atlas} survey \citep{Cappellari+11}. As part of their dynamical modeling of two-dimensional stellar kinematics maps and analysis of the stellar IMF, \cite{Cappellari+13} derived stellar mass-to-light ratios for 260 early-type galaxies by fitting their spectroscopic data with linear combinations of synthetic spectra of simple stellar populations spanning a grid of ages and metallicities. 

In the left panel of Figure~\ref{fig:litComp}, we plot our masses versus those from {\tt z0MGS} for the $\sim$290 galaxies cross-matched between the two catalogs. The small ($\sim 0.1$dex) systematic bias, with the {\tt z0MGS} masses being slightly lower than ours, is not surprising given the differences in data (optical versus UV-IR SEDs), photometry methods ({\sf Typhon}/{\sf GalFit} versus model magnitudes), models (FSPS versus \citealt{Bruzual+Charlot03}), and last but not least the fact that while in our analysis we use a single distance to all galaxies, {\tt z0MGS} used individual distances, an assumption that affects the mass determination. 

In the right panel of Figure~\ref{fig:litComp}, we plot the masses from our dustless modeling versus those from {\tt \atlas} for the $\sim$60 galaxies matched between the two catalogs.\footnote{The \cite{Cappellari+11} stellar population analysis assumed a Salpeter IMF so we have applied a correction (--0.3 dex) to put their \mstar\ values on the same scale as ours. No such a correction was necessary for the {\tt z0MGS} catalog because they adopt the same IMF as us \citep{Chabrier03}.} Once again, there is good agreement between the two datasets, aside from a small bias in our \mstar\ to higher values (median $\Delta \lesssim$ 0.05 dex). For the few objects with \mstar\ $\lesssim$ 10$^{10.1}$ M$_{\odot}$ in the \atlas\ catalog, a tail appears in the $\Delta$ distribution, with values reaching as large as $\sim$0.5 dex. Cross-referencing the least massive galaxies in this subsample against results from the stellar population analysis of \cite{McDermid+15} shows that they exhibit a combination of younger ages and low metallicities, whereas the bulk of the matched NGVS-\atlas\ catalog is described by older ages and/or higher metallicities. Based on this we suggest that the cause of this tail is that recent star formation in the associated objects, combined with the limited wavelength range sampled by the \atlas\ spectroscopic setup ($\sim$4800-5380 \AA),  results in their \sml\ values being biased low due to enhanced H$\beta$ line strengths. The \sml\ values we obtain for these objects would be more robust against the ``frosting'' phenomenon \citep{Trager+00} since our data sample galaxies' SEDs out to wavelengths as red as $\sim$1 $\mu$m and over their full volumes. 

In summary, these comparisons give us confidence that our mass estimates compare well to those based on panchromatic photometry (UV-IR) and more sophisticated modeling.

\section{The NGVS Catalog Tables} \label{sec:fin}

Photometric and structural parameters for all NGVS galaxies are presented in Tables \ref{tab:Tab7} to \ref{tab:Tab12}. A description of each column is given in the footnotes to the Tables, but we expand on a few points below. Additionally, we list below information that could be helpful to anyone using the NGVS catalog. Some of this information has also been included in the previous sections, but is repeated here for convenience.

\begin{enumerate}

\item \textit{Notes on the NGVS Nomenclature} (Table \ref{tab:Tab5}). The official NGVS name follows IAU conventions (https://cdsweb.u-strasbg.fr/Dic/iau-spec.html). The NGVS ``nickname" is just that, and is equal to ``NGVS" followed by a sequential number, with galaxies arranged in order of increasing RA. The exception is NGVS J123955.45+095520.8 (EVCC1010, see \S\ref{subsec:EVCC}), which was appended at the end after the main catalog was completed and was given the Nickname NGVS3690. 

\item \textit{Notes on the Galaxy Coordinates} (Table \ref{tab:Tab5}). The R.A. and Dec. reported for each galaxy are measured as follows: 1) if a $g$-band isophotal ({\sf Typhon}) analysis is available, R.A. and Dec. are defined as the isophotal center measured in the $g$-band and averaged between 1$\arcsec$ and one effective radius, where the effective radius is determined from the $g$-band curve of growth; 2) for the galaxies analyzed with {\sf GalFit}, the R.A. and Dec. are taken directly from the $g$-band {\sf GalFit} output, which is based on a single S\'ersic fit with fixed center. 
For NGVS J124848.54+140633.5 (NGVS3613, VCC2079), which is almost completely obscured by a bright star and therefore no analysis could be performed, the R.A. and Dec. are determined visually from the $g$-band image.

\item \textit{Notes on Membership Class and Heliocentric Radial Velocities}  (Table \ref{tab:Tab5}, \S\ref{sec:vcands} and \S\ref{subsec:spec}). For each galaxy, cluster membership is assessed based on its morphological and photometric properties, and subsequently updated, if needed, based on spectroscopic information, when available. Galaxies with spectroscopic information are classified as certain members (Membership Class = 0) if their radial velocity is $v < 3,500$ km~s$^{-1}$. Galaxies classified as members based on visual inspection but subsequently found to have $v > 3,500$ km~s$^{-1}$ are kept in the catalog for completeness (and also because, in some cases, the velocity information appears questionable) but identified as background sources (Membership Class = 4). Note that the several thousand galaxies that were deemed not to be members during the final visual inspection of the {\sf VCands} candidates (see \S \ref{sec:vcands}) are {\it not} included in the catalog. The Membership Class is therefore as follows:
    \begin{itemize}
    \item Membership Class 0 -- certain member: all galaxies that were either visually assessed to be certain members of the cluster, or have a measured radial velocity $v < 3,500$ km~s$^{-1}$;
    \item Membership Class 1 -- likely member based on visual assessment;
    \item Membership Class 2 -- possible member based on visual assessment;
    \item Membership Class 4 -- galaxies deemed to be certain, likely or possible members based on visual inspection, but for which, after the fact, a radial velocity estimate $v > 3,500$ km $s^{-1}$ was found. 
    \end{itemize}

\item \textit{Notes on the Morphological Classification}  (Table \ref{tab:Tab6} and \S\ref{subsec:morph}). Each galaxy is assigned a morphological classification based on a visual inspection of the images, the color maps, and the unsharped masked images. The classification includes two distinct codes, the first meant to capture the overall morphology and structure of the galaxy (for instance, whether the galaxy has a NSC, spiral arms, tidal stream, shells, etc..), the second to capture any indication of star formation, as well as its appearance.  The codes are as follows:
    \begin{itemize}
    \item \textit{Structure/Morphology codes}: E: elliptical; ES: elliptical/spiral transition; S: spiral; SI: spiral/irregular transition; I: irregular; EI: elliptical/irregular transition;
    \item \textit{Structure/Morphology sub-codes} (a ``:" indicates uncertainty): sp - spiral arms; di - disk; b - bar; sh - shells; du - dust; st - stellar stream; r - ring; fi - filaments; pm - post merger remnant;
    \item \textit{Star formation code}: Q: Quiescent; I: Intermediate; A: Active;
    \item \textit{Star formation sub-codes} (a ``:' indicates uncertainty): g - global; c - core; f - filamentary; p - patchy; e - extended; d - SF in disk; sp - SF in spiral arm; sh - SF in shell; r - SF in ring.
    \end{itemize}

\item \textit{Notes on Curve of Growth (CoG) non-Parametric Parameters}  (Table \ref{tab:Tab7} and \S\ref{subsec:typhon}). CoG parameters are only available for galaxies analyzed by {\sf Typhon} and are (generally) measured independently in each filter. Note that in some filters (most notably $u^*$ or $z$) some galaxies might have too low surface brightness for a reliable analysis, and therefore no values are reported. The CoG magnitude is estimated as the (background-subtracted) mean of the fluxes enclosed within isophotes with semi-major axis between $r_{0.01}$ and $1.5\times r_{0.01}$, where $r_{0.01}$ is the radius within which the enclosed magnitude is within 1\% of the total magnitude (extrapolated to infinity) and is determined based on the S\'ersic index, $n$, that best approximates the profile. The CoG geometric effective radius is defined as the radius enclosing half the total CoG flux, and is measured non-parametrically from the surface brightness profile. Note that this is a \textit{geometric} radius, i.e. it is equal to the radius measured along the semi-major axis multiplied by $\sqrt{(1-\epsilon)}$, where $\epsilon$ is the ellipticity averaged between 1$\arcsec$ and one effective radius (measured in the $g$-band). The central surface brightness is measured non-parametrically from the surface brightness profile at 0\psec187 arcsec (1 MegaCam pixel). The surface brightness at one effective radius and the surface brightness averaged within one effective radius are measured non-parametrically from the surface brightness profiles. Finally, the concentration, $C_{\rm 80/20}$, defined as the ratio of the radii encompassing 80\% and 20\% of the total flux, is also measured non-parametrically. Note that the concentration can be easily in error for faint galaxies, since it can be biased by even small fluctuations in the background. No errors are available for CoG parameters, except for the magnitudes, for which errors are calculated directly using the photometric properties (sky background, Poisson noise, etc.) of the images (see \S\ref{subsec:SED_data}). 

\item \textit{Notes on the Fit Model} (Table \ref{tab:Tab6}, \S\ref{subsec:typhon} and \S\ref{subsec:galfit}). This code identifies the procedure used to determine structural and photometric parameters:
\begin{itemize}
    \item T for the {\sf Typhon} analysis;
    \item G for the {\sf GalFit} analysis.
\end{itemize} 
This is followed by a code that identifies the model used to fit the 1D or 2D profile of the galaxy: 
    \begin{itemize}
    \item S for a single S\'ersic law; this applies to both {\sf Typhon} and {\sf GalFit};
    \item D for a double S\'ersic law; only used in the {\sf Typhon} analysis;
    \item C for a core-S\'ersic law; only used in the {\sf Typhon} analysis;
    \item S+P for a single S\'ersic law with the addition of a central PSF; only used in the {\sf GalFit} analysis. 
    \end{itemize}
A total of 897 galaxies  have been analyzed with {\sf Typhon}: all galaxies in the central 3.71 deg$^2$ (the core of the Virgo cluster), all galaxies brighter than $g\sim$16 mag, and the occasional galaxy of interest. There are 2792 galaxies with only {\sf GalFit} analysis, and an additional 631 galaxies processed with both {\sf Typhon} and {\sf GalFit}. When both {\sf Typhon} and {\sf GalFit} analysis are available, only the former is reported in the tables.

\item \textit{Notes Regarding the Parametric Photometric and Structural Parameters}  (Tables \ref{tab:Tab8}, \ref{tab:Tab9}, and \ref{tab:Tab10}, \S\ref{subsec:typhon} and \S\ref{subsec:galfit}). Up to three sets of parameters are reported for each galaxy: ``outer", representing the main body of the galaxy, ``inner", representing an NSC, and ``total" representing the combination of outer and inner. For galaxies fitted with a single component, (Model Code = T/S, T/C, and G/S), outer and total parameters are identical, and no parameters are reported for the inner component. For galaxies fitted with two components, the parameters of the outer component correspond to the S\'ersic fit for the G/S+PSF analysis, and the S\'ersic fit with the larger effective radius for the T/D analysis (but see exceptions below). The parameters of the inner component correspond to the PSF fit for the G/S+PSF analysis, and the S\'ersic fit with the smaller effective radius for the T/D analysis (but see exceptions below). For the total component, the magnitude corresponds to the sum of the fluxes of the two separate components; the effective radius is measured non parametrically from the profile as the radius enclosing half of the total flux; the surface brightness at, and averaged within, the effective radius, are also measured directly from the profile. 
More specifically, the following parameters are available:
    \begin{itemize}
    \item Parameters for the outer component, for both {\sf Typhon} and {\sf GalFit}, include, for all galaxies: magnitude; S\'ersic index $n$; geometric effective radius $R_e$; surface brightness at $R_e$, $\mu_{R_e}$; and average surface brightness within $R_e$, $\langle\mu_{R_e}\rangle$. Note that, by construction, {\sf Typhon} surface brightness profiles are expressed as a function of geometric, rather than semi-major axis radius, while {\sf GalFit} outputs semi-major axis effective radii. Therefore the {\sf GalFit} effective radii reported in the Tables have been multiplied by $\sqrt{(1-\epsilon)}$, where $\epsilon$ is the (radially invariant)  best fitting ellipticity from the {\sf GalFit} output. The errors returned by the fitting codes are reported in the Tables except when parameters ($n$ and/or $R_e$) are held fixed during the fitting procedure. However, these errors should be used with extreme caution, for various reasons, including the fact that errors reported by {\sf Typhon} and {\sf GalFit} are not necessarily comparable.  

    In addition, for the outer component we report in Table \ref{tab:Tab12} ellipticity, $\epsilon$; major axis position angle, $PA$; and (for {\sf Typhon} only) deviations of the isophotes from pure ellipses, $B4$. For the {\sf Typhon} analysis, these are averaged between 1$\arcsec$ and $R_e$ (measured in the $g$-band), with error calculated as the standard deviation in the mean; no error is reported if the parameter was held fixed while fitting the isophotes. For {\sf GalFit}, $\epsilon$ and $PA$ are invariant with radius, and the values reported are representative of the galaxy as a whole, with errors as given by the code. 
    
    \item Parameters for the inner component, for those galaxies with a two component {\sf Typhon} fit (but see exceptions below), include: magnitude, $R_e$, $\mu_{R_e}$, and $\langle\mu_{R_e}\rangle$ (note that the S\'ersic index $n$ was always held equal to 2.0, and the ellipticity was always fixed to 0, making the {\it PA} meaningless). For galaxies with a two component {\sf GalFit} fit, only the magnitude is available, as the fits used a PSF in fitting the inner profile. Again, any errors  should be used with caution.
    
    \item Parameters for the total component include: magnitude, $R_e$, $\mu_{R_e}$, and $\langle\mu_{R_e}\rangle$ for both {\sf Typhon} and {\sf GalFit}. Since total parameters include the inner component, the total magnitude, $\mu_{R_e}$, and $\langle\mu_{R_e}\rangle$ should be brighter than the corresponding values for the outer component, while $R_e$ should be smaller. This is generally the case, but there are a few exceptions, since outer parameters rely on the assumption that the surface brightness profile follows a S\'ersic law, while the total parameters are measured non-parametrically.  
    \end{itemize}

The exception to the above are morphologically complex galaxies with a two component {\sf Typhon} fit (Model Code T/D), but for which the two S\'ersic profiles do not actually represent structurally distinct components within the galaxy (see \S\ref{subsec:nuclei}). In this case, total parameters are computed and the outer and inner parameters are not reported separately. Note that in this case, no S\'ersic index is available (since the overall shape of the galaxy does not follow a single S\'ersic profile, see also following notes on the nuclear classification).

\item \textit{Notes on the Relation between the Choice of Parametric Fits and NSC Parameters (Table \ref{tab:Tab9})}. As mentioned in \S \ref{subsec:nuclei}, not all galaxies with a {\sf Typhon} double-S\'ersic fit are classified as nucleated; likewise, there are some nucleated galaxies fitted with a single S\'ersic model. In more detail:

    \begin{itemize}
    \item Galaxies classified as nucleated (Nuclear Code = 1 or 2), but with a single component fit (Model Code = T/S, T/C or G/S). These are galaxies for which the presence of a NSC is established based on visual inspection of the images and of the surface brightness profiles, but it was not possible to obtain a reliable two-component fit. There can be several explanations for this: for some early-type galaxies with steep surface brightness profiles, the presence of a NSC is inferred but cannot be constrained by a parametric fit; for some morphologically complex galaxies, a double S\'ersic fit might not provide a significantly better fit than a single S\'ersic; or in some cases, the NSC is simply too faint for a reliable fit. There are 212 galaxies in this category, of which 46 were fit with {\sf Typhon}, and 166 with {\sf GalFit}. 

    \item Galaxies classified as nucleated and with a two component fit (Model Code = T/D; there are no {\sf GalFit}-fitted galaxies in this category), but for which the inner component represents a morphological structure other than the NSC (e.g. a bulge, bar or disk), or is simply needed to better fit a complex profile.  For the 92 galaxies in this category, the individual components (inner and outer S\'ersic) are not listed separately, as they do not necessarily measure a physical structure, and only total values are reported. These galaxies are identified in Table \ref{tab:Tab6} by a Nuclear Code equal to 1* or 2*.

    \item Galaxies classified as non-nucleated (Nuclear Code = 0) but with a two-component fit (Model Code = T/D; there are no {\sf GalFit}-fitted galaxies in this category). These fall in two categories, for a total of 28 galaxies: 1) Late-type, irregular, or early-type transition galaxies for which the central component of the double S\'ersic measures a central bulge/bar/disk/etc. For these galaxies (15 in total), the two components are not listed separately and only total values are reported. 2) Galaxies with an often resolved, elongated object that is generally slightly off-center (and sometimes very bright) and is, in all likelihood, a background galaxy. For these galaxies (13 in total), only the outer component of the double S\'ersic fit is included; the inner component is omitted.

    \item Galaxies for which the presence of a NSC cannot be established (Nuclear Code = 3), often because of the presence of dust or contamination by a nearby bright star, but with a double S\'ersic profile (Model Code = T/D). These are, again, galaxies for which the inner S\'ersic component measures something other than a NSC. For these galaxies (23 in total), the two components are not listed separately and only total values are reported. 
    \end{itemize}

To conclude, a Nuclear Code is assigned to each galaxies as follows:
    \begin{itemize}
        \item 0: non-nucleated
        \item 1 and 1*: nucleated. An * indicates that the galaxy is nucleated but the “inner” parameters fit a component other than the NSC and are therefore not reported in the catalog. 
        \item 2 and 2*: probable NSC. An * indicates that  the galaxy is probably nucleated but the “inner” parameters fit a component other than the NSC and are therefore not reported in the catalog. 
        \item 3: presence of NSC is impossible to assess due to dust or foreground/background contaminants (bright foreground stars, etc.)
    \end{itemize}
An ``off" following the Nuclear Code indicates that the NSC is offset from the isophotal center of the galaxy. Note that occasionally NSC parameters are missing in one or more filters. This is generally the case of galaxies with very faint NSCs.

\item \textit{Notes on ($u^*-g$), ($g-i$), and ($g-z$) Colors} (Table \ref{tab:Tab11}, \S\ref{subsec:typhon} and \S\ref{subsec:galfit}). For galaxies analyzed with {\sf Typhon}, the colors are calculated from dedicated {\sf Typhon} fits where, when fitting the $u^*$-, $i$- and $z$-band images, the effective radius and S\'ersic index of all components are fixed to the values determined in the $g$-band. In addition, all fits use the same “master mask” that includes all pixels masked in any band. The colors are then calculated as the difference of the magnitudes measured directly by integrating the surface brightness profiles from 1$\arcsec$ out to the $g$-band geometric effective radius. 

For {\sf GalFit}, $n$ and $R_e$ for the S\'ersic fit in the $u^*$-,$i$-, and $z$-band, were fixed to the values measured in the $g$-band. The color is then simply the difference between the total magnitudes as reported by {\sf GalFit}, each corrected for extinction. 

Note that the colors are measured differently whether the {\sf Typhon} and {\sf GalFit} analysis is used. For {\sf Typhon}, the colors are measured within the $g$-band geometric effective radius. For spiral galaxies, this means that the color can be weighted more heavily towards the bulge rather than the disk. Additionally, the color might not be representative of the color of the entire galaxy if a color gradient is present. Note that excluding the central 1\arcsec~ does not appear to affect the colors in any appreciable way. For {\sf GalFit}, the color represents the color of the entire galaxy. However, analysis of the color-magnitude relation shows that {\sf GalFit} and {\sf Typhon} produce CMRs with similar scatter and mean, so there appear to be no significant systematic differences between the two datasets. Note that all colors are dereddened using the $E(B-V)$ extinction tabulated in Table \ref{tab:Tab11}.

\end{enumerate}

\section{An Overview of Science Results} \label{sec:results}

Given its areal coverage, depth, resolution, and cadence, the NGVS data are conducive to studies ranging from an investigation of objects in the solar system, to the detection of high redshift clusters. To date, the NGVS team has published 40 refereed papers based predominantly or exclusively on NGVS data, and 27 additional papers making extensive use of NGVS data in combination with other surveys. The following sections summarize some of the science results from the NGVS, in Virgo and beyond. 

\subsection{Catalogs and Data}

An introduction to the survey itself, including its observational design and scientific motivation, was presented in NGVS-I. That paper also introduced the {\it Elixir-LSB} pipeline --- a ``step-dither" data acquisition and reduction procedure developed specifically by the NGVS team to maximize the detection limit for LSB features. \citet{Munoz+14} described an IR extension to the main survey which featured deep $K_s$-band imaging in the central $\sim$4 deg$^2$ of the cluster using the WIRCam instrument on CFHT (5$\sigma$ limiting AB magnitude of 24.4 mag). Their analysis highlighted the power of the $u^*iK_s$ diagram for the clean separation of foreground stars, GCs and background galaxies. A catalog of $\sim$400 candidate member galaxies in the cluster core was presented in NGVS-XIV. This paper also described the suite of simulations used to assess survey completeness and outlined the procedures for identifying objects and assessing cluster membership.

\subsection{Luminosity and Mass Functions}

\citet{Grossauer+15} used the catalog of \citet{Ferrarese+20} to explore the stellar-to-halo-mass ratio (SHMR) of dwarf galaxies, as it would have appeared prior to cluster infall. Comparing the observed stellar mass function to simulations of cluster formation, they suggest that the trends in the SHMR characterized previously for high-mass galaxies  extend in a scale-invariant way down to galaxy masses of $\log{M_*/M_{\odot}} \sim 5$.

\citet{Morgan+25} separated Virgo galaxies into three samples based on their star formation properties (i.e., star forming, quiescent and intermediate) and constructed their stellar mass functions (SMFs). They found the shape of the SMF to be universal throughout the cluster, from the core to the outskirts. Comparing to infall models, they concluded that at least some of the quiescent galaxies were likely preprocessed outside the virial radius of the cluster, possibly in groups prior to infall.

A measurement of the  Virgo luminosity function was a primary motivation for the NGVS, with previous investigations having shown a surprising amount of scatter (see, e.g., \S{5.1.1} of \citealt{Ferrarese+12}). First results were presented by \citet{Ferrarese+16} who used a catalog of 404 galaxies in the central 3.71 deg$^2$ to construct a luminosity function down to a 50\% completeness limit of $M_g = -9.13$ mag. Somewhat unexpectedly, the derived faint-end slope of $\alpha$ = --1.33$\pm$0.02 was only slighter steeper (1.7$\sigma$) than that obtained from an analyis of the Local Group ($\alpha$ = --1.21$\pm$0.05). A comprehensive analysis of the luminosity and stellar mass functions across the entire cluster is presented in Todd et al. (2026, in preparation), including variations as a function of color, morphology and environment.

\subsection{Galaxy Morphologies and Star Formation}

A customized classification scheme based on the structural and star-formation properties of member galaxies was described in \citet[][see also \S\ref{subsec:morph}]{Kurzner+25}. This paper explored the morphology-density relation and presented catalogs of rare or morphologically interesting galaxy types, including structurally extreme galaxies, possible UCD transition objects, and candidate post-merger systems.

The Virgo color-magnitude relation (CMR) for galaxies in the cluster core was presented by \citet{Roediger+17}. This CMR, spanning more than five orders of magnitude in stellar mass, revealed the red sequence to flatten in all colors at the low-mass end, beginning around $\log{(M_*/M_{\odot})} \sim 7.6$. This may indicate that the central (mostly low-mass) galaxies were quenched coevally, possibly via pre-processing in small groups. By overall number, these quenched, low-mass galaxies dominate in the cluster core.

The GALEX Ultraviolet Virgo Cluster Survey ({\tt GUViCS}) carried out UV imaging of Virgo, often analyzing the UV data in combination with optical data from the NGVS \citep[e.g.,][]{Boselli+11}. Notable outcomes include the UV luminosity function \citep{Boselli+11,Boselli+16}, UV/optical source catalogs \citep{Voyer+14} and a characterization of star formation efficiency in ram-pressure stripped gas \citep{Boissier+12}.

An ambitious study of star formation processes within the cluster is described in \citet{Boselli+18}. This  program (Virgo Environmental Survey Tracing Ionised Gas Emission = {\tt VESTIGE}) combined NGVS optical imaging with  narrow-band H$\alpha$+[NII] imaging, also acquired with MegaCam. Science results from this CFHT Large Program span a wide range of topics related to star formation in the cluster environment: e.g., ram pressure stripping and the fate of the stripped gas; environmental quenching; galaxy mergers; star formation properties as a function of galaxy type and stellar mass; the H$\alpha$ luminosity function and scaling relations. More details may be found in \citet{Boselli+25}.

\citet{Sanchez-Janssen+16} studied the intrinsic shapes of low-mass galaxies and concluded they are best described as a family of thick, nearly oblate spheroids with mean intrinsic axis ratios 1 : 0.94 : 0.57. They also reported the intrinsic flattening in this low-mass regime to be nearly independent of the environment, aside from a hint that objects may be slightly rounder in denser environments.

\subsection{Mergers and Interactions}

\citet{Arrigoni+12} used imaging from NGVS and GALEX to study stripped gas and HII regions in VCC1249 (NGVS1968) --- a low-mass galaxy interacting with M49, the brightest member of the cluster. They concluded that both ram-pressure stripping and tidal interactions are needed to explain this close interaction. \citet{Paudel+13} used NGVS imaging to identify and study an infalling group in the outskirts of the cluster. This group contains the edge-on spiral galaxy NGC 4216 and a number of disturbed low-mass galaxies and stellar streams --- signs of ongoing interactions. 

Three low-mass galaxies with shell-like features were reported in \citet{Paudel+17}. Numerical simulations and follow-up observations \citep{Zhang+20a,Zhang+20b} showed these objects to be the result of nearly equal-mass mergers. An expanded sample of post-merger candidates, including many in the dwarf galaxy mass regime, was presented in \citet{Kurzner+25}. Some first results from a dedicated survey to study these dwarf-dwarf mergers using optical, HI, H$\alpha$ and UV data are presented in \citet{Sun+20a}, \citet{Sun+20b}, and \citet{Li+26} --- the Atomic gas in Virgo Interacting Dwarf galaxies ({\tt AVID}) survey. 

\subsection{Extreme Galaxies: In Virgo and Beyond} 

A sample of structurally compact galaxies with elliptical morphologies was selected by \cite{Guerou+15} from the NGVS and targeted with the Gemini GMOS integral field unit \citep[IFU; ][]{Guerou+15}. From an analysis of stellar kinematics and stellar populations, these objects were found to show a range of kinematic properties ---  from non-rotating systems to strongly rotating objects, often associated with underlying disky isophotes --- and fully half of the sample contained a centrally concentrated younger, more metal-rich stellar population (see also \citealt{Kurzner+25}). These findings support a scenario where long-lived interactions with their environment, including ram-pressure stripping and gravitational tidal forces, may be responsible for the compact nature of these galaxies. 

At the other extreme in surface brightness, \citet{Lim+20} found  ultra-diffuse galaxies (UDGs) to occupy the extended LSB tails of the main locus of Virgo galaxies \citep[rather than being a distinct population, see also ][]{Junais+22}. On the other hand, \citet{Lim+20} found the galaxies to exhibit a diversity in morphology and GC content that suggests no single process gives rise to all objects within this class. The GC content and dynamical properties of Virgo UDGs were examined in a series of papers \citep{Toloba+16,Toloba+18,Toloba+23} that similarly pointed to multiple formation channels and an apparent wide range in dark matter content.

A pan-chromatic study of the extended star-forming galaxy Malin~1 (which is located at a distance of $\sim$366~Mpc but happens to fall inside the NGVS footprint) was presented in \citet{Boissier+16} who combined GALEX and NGVS imaging to characterize the stellar mass, metallicity and extended star formation history of this ultra-LSB disk galaxy. A study of the hyper-luminous starburst galaxy at z = 4.72, discovered in the {\it Herschel} Reference Survey and magnified by a lensing galaxy pair at z = 1.48, was presented in \citet{Ciesla+20}.

\subsection{Cluster Structure and Distance Scale}

For more than four decades, the prevailing view of Virgo's spatial extent and internal structure --- including its substructures --- was that of \citet{Binggeli+87} based on data from the photographic catalog of \citet{Binggeli+85}. Because the NGVS has increased the number of cataloged members by over a factor 2, the cluster's internal structure is reexamined in Todd et al. (2026b). This study utilizes a data clustering algorithm designed to quantitatively identify density-based clustering (i.e., Ordering Points To Identify the Clustering Structure = OPTICS; \citealt{Ankerst+99}). The analysis also incorporates SBF distances and updated radial velocity information for $\sim$300 and $\sim$1000 galaxies, respectively, and presents updated structural parameters for the cluster as a whole as well as its main substructures.

NGVS multi-band imaging was used to derive distances for a subset of the cluster's brightest galaxies using the SBF method. A calibration of SBF magnitudes was presented in \citet{Cantiello+18} who, after comparing the ground-based fluctuation magnitudes to HST measurements \citep{Mei+07,Blakeslee+09}, presented preliminary NGVS SBF distances for 89 galaxies. Final SBF distances for 278 galaxies spanning the magnitude range $9 \lesssim g \lesssim 19$ were presented in \citet{Cantiello+24} who used these data to characterize the spatial structure of cluster and estimate distances for its main substructures.

\subsection{NSCs and UCDs}

A detailed study of the masses and stellar content of $\sim$50 NSCs was presented in \citet{Spengler+17}. Their analysis combined NGVS and HST imaging over 10 passbands, from the UV to IR, as well as IFU and long-slit spectroscopy from the Gemini and Keck telescopes. A notable outcome from this study was the finding that NSCs in the most massive and structurally complex galaxies may have formed through predominantly dissipative processes. 

Working from a sample of $\sim$400 galaxies in the cluster core, \citet{Sanchez-Janssen+19a} showed that: (1) at low-masses, nucleation is more frequent in denser environments; and (2) there exists a nonlinear relation between the stellar masses of NSCs and those of their host galaxies (with NSCs being more prominent in high- and low-mass galaxies). The intrinsic shapes of these same galaxies was explored in \citet{Sanchez-Janssen+19b} who showed that flattening depends both on luminosity and on the presence of a NSC. A comprehensive study of NSCs in the survey, including the dependence of their properties on galaxy mass, morphology and environment, is presented in Kurzner et al. (2026, in preparation).

The connection between UCDs and NSCs is explored in a series of papers beginning with \citet{Liu+15b,Liu+15a} who presented evidence that UCDs follow a morphological sequence ordered by the prominence of their outer, low surface brightness envelope, ultimately merging with the sequence of nucleated low-mass galaxies. This evolutionary connection between NSCs and UCDs was later confirmed using an expanded sample of NSCs and UCDs with HST imaging by \citet{Wang+23} and Wang et al. (2026, in preparation). A catalog of more than 600 UCD candidates selected from the NGVS was presented by \citet{Liu+20}.

\subsection{Globular Cluster Systems} 

\citet{Longobardi+18} explored the kinematics of intracluster GC candidates selected from the NGVS imaging and having radial velocities from a variety of 4m- to 10m-class telescopes. Spectra from these programs were used by \citet{Sun+19} to search for faint [O~III] emission in individual GCs, allowing an estimate of the occurrence rate of planetary nebulae in globular cluster environments. Somewhat surprisingly, the inferred luminosity-specific frequency of PNe in the Virgo cluster GCs was found to be 5--6 times lower than that for Galactic GCs.

\citet{Durrell+14} selected GC candidates over the NGVS footprint using colors, magnitudes and inverse concentrations. After correcting for background contamination, they mapped the spatial distribution of GC candidates (including the distinct blue and red subpopulations) over an area of $\sim$100 deg$^2$. They estimated the total number of GCs within Virgo's virial radius to be $N_{\rm GC} = 67,300 \pm 14,400$ with $\sim$35\% of these located in M87 and M49 alone. These numbers translate to a GC-to-baryonic mass fraction of $f_b = (5.7 \pm 1.1) \times 10^{-4}$ and a GC-to-total cluster mass formation efficiency $\epsilon_t = (2.9 \pm 0.5) \times 10^{-5}$.

The distribution of gravitating mass in the core region was computed by \citet{Zhu+14} who combined radial velocities for nearly one thousand GCs extending to $\sim$180 kpc with SAURON IFU data within the central 2.4 kpc. Fitting the kinematic data with a Made-to-Measure model, they derived a stellar mass-to-light ratio of $M/L_I$ = $6.0 \pm 0.3$ in solar units, with a dark matter potential scale velocity of $591 \pm 50$ km~s$^{-1}$ and scale radius of $42 \pm 10$ kpc. Within 30 kpc, their estimated mass was found to be consistent with that obtained from X-ray (Chandra and XMM) observations, but about 20\% smaller within 120~kpc. Their analysis was complemented by \citet{Li+20}, who used applied chemo-dynamical, axisymmetric Jeans equation modeling to an expanded GC radial velocity catalog extending to $\sim$430 kpc, and \citet{Zhang+15,Zhang+18} who explored the kinematics, stellar populations and orbital structure of UCDs in the cluster core. A kinematic analysis of the GC system in the M49 region was presented in \citet{Taylor+21}.

The stellar populations of GCs were explored in a series of papers \citep{Powalka+16a,Powalka+16b,Powalka+17,Powalka+18} that analyzed their integrated colors and compared them to the predictions of widely used population synthesis models. While certain color combinations were useful for estimating metallicities, or constraining ages, no single model was found to adequately match the data in all passbands over the UV, optical and IR region. \citet{Ko+22} explored radial and azimuthal variations in the mean age, total metallicity, [Fe/H], and $\alpha$-element abundance of GCs in the core region using co-added spectra. They found the blue GCs to have a steep radial gradient in [Z/H] within $\sim$165 kpc, with roughly equal contributions from [Fe/H] and [$\alpha$/Fe], and flat gradients beyond. The red GCs, by contrast, showed a much shallower gradient in [Z/H] driven entirely by [Fe/H].

\citet{Lim+24} studied the size and density structure of GC systems belonging to more than one hundred early-type galaxies from the NGVS, MATLAS \citep{Duc+15} and ACSVCS \citep{Cote+04} surveys. By fitting S\'ersic profiles to the GC populations of these galaxies, they characterized the relationship between the total numbers and effective radii of GC systems to their host galaxies, including the dark matter halos. Catalogs for the GC samples used in this analysis are available in \citet{Lim+25}. Zhu et al. (2026, in preparation) explored the evolution of GC systems in low-mass galaxies chosen to span a range in time since accretion into the cluster.

\subsection{Photo-zs, Background Clusters, Weak Lensing}

\citet{Raichoor+14} used the {\it Le Phare} \citep{Ilbert+06} and BPZ \citep{Benitez+00} codes to measure photometric redshifts for more than half a million galaxies in the NGVS footprint, after  calibrating against a sample of $\sim$ 83,000 galaxies with spectroscopic redshifts.

\citet{Licitra+16b} presented catalogs of galaxy cluster candidates up to redshifts of $z\sim1.1$ identified with the REDGOLD cluster detection algorithm \citep{Licitra+16a}. From a comparison to external catalogs, including those based on X-ray observations, they concluded that their catalogs are $\sim$70\% complete and $\sim$80\% pure at $z \lesssim$ 1 for clusters with masses above 10$^{14}M_{\odot}$. 

Stacked weak lensing cluster masses for a subset of the clusters in this sample were presented in \citet{Parroni+17}.  \citet{Schrabback+21} combined lensing data from the NGVS and four other weak lensing surveys (KiDS/KV450, CFHTLenS, CS82, and RCSLenS) to tighten observational constraints on galaxy-scale halo ellipticity for photometrically selected lens samples. 

\subsection{Milky Way Halo and Disk}

Straddling the constellations of Virgo and Coma Berenices, the NGVS footprint intersects two of the most prominent substructures in the Galactic halo: the Virgo Overdensity and Sagittarius Stream. \citet{Lokhorst+16} studied halo main-sequence star candidates selected from the NGVS and characterized the structure of these two features on the sky, as well as along the line-of-sight. This analysis clearly demonstrated that the Sagittarius Stream slices across the NGVS field {\it twice} --- at distances of $\sim25$~kpc and $\sim$40~kpc, with a density maximum at $\sim35$~kpc.  

\citet{Fantin+17} identified several hundred white dwarf candidates in the NGVS based on a combination of optical and UV color-color diagrams, as well as NGVS-SDSS proper motion selections. Their sample was compared to the predictions of the TRILEGAL and Besan{\c c}on models and used to compute number densities for the white dwarf populations associated with the disk and halo.

\citet{Feng+26} analyzed the multi-epoch $u^*,g,i,z$ images used to generate the NGVS stacks and carried out a search for variable stars. By fitting template light-curves based on Sloan Digital Sky Survey Stripe data, they identified 180 RR Lyrae spanning the range $\sim$20~kpc to $\sim$300~kpc. They showed that the halo stellar density distribution is consistent with an $\rho \propto r^{4.09\pm0.10}$ power-law radial profile over most of this distance range, with no signs of a break. A kinematic analysis for 55 of these stars based on Keck/ESI spectra was presented in \citet{Feng+26}.

\subsection{Outer Solar System}

Individual $g$- and $i$-band frames from the NGVS were used to carry out a search for faint, moving objects at moderate ecliptic latitudes. In all, 91 trans-Neptunian objects and Centaurs were discovered in a 76 deg$^2$ region inside the NGVS footprint. Notably, this sample included 2010~GB174, a new member of the Inner Oort Cloud \citep[IOC;][]{Chen+13} with a semi-major axis of a $\simeq$ 350.8 AU, an inclination of i $\simeq$ 21\fd6, and a pericenter of q $\sim$ 48.5 AU. This Sedna-like object was one of six IOC members with $a > 250$ AU used by \citet{Batygin+16} to report evidence for a distant giant planet based on a clustering of perihelion positions and alignment of orbital planes.

\section{Summary and Conclusions} \label{sec:end}

We have presented the final catalog of galaxies in a 104 deg$^2$ area spanning the Virgo cluster from its core out to one virial radius, based on data from the CFHT/MegaCam Next Generation Virgo Cluster Survey.  The catalog comprises 3680 galaxies considered to be {\it bona fide} members of the cluster, with magnitudes in the range $g = 8.42$ to $g = 24.41$ ($M_g = -22.67$ to $M_g = -6.68$). Of these, 1483 are identified in the VCC, and an addition 97 are included in the EVCC. The remaining 2100 galaxies are new detections. An extensive set of simulations shows the NGVS catalog to be complete down to $g = 18.6$ mag ($M_g=-12.5$ mag, corresponding to a stellar mass \mstar $\sim 1.6\times10^7$ M$_{\odot}$ for an old stellar population) and 50\% complete at $g = 22.0$ mag ($M_g=-9.1$ mag, \mstar $\sim 6.2\times10^5$ M$_{\odot}$). For comparison, the VCC, which is 50\% complete at $g = 18.8$ mag ($M_g=-12.3$ mag), is a full three magnitudes shallower than the NGVS: indeed 1369 NGVS galaxies are fainter than the faintest VCC galaxy. The improvement over the VCC is equally dramatic in surface brightness, with the NGVS detecting galaxies to a surface brightness limit (averaged within an effective radius) of $\langle\mu_e\rangle \sim 29$ mag arcsec$^{-2}$, at least three magnitudes below the VCC limit. 

For each NGVS galaxy, we list photometric and structural parameters, nuclear and morphological classification, as well as stellar masses and, when available, radial velocities and SBF distances. The following are the main points discussed in the previous sections:

\begin{itemize}

\item
The NGVS provides contiguous coverage of Virgo from its core out to one virial radius, for a total area of 104 deg$^2$ (8.63 Mpc$^2$ at the 16.5 Mpc distance of Virgo, see Figure \ref{fig:fig1}). The entire footprint was imaged in the $u^*$-,$g$-,$i$- and $z$-band, while the $r$-band coverage was limited to the 3.71 deg$^2$ surrounding M87. In the $g$-band, the data reaches point- and extended-source limits of 25.9 mag (10$\sigma$) and 29 mag arcsec$^{-2}$ (2$\sigma$), respectively (see Tables \ref{tab:Tab1} and \ref{tab:Tab2}).  All images were acquired in sub-arcsecond seeing (see Tables \ref{tab:Tab2} and \ref{tab:Tab4}), with the best seeing achieved in the $i$-band (mean seeing FWHM 0.57\err 0.06).  See \S2 for more information.

\item
The NGVS depth makes it possible to detect dwarf galaxies in Virgo to a completeness limit comparable to what is achieved for classical dwarf spheroidal galaxies in the Local Group; for compact and unresolved sources, the NGVS samples the brightest 90\% of the luminosity function of GCs in Virgo at S/N $\geq$ 10. Additionally, in the $i$-band, the NGVS resolution is sufficient to marginally resolve the brightest and largest GCs, UCDs and NSCs.

\item
A dedicated code, {\sf VCands}, was developed by the NGVS team for the detection of objects in the NGVS footprint and the initial selection of possible Virgo galaxies. {\sf VCands} was designed specifically to ensure the detection of low surface brightness, extended objects that might elude standard detection algorithms, and was optimized based on a set of 393 galaxies visually identified as {\it bona fide} Virgo members. Over 11,000 objects, identified by {\sf VCands} as potential galaxies belonging to Virgo based on a set of diagnostic plots constructed using both structural and photometric parameters, were inspected by eye to assess Virgo membership. After correlation with existing spectroscopic catalogs, the final NGVS catalog of Virgo galaxies comprises 3690 objects, of which 3680 deemed {\it bona fide} members of the cluster: 1690 (46\%) are classified as certain members (Membership Class 0), 802 (22\%) as likely members (Membership Class 1), and 1188 (32\%) as possible members (Membership Class 2). Of the certain members, 996 are spectroscopically confirmed. See \S 3 for further details. 

\item
The NGVS sample is both highly pure and highly complete. Simulations employing artificial galaxies show that surface brightness is the principal driver in determining completeness: in the $g$-band, essentially all galaxies become undetectable once the average surface brightness within an effective radius falls below 29 mag arcsec$^{-2}$. Comparisons with the EVCC shows that only one spectroscopically-confirmed EVCC galaxy was missed in the original NGVS catalog — i.e. 0.03\% of the sample. This argues that the NGVS is highly complete, more so that the artificial galaxy tests would indicate, at least for galaxies brighter than $g\sim 20.5$ mag (the limit of the EVCC). Likewise, a literature search for heliocentric radial velocities for all galaxies in the NGVS catalog reveals that 10 (0.3\% of the sample) have radial velocities that places them beyond the 3,500 km s$^{-1}$ boundary adopted for Virgo, although in many cases only marginally. This argues that the NGVS sample has high purity, at least for galaxies brighter than $g\sim18$ mag, below which very few radial velocities are available. See \S 4 for further details.

\item
The presence of an NSC was assessed visually, and a morphological class was assigned to each galaxies according to the criteria discussed in \citet{Kurzner+25}. Galaxies are classified as non-nucleated (Nuclear Code 0), certainly nucleated (Nuclear Code 1),  and possibly nucleated (Nuclear Code 2); a special code (Nuclear Code 3) is reserved to galaxies for which the presence of a nucleus cannot be assessed. The morphological classification includes one code that captures the overall morphology and structure, and a second code that captures the degree to which the galaxy is actively forming stars. Both codes are subdivided into subcodes that identify the presence of disks, bars, spiral arms, and other structural features, as well as the appearance of star forming regions. See \S \ref{subsec:nuclei} and \S \ref{subsec:morph} for further details.

\item
Structural parameters were derived using a full isophotal analysis followed by parametric fits to the surface brightness profile for 897 of the NGVS galaxies, including all galaxies brighter than $g \sim 16$ mag, all galaxies in the 3.71 deg$^2$ surrounding M87, and the occasional galaxy of interest. For 3422 galaxies, of which 631 in common with the previous sample, structural parameters were derived using 2D {\sf GalFit} fits. We found no indication of  biases or systematic differences between the parameters derived using the two methods. Based on these analyses, the NGVS galaxy catalog includes, for each galaxy, magnitude, effective radius, S\'ersic index (for the outer component in the case of nucleated galaxies), central surface brightness, surface brightness at, and averaged within, the effective radius, integrated colors, and, for nucleated galaxies, an NSC magnitude. For galaxies for which a full isophotal analysis was performed, we also list effective radius, surface brightness at, and averaged within, an effective radius of the inner (NSC) component, as well as non parametric parameters,  including curve of growth total magnitude, effective radius, central surface brightness, surface brightness at, and averaged within, the effective radius, and concentration. See \S \ref{sec:par} and \S\ref{sec:fin} for further details.

\item
Masses were derived via SED fitting assuming a delayed star formation history and non-evolving metallicity. We derive both a dust-free solution, and a solution assuming a two component dust model, although we prefer the former, as we find that the limited spectral coverage of the data does not allow to meaningfully constrain the presence of dust. Comparison with mass estimates available in the literature for 290 galaxies with masses derived using UV-IR SEDs, and 60 galaxies with masses derived using optical spectroscopy, show very good agreement. See \S\ref{sec:mass} for further details.

\item
Finally, we list some of the most notable science results derived using NGVS data. These range from a study of trans-Neptunian objects, to tomography of the Milky Way halo, to the detection of high redshift ($z < 1$) galaxy clusters. See \S\ref{sec:results} for further details.
\end{itemize}

We conclude by highlighting three papers that will soon be published based on the full NGVS galaxy catalog. The work of \citet{Ferrarese+16}, who explored the galaxy luminosity and mass function in the 3.71 deg$^2$ surrounding M87, will be expanded to the full cluster in Todd et al. (2026, in preparation), who will investigate the dependence of the luminosity and mass function on color, morphology and environment. The spatial structure of the cluster was discussed in \citet{Cantiello+24} based on SBF distance to 278 cluster members; updated structural parameters for the cluster as a whole as well as its main substructures will be presented in Todd et al. (2026b, in preparation) based on the full galaxy catalog and employing a data clustering algorithm designed to quantitatively identify density-based clustering. Finally, \citet{Sanchez-Janssen+19a} analyzed the nucleation fraction as well as correlation between NSC and overall galaxy properties in the core of Virgo. This work is expanded to the full cluster in Kurzner et al. (2026, in preparation) who will present a comprehensive study of NSCs, including the dependence of their properties on galaxy mass, morphology and environment. 

\section{Acknowledgments}

Based on observations obtained at the Canada-France-Hawai'i Telescope (CFHT) which is operated by the National Research Council of Canada, the Institut National des Sciences de l'Univers of the Centre National de la Recherche Scientifique of France, and the University of Hawai'i. CFHT is located on Maunakea on Hawai'i Island, a mountain of considerable cultural, natural, and ecological significance. Maunakea is a sacred site to Native Hawaiians, also known as Kānaka 'Oiwi. We would like to thank the Canada-France-Hawai'i Telescope (CFHT) Operations and Software Groups for their contributions and diligence in maintaining observatory operations; the CFHT Astronomy Group for their observation coordination and data acquisition efforts; and the CFHT Finance \& Administration Group for their contributions to the management and administration of the observatory. Based on observations obtained with MegaPrime/MegaCam, a joint project of CFHT and CEA/DAPNIA.

This research used the Canadian Advanced Network For Astronomy Research (CANFAR) operated in partnership by the Canadian Astronomy Data Centre and The Digital Research Alliance of Canada with support from the National Research Council of Canada the Canadian Space Agency, CANARIE and the Canadian Foundation for Innovation.

ET is thankful for the support from NSF-AST-2206498 grant. RS acknowledges financial support from FONDECYT Regular 2023 project No. 1230441 and also gratefully acknowledges financial support from ANID-MILENIO NCN2024\_112. CL acknowledges support from the National Natural Science Foundation of China (NSFC, Grant No. 12595311, 12173025), the National Key R\&D Program of China (2023YFA1607800, 2023YFA1607804), the 111 Project (No. B20019), and the Key Laboratory for Particle Physics, Astrophysics and Cosmology, Ministry of Education. THP gratefully acknowledges support from the Chilean National Agency for Research and Development (ANID) in form of the CATA-Basal grant (FB210003). MC acknowledges support from ASI–INAF grant no. 2024-10-HH.0 (WP8420), the ESO Scientific Visitor Programme, and INAF GO-grant no. 12/2024. JET acknowledges support from the Natural Sciences and Engineering Research Council of Canada (NSERC) through a Discovery Grant. 

This paper is dedicated to KBB: Never give up! Never surrender!

\clearpage
\bibliography{sample631}{}
\bibliographystyle{aasjournal}

\clearpage
\appendix

\section{r-band Short exposures}\label{app:short}
As mentioned in \S\ref{sec:data}, full depth $r$-band exposures were only acquired for the 4 deg$^2$ surrounding M87 ({\sf NGVS+0+0, NGVS+0+1, NGVS-1+0}, and {\sf NGVS-1+1}, see Figure \ref{fig:fig1}). Shallower exposures were acquired for additional fields, and are listed in Table \ref{tab:rband}, although not used in this paper.

\section{Notes on Individual Galaxies}\label{app:notes}
Brief notes on individual galaxies are given in Table \ref{tab:notes}.

\section{Additional Parameters for galaxies fitted by a core-S\'ersic Law}\label{app:CS}
Table \ref{tab:coresersic} lists additional parameters derived for the five galaxies best fitted with a core-S\'ersic profile \citep{Graham+03}. The functional form of the profile is:
$$I(r) = I_b  \left[ 1+ \left(\frac{R_b}{r} \right)^{\alpha} \right]^{\gamma/\alpha} {\rm exp} \left[ b_n \left( \frac{R^{\alpha}+R_b^{\alpha}}{R_e^{\alpha}}\right)^{1/(\alpha n)}\right],$$ 
where the S\'ersic index $n$ and the effective radius $R_e$ are given in Table \ref{tab:Tab8} (note that $b_n$ is a function of $n$ and not an independent parameter).

\section{Consistency Checks}\label{app:consistency}

\begin{figure}
    \centering
    \includegraphics[width=0.45\textwidth]{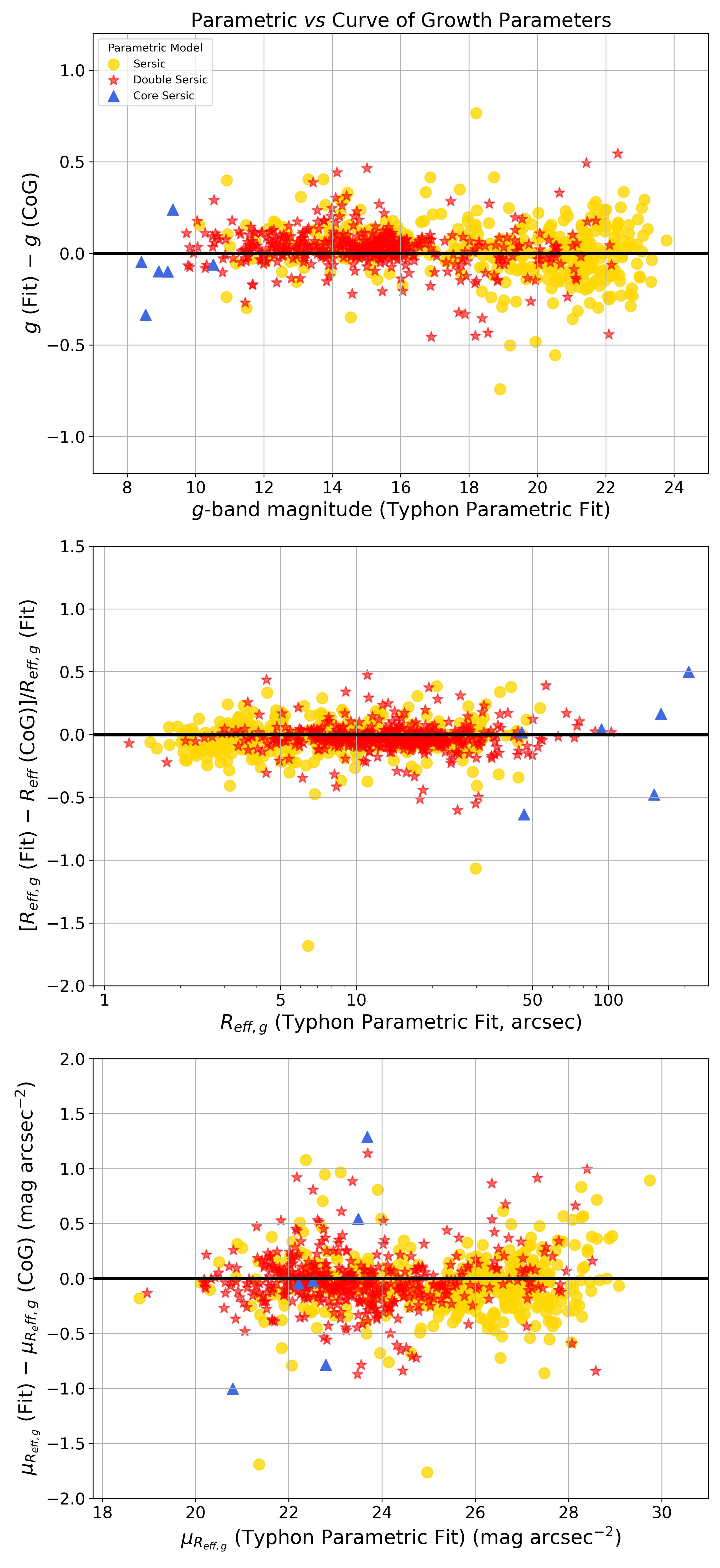}
    \includegraphics[width=0.45\textwidth]{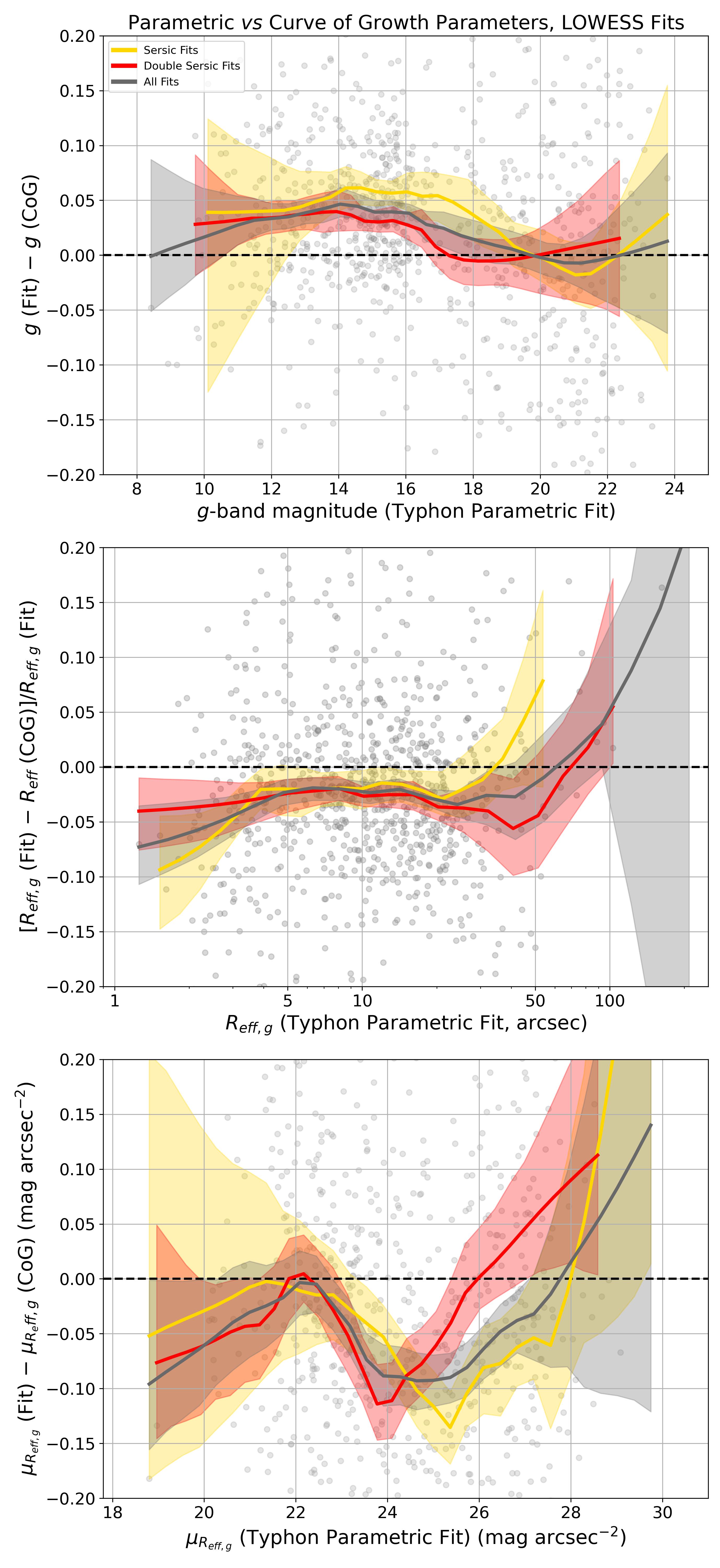}
    \caption{{\it Left panels}: Comparison of $g$-band magnitude, geometric effective radius $R_{\rm eff}$, and surface brightness at $R_{\rm eff}$, $\mu_{R_{\rm {eff}}}$, derived from {\sf Typhon}'s parametric fits to the surface brightness profile (identified with the suffix `Fit'), and from the curve-of-growth analysis (identified with the suffix `CoG'), as described in \S \ref{subsec:typhon}. The 897 galaxies analyzed with {\sf Typhon} are color-coded according to the components included in fitting the profile: S{\'e}rsic,  double S{\'e}rsic, or core-S{\'e}rsic, as shown in the legend. See text for further details.
    {\it Right panels}: as the left panels, but showing Locally Weighted Scatterplot Smooth (LOWESS) fits and associated 95\% confidence limits to the data. The fits are performed for the entire sample, and separately for galaxies fitted with a S{\'e}rsic and double S{\'e}rsic profile, as shown in the legend. Data points are shown as gray dots; note, however, that to better show the fits, the y-axis has been expanded compared to the left panels and some points fall outside the range plotted.}
    \label{fig:fig13}
\end{figure}

\begin{figure}
    \centering
    \includegraphics[width=0.45\textwidth]{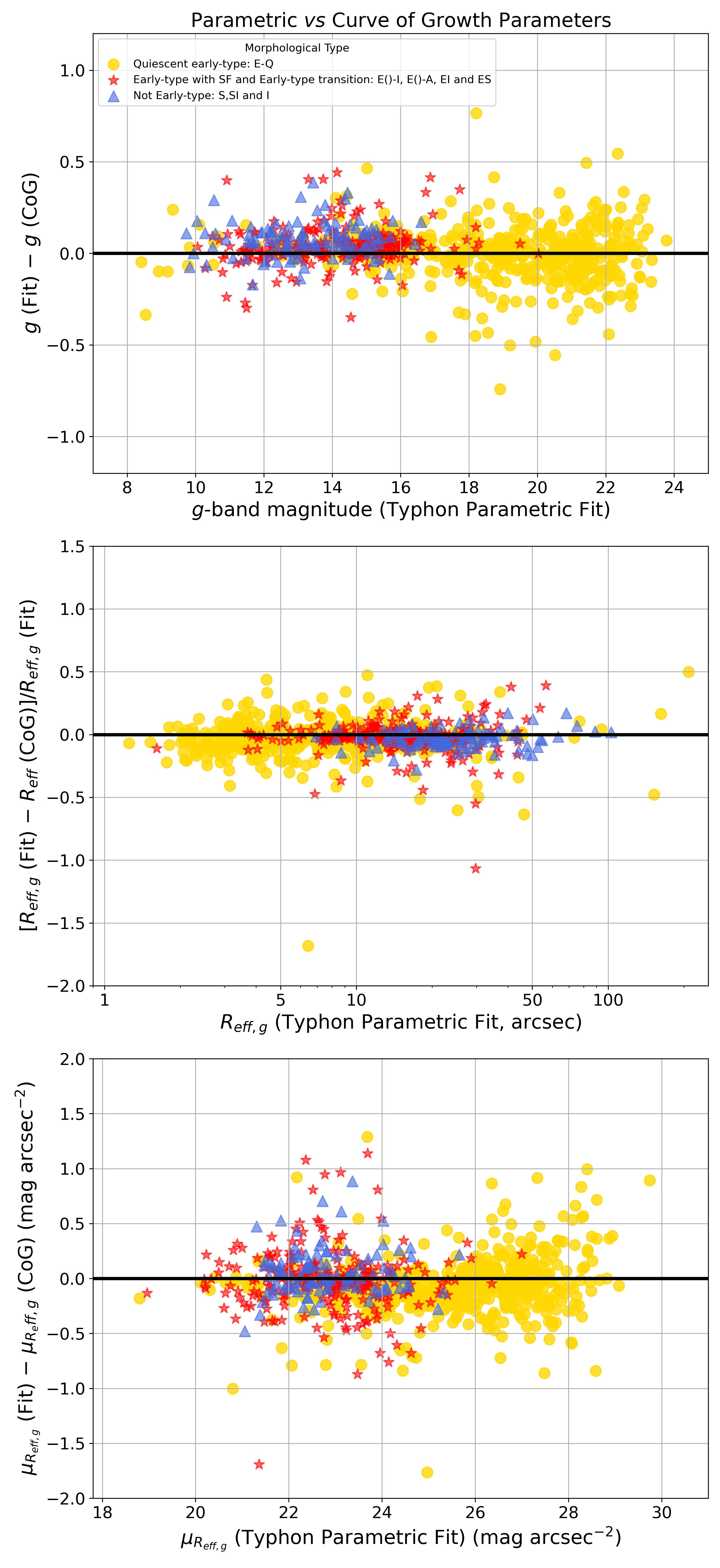}
    \includegraphics[width=0.45\textwidth]{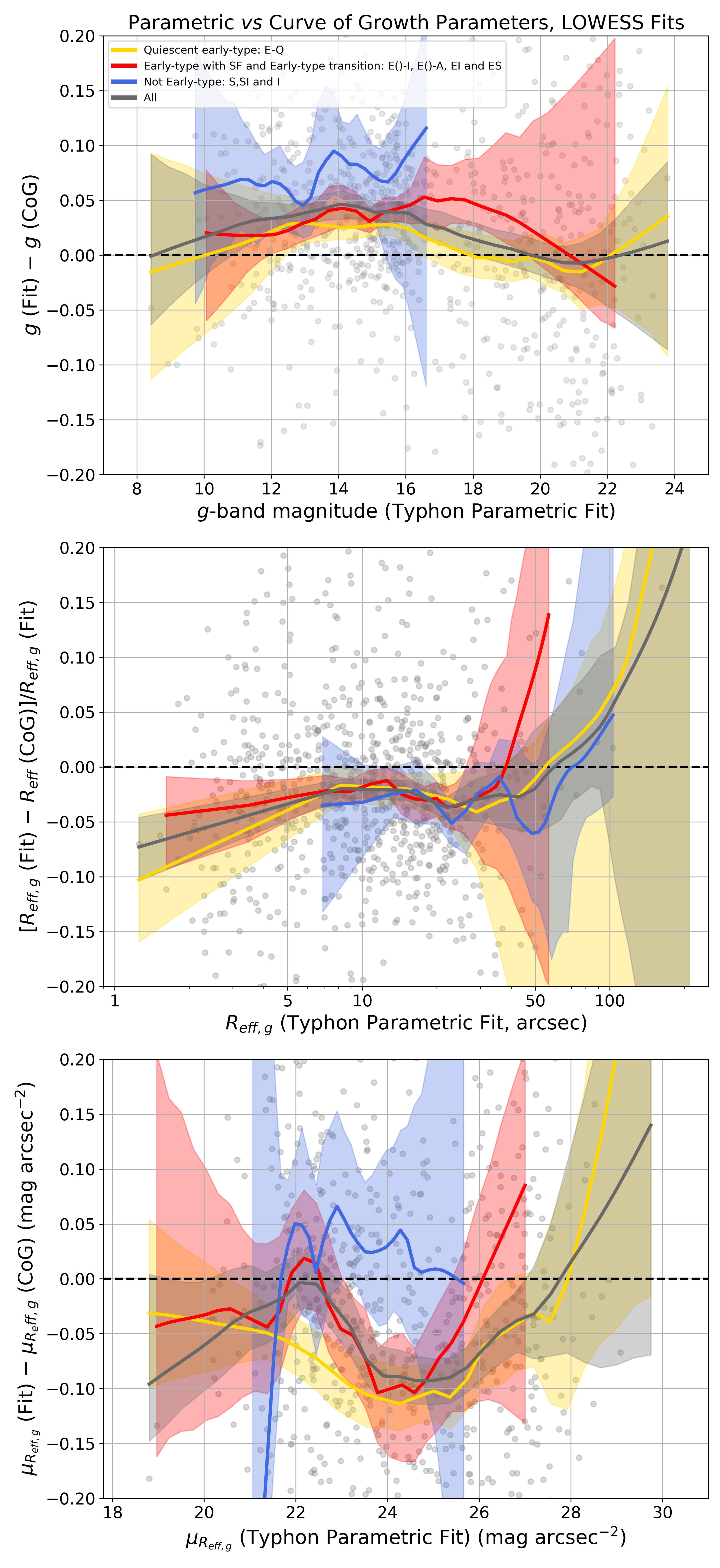}
    \caption{{\it Left panels}: As Figure \ref{fig:fig13}, but with galaxies color-coded according to morphology: quiescent early-type galaxies (i.e. E-Q morphology); early-type galaxies with some level of star formation (E-I and E-A as well as early-type to irregular or spiral transition objects (EI and ES); and all Spiral (S), Irregular (I) or transition (SI) galaxies (see \S\ref{subsec:morph}). See text for further details.
    {\it Right panels}: as the left panels, but showing LOWESS fits and associated 95\% confidence limits to the data. The fits are performed using the entire sample, and separately for galaxies of different morphologies, as shown in the legend. Data points are shown as gray dots; note, however, that to better show the fits, the y-axis has been expanded compared to the let panels and some points fall outside the range plotted.}
    \label{fig:fig14}
\end{figure}

\begin{figure}
    \centering
    \includegraphics[width=0.45\textwidth]{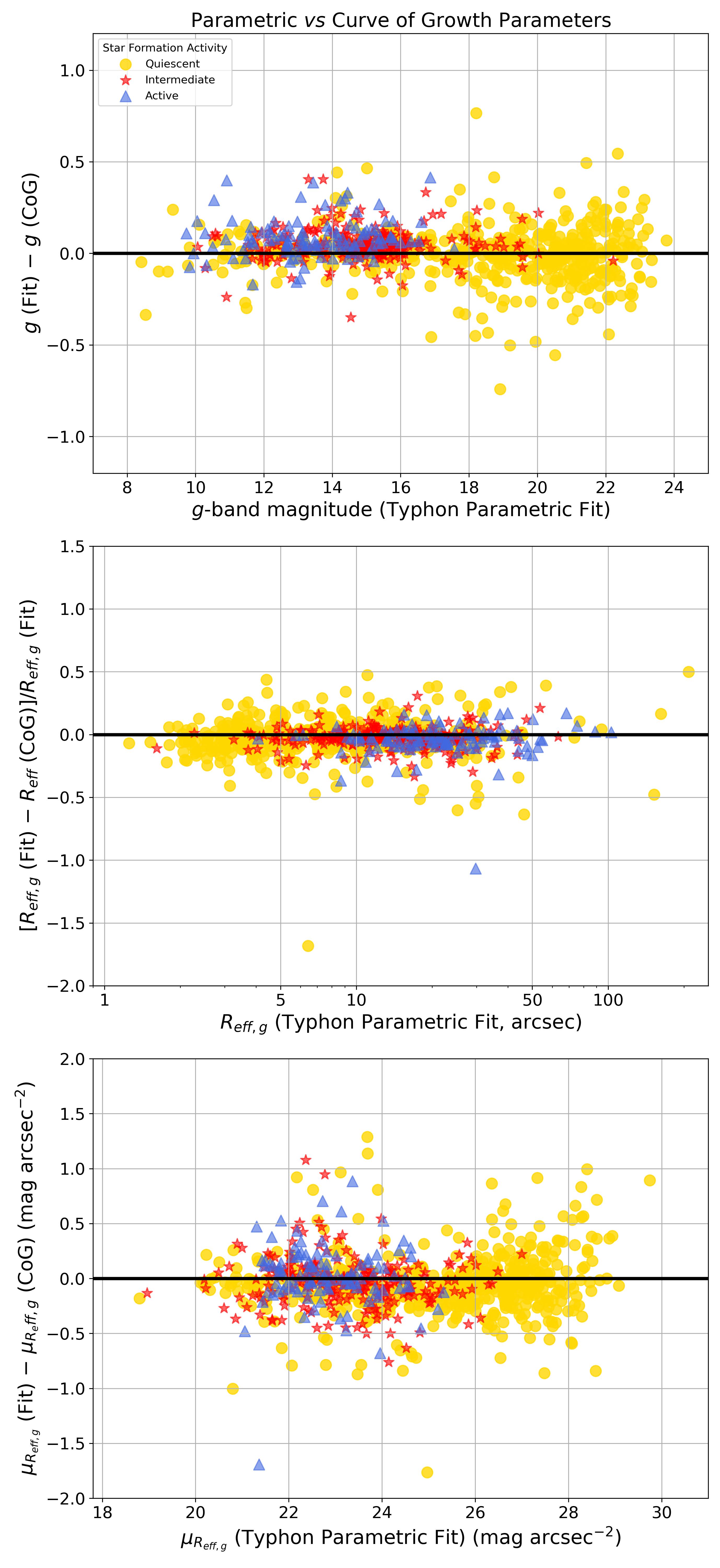}
    \includegraphics[width=0.45\textwidth]{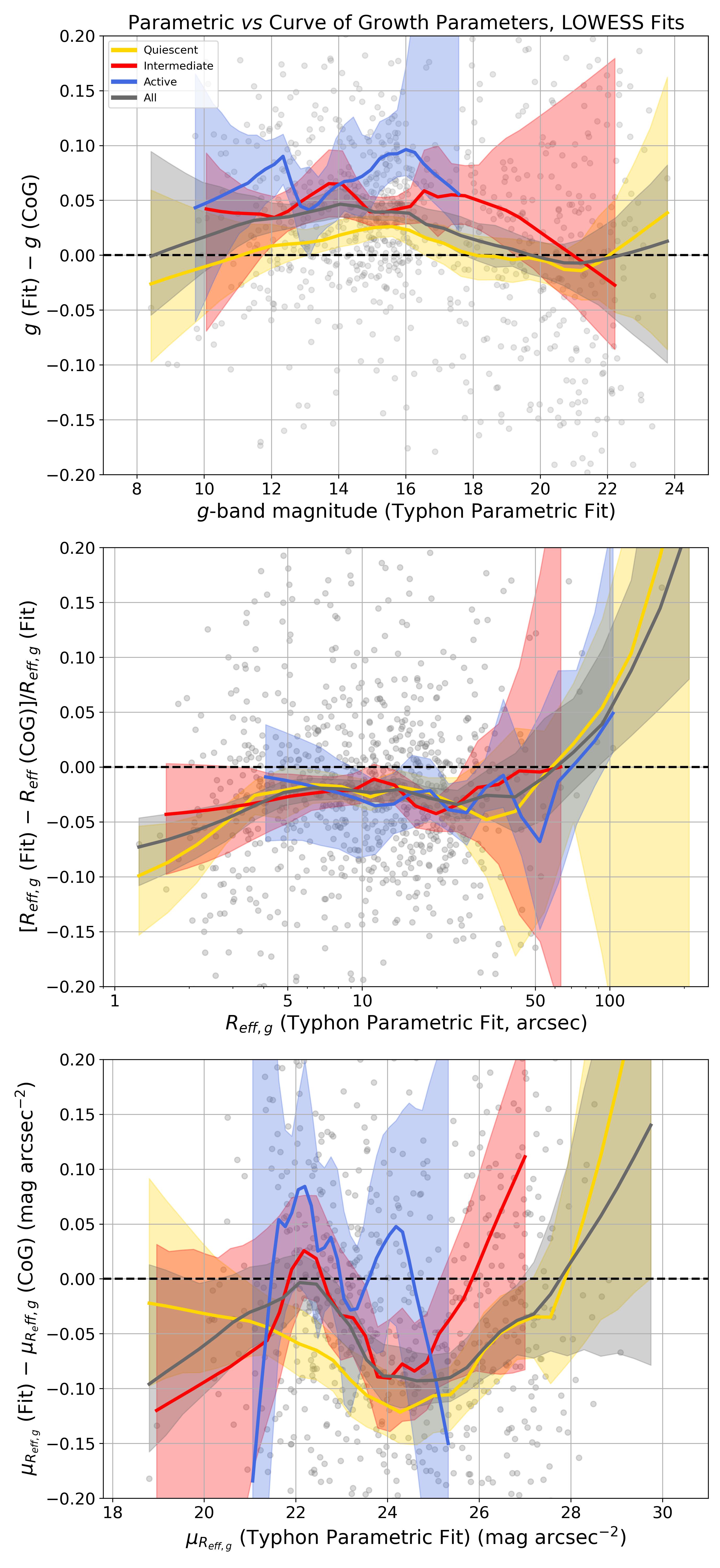}
    \caption{{\it Left Panels:} as Figure \ref{fig:fig13}, but with galaxies color-coded according to global star formation activity: Quiescent, Intermediate, and Active (see \S\ref{subsec:morph}). See text for further details.
    {\it Right Panels:} as the left panels, but showing LOWESS fits and associated 95\% confidence limits to the data. The fits are performed for the entire sample, and separately for galaxies of different star formation activity, as shown in the legend. Data points are shown as gray dots;  note, however, that to better show the fits, the y-axis has been expanded compared to the left panels and some points fall outside the range plotted.}
    \label{fig:fig15}
\end{figure}

\begin{figure}
    \centering
    \includegraphics[width=0.45\textwidth]{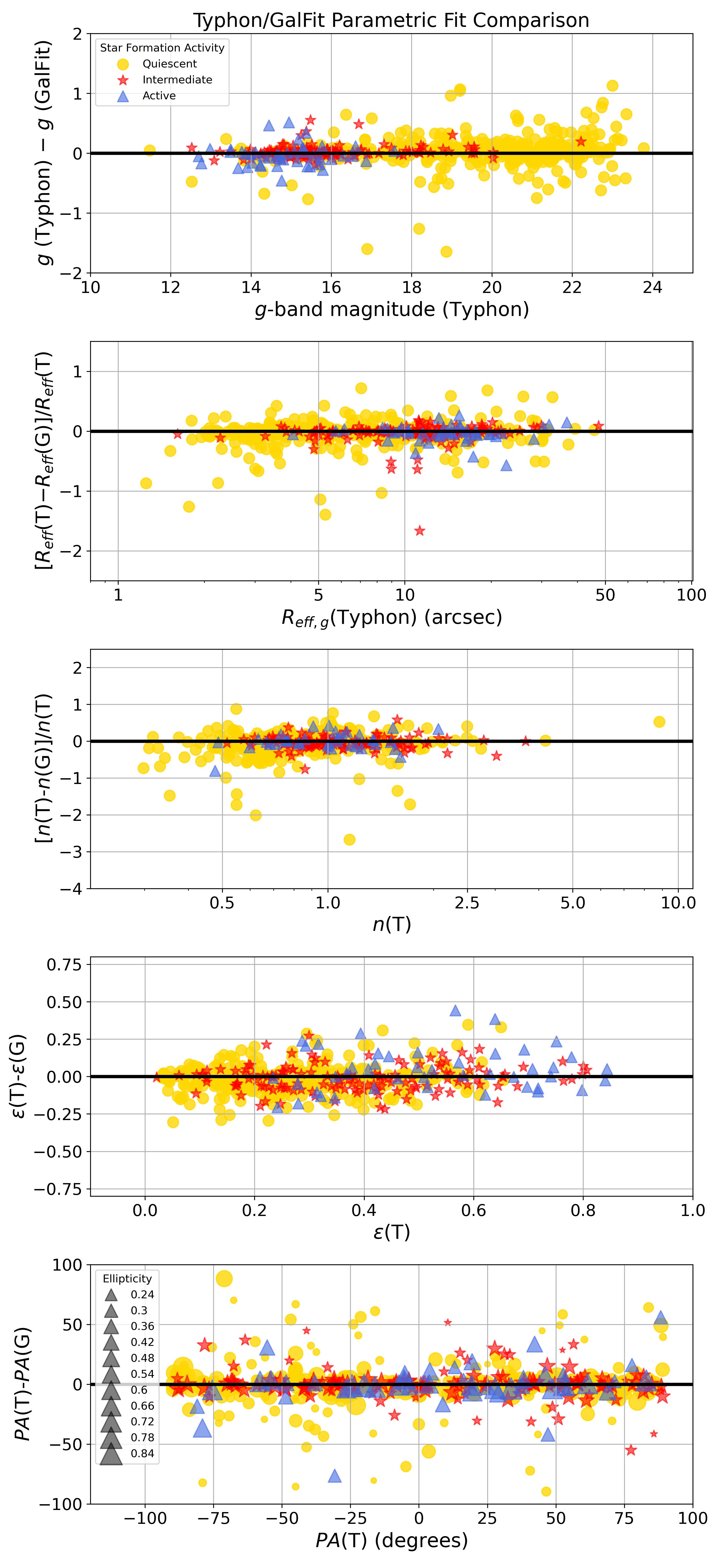}
        \includegraphics[width=0.45\textwidth]{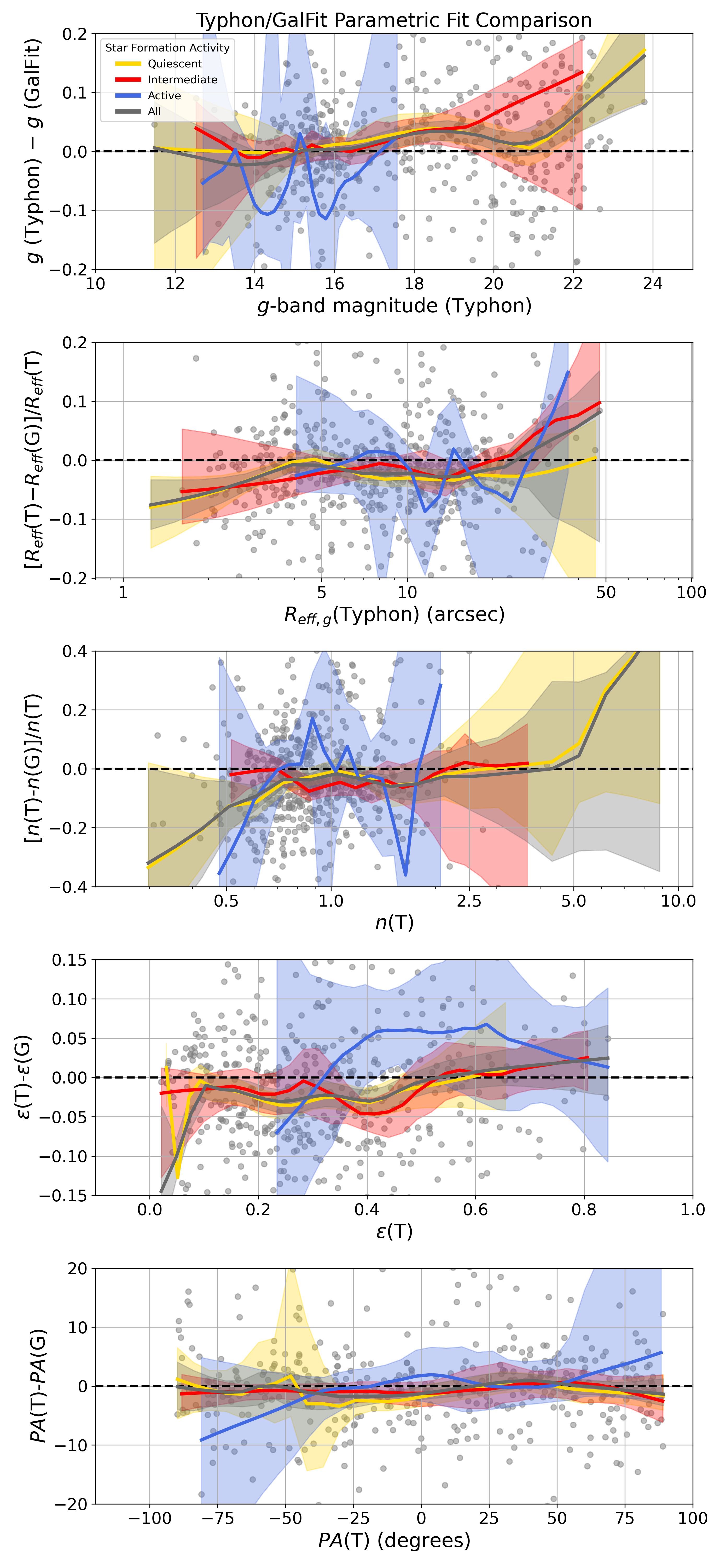}
    \caption{{\it Left Panels}: Comparison of $g$-band magnitude, geometric effective radius $R_{\rm eff}$, S{\'e}rsic index, ellipticity, $\epsilon$, and major axis position angle, $PA$, derived from {\sf Typhon}'s parametric fits to the surface brightness profile, and from the {\sf GalFit} analysis, as described in \S \ref{subsec:typhon} and \S \ref{subsec:galfit}. A total of 631 galaxies were analyzed using both methods. Galaxies are color-coded according to the level of star formation (Quiescent, Intermediate, and Active) (see \S \ref{subsec:morph}). In the bottom panel, galaxies are plotted with size proportional to the ellipticity, as shown in the legend. See text for further details.
    {\it Right Panels:} as the left panels, but showing LOWESS fits and associated 95\% confidence limits to the data. The fits are performed for the entire sample, and separately for galaxies with varying level of star formation, as shown in the legend. Data points are shown as gray dots; note, however, that to better show the fits, the y-axis has been expanded compared to the left panels and some points fall outside the range plotted.}
    \label{fig:fig16}
\end{figure}

As mentioned in \S \ref{subsec:typhon}, for the galaxies analyzed with {\sf Typhon}, two sets of parameters are available: non parametric derived from the curve of growth analysis, and parametric derived from a functional fit to the surface brightness profile. Figures \ref{fig:fig13} to \ref{fig:fig16} show a comparison between the two approaches for $g$-band magnitude, effective radius, and effective surface brightness. In the figures, galaxies are color-coded according to different criteria: the function used in fitting the surface brightness profiles (Figure \ref{fig:fig13}), the galaxy morphological type (Figure \ref{fig:fig14}), and the level of star formation activity estimated visually from the $u^*,g,i$ and $z$-band images, as well as unsharped masked images (Figure \ref{fig:fig15}, see also \S\ref{subsec:morph}). There is a small bias in the measured magnitude and effective surface brightness for galaxies in the intermediate magnitude range ($12 \lesssim g \lesssim 16$ mag); this seems to correlate mostly with the level of star formation, with galaxies classified as ``active" being the most discrepant (Figure \ref{fig:fig15}), in the sense of their total magnitude being brighter when measured non-parametrically. This is easily understood as the surface brightness profile, and the parametric fit to it, will effectively ``smooth over" star forming regions, thus leading to fainter magnitudes than those based on a curve of growth. For this reason, when available, we deem the curve-of-growth parameters to be more accurate. 

Also as mentioned in \S \ref{subsec:galfit}, 631 galaxies with magnitude ranging from $g \sim 12$ to $g \sim 24$ were analyzed with both {\sf Typhon} and {\sf GalFit}, with the goal of determining whether there are any biases or systematic differences between the parameters derived based on the two methods.  A comparison between {\sf Typhon} and {\sf GalFit} magnitudes, effective radii, effective surface brightnesses, ellipticities, and major axis position angles is shown in Figure \ref{fig:fig16}, with galaxies color-coded according to the level of star-formation activity. There is no discernible systematic difference between the parameters derived by the two methods.

\clearpage

\pagebreak

\pagebreak

\small
\startlongtable
\begin{deluxetable*}{llccc}
\tablecaption{NGVS CFHT/MegaCam Short Exposures: Summary\tablenotemark{$\dagger$}} \label{tab:Tab1}
\tablehead{
\multicolumn{1}{c}{Filter} &       
\multicolumn{1}{c}{Observing Dates} &             
\multicolumn{1}{c}{No. of Fields} &             
\multicolumn{1}{c}{Average Seeing} &             
\multicolumn{1}{c}{Point Source Limit}\\
\multicolumn{1}{c}{} &       
\multicolumn{1}{c}{(Y-M-D)} &            
\multicolumn{1}{c}{} &            
\multicolumn{1}{c}{(arcsec)} &              
\multicolumn{1}{c}{(mag arcsec$^{-2}$)} \\
\multicolumn{1}{c}{(1)} &       
\multicolumn{1}{c}{(2)} &            
\multicolumn{1}{c}{(3)} &            
\multicolumn{1}{c}{(4)} &            
\multicolumn{1}{c}{(5)}
}          
\startdata
\phantom{0l}$u^*$& 2008 Mar 14 $-$ 2013 Jan 20 & \phantom{1}59 & 0.90 \err 0.10 & 24.3\\
\phantom{0l}$g$ & 2008 Mar \phantom{1}5 $-$ 2013 Mar \phantom{l}9 & 117 & 0.78 \err 0.11 & 23.2\\
\phantom{0l}$r$ & 2008 Mar \phantom{1}1 $-$ 2009 Apr 26 & \phantom{11}6 & 0.86 \err 0.25 & 22.2\\
\phantom{0l}$i$ & 2008 Mar \phantom{1}3 $-$ 2013 Mar \phantom{l}5 & \phantom{1}67 & 0.57 \err 0.06 & 21.7\\
\phantom{0l}$z$ & 2008 Mar \phantom{1}1 $-$ 2013 Jan 14 & 117 & 0.79 \err 0.14 & 21.2\\
\enddata 
\tablenotetext{\dagger}{Columns are as follows: (1) CFHT/MegaCam filter in which the observations were acquired; (2) the range of dates when the observations took place; (3) the number of fields observed; (4) the mean and standard deviation of the seeing, taken over all exposures; (5) the 5$\sigma$ point sources limit of the data. Note that $u^*$-, $r$-, and $i$-band short exposures were not acquired for all NGVS fields (see Table \ref{tab:Tab3} and Figures~\ref{fig:fig2}~and~\ref{fig:fig3}).}
\end{deluxetable*}
\normalsize
\vskip .4in

\small
\startlongtable
\begin{deluxetable*}{llcccc}
\tablecaption{NGVS CFHT/MegaCam Long Exposures: Summary\tablenotemark{$\dagger$}} \label{tab:Tab2}
\tablehead{
\multicolumn{1}{c}{Filter} &       
\multicolumn{1}{c}{Observing Dates} &             
\multicolumn{1}{c}{No. of Fields} &             
\multicolumn{1}{c}{Average Seeing} &             
\multicolumn{1}{c}{Point Source Limit} &          
\multicolumn{1}{c}{Extended Source Limit} \\
\multicolumn{1}{c}{} &       
\multicolumn{1}{c}{(Y-M-D)} &            
\multicolumn{1}{c}{} &            
\multicolumn{1}{c}{(arcsec)} &          
\multicolumn{1}{c}{(mag)} &          
\multicolumn{1}{c}{(mag arcsec$^{-2}$)} \\
\multicolumn{1}{c}{(1)} &       
\multicolumn{1}{c}{(2)} &            
\multicolumn{1}{c}{(3)} &            
\multicolumn{1}{c}{(4)} &            
\multicolumn{1}{c}{(5)} &          
\multicolumn{1}{c}{(6)} 
}
\startdata       
\phantom{0l}$u^*$ & 2009 Feb 27 $-$ 2013 Jun 12  & 117 & 0.88 \err 0.06 & 26.3 & 29.3 \\
\phantom{0l}$g$ & 2009 Feb 22 $-$ 2011 Mar 31  & 117 & 0.83 \err 0.09 & 25.9 & 29.0 \\
\phantom{0l}$r$\tablenotemark{$\ddagger$} & 2008 Mar \phantom{l}5 $-$ 2010 Jun 14  & \phantom{11}4 & 0.76 \err 0.13 & 26.3 & 29.3 \\
\phantom{0l}$i$ & 2009 Feb 18 $-$ 2012 Mar 29  & 117 & 0.58 \err 0.03 & 25.1 & 27.4 \\
\phantom{0l}$z$ & 2009 Feb 22 $-$ 2012 Mar 30  & 117 & 0.78 \err 0.07 & 24.8 & 26.0 \\
\enddata 
\begin{center}\begin{minipage}{0.9\linewidth}
\tablenotetext{\dagger}{Columns are as in Table \ref{tab:Tab1}}
\tablenotetext{\ddagger}{Full-depth $r$-band images were  obtained only for the `core region': approximately four square degrees roughly centered on M87. For completeness, $r$-band observations are listed in Tables~\ref{tab:Tab1}~and~\ref{tab:Tab2}, but no $r$-band data are included in this paper. For $r$-band photometry for the galaxies in the core region, see NGVS-XIV.}
\end{minipage}
\end{center} 
\end{deluxetable*}
\normalsize

\small
\begin{longrotatetable}
\startlongtable
\begin{deluxetable*}{lcccccccccc}
\tablecaption{NGVS CFHT/MegaCam Short Exposures: Details of the Observations\tablenotemark{$\dagger$}} \label{tab:Tab3}
\tablehead{
\multicolumn{3}{c}{} &  
\multicolumn{2}{c}{$u^*$-band} &  
\multicolumn{2}{c}{$g$-band} &  
\multicolumn{2}{c}{$i$-band} &   
\multicolumn{2}{c}{$z$-band} \\
\multicolumn{3}{c}{} &  
\multicolumn{2}{c}{\rule{100pt}{0.5pt}} &  
\multicolumn{2}{c}{\rule{100pt}{0.5pt}} &  
\multicolumn{2}{c}{\rule{100pt}{0.5pt}} &   
\multicolumn{2}{c}{\rule{100pt}{0.5pt}} \\
\multicolumn{1}{c}{Field} &       
\multicolumn{1}{c}{RA} &             
\multicolumn{1}{c}{DEC} &          
\multicolumn{1}{c}{Exposures} &  
\multicolumn{1}{c}{FWHM} &      
\multicolumn{1}{c}{Exposures} & 
\multicolumn{1}{c}{FWHM} &    
\multicolumn{1}{c}{Exposures} &
\multicolumn{1}{c}{FWHM} &       
\multicolumn{1}{c}{Exposures} &  
\multicolumn{1}{c}{FWHM} \\
\multicolumn{1}{c}{} &       
\multicolumn{1}{c}{(h:m:s)} &            
\multicolumn{1}{c}{(d:m:s)} &          
\multicolumn{1}{c}{N/t(s)} &  
\multicolumn{1}{c}{(arcsec)} &      
\multicolumn{1}{c}{N/t(s)} &  
\multicolumn{1}{c}{(arcsec)} &      
\multicolumn{1}{c}{N/t(s)} &  
\multicolumn{1}{c}{(arcsec)} &      
\multicolumn{1}{c}{N/t(s)} &  
\multicolumn{1}{c}{(arcsec)} \\
\multicolumn{1}{c}{(1)} &       
\multicolumn{1}{c}{(2)} &            
\multicolumn{1}{c}{(3)} &          
\multicolumn{1}{c}{(4)} &  
\multicolumn{1}{c}{(5)} &      
\multicolumn{1}{c}{(6)} &  
\multicolumn{1}{c}{(7)} &      
\multicolumn{1}{c}{(8)} &  
\multicolumn{1}{c}{(9)} &      
\multicolumn{1}{c}{(10)} &  
\multicolumn{1}{c}{(11)} 
}                         
\startdata
NGVS+0+0 & 12:32:11 & +12:00:37 & 5/250 & 1.01\arcsec & 5/60 & 0.72\arcsec & 5/40 & 0.53\arcsec & 5/65 & 1.04’’\\ 
NGVS+0+1 & 12:32:11 & +12:56:37 & 5/250 & 1.12\arcsec & 5/60 & 0.60\arcsec & 5/40 & 0.76\arcsec & 5/65 & 0.94\arcsec \\
NGVS+0+2 & 12:32:11 & +13:54:35 & 5/250 & 0.94\arcsec & 5/60 & 0.66\arcsec & 5/40 & 0.66\arcsec & 5/65 & 0.86\arcsec \\
NGVS+0+3 & 12:32:11 & +14:51:35 & \nd & \nd & 5/60 & 0.72\arcsec & 5/40 & 0.62\arcsec & 5/65 & 0.82\arcsec \\
NGVS+0+4 & 12:32:12 & +15:48:36 & 5/250 & 0.65\arcsec & 5/60 & 0.70\arcsec & \nd & \nd & 5/65 & 0.92\arcsec \\
NGVS+0+5 & 12:32:11 & +16:45:34 & 5/250 & 0.71\arcsec & 5/60 & 0.59\arcsec & 5/40 & 0.52\arcsec & 5/65 & 0.63\arcsec \\
NGVS+0+6 & 12:32:11 & +17:42:33 & \nd & \nd & 5/60 & 0.83\arcsec & \nd & \nd & 5/65 & 0.74\arcsec \\
NGVS+0$-$1 & 12:32:11 & +11:03:34 & 5/250 & 0.98\arcsec & 5/60 & 0.68\arcsec & 5/40 & 0.63\arcsec & 5/65 & 0.63\arcsec \\
\enddata 
\begin{center}\begin{minipage}{0.8\linewidth}
\tablenotetext{\dagger}{Columns are as follows: (1) the name of the NGVS field, with the two numbers roughly indicating the location of the field measured in degrees East and North of M87, with NGVS+0+0 being the field that includes M87; (2) the J2000 Right Ascention and (3) Declination of the field's center; (4) the number of exposures and the total exposure time, in seconds, in the $u^*$-band; (5) the seeing FWHM, averaged across the full 1 deg$^{2}$ stacked image, in the $u^*$-band. Columns (6) to (10) are equivalent to columns (4) and (5), but for the $g-$, $i-$, and $z$-band. For details about the $r$-band data acquired for the $\sim 4$ deg$^2$ roughly centered on M87,  see NGVS-XIV.}
\end{minipage}\end{center}
\end{deluxetable*}
\end{longrotatetable}
\normalsize

\small
\begin{longrotatetable}
\startlongtable
\begin{deluxetable*}{lcccccccccc}
\tablecaption{NGVS CFHT/MegaCam Long Exposures: Details of the Observations\tablenotemark{$\dagger$}} \label{tab:Tab4}
\tablehead{
\multicolumn{3}{c}{} &  
\multicolumn{2}{c}{$u^*$-band} &  
\multicolumn{2}{c}{$g$-band} &  
\multicolumn{2}{c}{$i$-band} &   
\multicolumn{2}{c}{$z$-band} \\
\multicolumn{3}{c}{} &  
\multicolumn{2}{c}{\rule{100pt}{0.5pt}} &  
\multicolumn{2}{c}{\rule{100pt}{0.5pt}} &  
\multicolumn{2}{c}{\rule{100pt}{0.5pt}} &   
\multicolumn{2}{c}{\rule{100pt}{0.5pt}} \\
\multicolumn{1}{c}{Field} &       
\multicolumn{1}{c}{RA} &             
\multicolumn{1}{c}{DEC} &          
\multicolumn{1}{c}{Exposures} &  
\multicolumn{1}{c}{FWHM} &      
\multicolumn{1}{c}{Exposures} & 
\multicolumn{1}{c}{FWHM} &    
\multicolumn{1}{c}{Exposures} &
\multicolumn{1}{c}{FWHM} &       
\multicolumn{1}{c}{Exposures} &  
\multicolumn{1}{c}{FWHM} \\
\multicolumn{1}{c}{} &       
\multicolumn{1}{c}{(h:m:s)} &            
\multicolumn{1}{c}{(d:m:s)} &          
\multicolumn{1}{c}{N,t(s)} &  
\multicolumn{1}{c}{(arcsec)} &      
\multicolumn{1}{c}{N/t(s)} &  
\multicolumn{1}{c}{(arcsec)} &      
\multicolumn{1}{c}{N/t(s)} &  
\multicolumn{1}{c}{(arcsec)} &      
\multicolumn{1}{c}{N/t(s)} &  
\multicolumn{1}{c}{(arcsec)} \\
\multicolumn{1}{c}{(1)} &       
\multicolumn{1}{c}{(2)} &            
\multicolumn{1}{c}{(3)} &          
\multicolumn{1}{c}{(4)} &  
\multicolumn{1}{c}{(5)} &      
\multicolumn{1}{c}{(6)} &  
\multicolumn{1}{c}{(7)} &      
\multicolumn{1}{c}{(8)} &  
\multicolumn{1}{c}{(9)} &      
\multicolumn{1}{c}{(10)} &  
\multicolumn{1}{c}{(11)} 
}       
\startdata
NGVS+0+0\tablenotemark{$\ddagger$} & 12:32:11 & +12:00:37 & 11/6402 & 1.42\arcsec & 5/3170 & 1.22\arcsec & 5/2055 & 1.33\arcsec & 8/4400 & 1.31\arcsec \\
NGVS+0+0 & 12:32:11 & +12:00:37 & \phantom{1}8/4646 & 0.80\arcsec & 5/3170 & 0.62\arcsec & 5/2055 & 0.54\arcsec & 8/4400 & 0.97\arcsec \\
NGVS+0+1\tablenotemark{$\ddagger$} & 12:32:11 & +12:56:37 & 11/6402 & 1.41\arcsec & 5/3170 & 1.36\arcsec & 5/2055 & 1.28\arcsec & 8/4400 & 1.28\arcsec \\
NGVS+0+1 & 12:32:11 & +12:56:37 & \phantom{1}8/4646 & 1.09\arcsec & 5/3170 & 0.63\arcsec & 5/2055 & 0.68\arcsec & 8/4400 & 1.02\arcsec \\
NGVS+0+2 & 12:32:11 & +13:54:35 & 11/6402 & 0.96\arcsec & 5/3170 & 0.97\arcsec & 5/2055 & 0.59\arcsec & 7/3850 & 0.68\arcsec \\
NGVS+0+3 & 12:32:11 & +14:51:35 & 11/6402 & 0.94\arcsec & 5/3170 & 0.86\arcsec & 5/2055 & 0.60\arcsec & 7/3850 & 0.71\arcsec \\
NGVS+0+4 & 12:32:12 & +15:48:36 & 11/6402 & 0.86\arcsec & 5/3170 & 0.91\arcsec & 5/2055 & 0.54\arcsec & 7/3850 & 0.78\arcsec \\
NGVS+0+5 & 12:32:11 & +16:45:34 & 10/5820 & 0.83\arcsec & 5/3170 & 0.81\arcsec & 5/2055 & 0.57\arcsec & 8/4400 & 0.78\arcsec \\
\enddata 
\begin{center}\begin{minipage}{0.8\linewidth}
\tablenotetext{\dagger}{Columns are as in Table \ref{tab:Tab3}}
\tablenotetext{\ddagger}{These data were acquired between 2008 March 1 and 2008 March 12 prior to the start of the main survey, using a standard observing strategy that is not conducive to the {\sf Elixir-LSB} processing.}
\end{minipage}\end{center}
\end{deluxetable*}
\end{longrotatetable}
\normalsize

\small
\begin{longrotatetable}
\startlongtable
\begin{deluxetable*}{lccccclll}
\tablecaption{NGVS Galaxy catalog: General Properties I\tablenotemark{$\dagger$}} \label{tab:Tab5}
\tablehead{
\multicolumn{1}{c}{Nickname} &       
\multicolumn{1}{c}{Official Name} &             
\multicolumn{1}{c}{R.A.} &          
\multicolumn{1}{c}{Dec.} &
\multicolumn{1}{c}{Membership} &       
\multicolumn{1}{c}{VCC ID} &            
\multicolumn{1}{c}{$v$\err $\delta v$} &          
\multicolumn{1}{c}{$v$ Ref.} &            
\multicolumn{1}{c}{($m-M$) \err $\sigma$} \\
\multicolumn{1}{c}{} &       
\multicolumn{1}{c}{} &            
\multicolumn{1}{c}{(deg)} &          
\multicolumn{1}{c}{(deg)} &  
\multicolumn{1}{c}{Class} &      
\multicolumn{1}{c}{} &  
\multicolumn{1}{c}{(km s$^{-1}$)} &      
\multicolumn{1}{c}{} &      
\multicolumn{1}{c}{Mag} \\
\multicolumn{1}{c}{(1)} &       
\multicolumn{1}{c}{(2)} &            
\multicolumn{1}{c}{(3)} &          
\multicolumn{1}{c}{(4)} &  
\multicolumn{1}{c}{(5)} &      
\multicolumn{1}{c}{(6)} &  
\multicolumn{1}{c}{(7)} &      
\multicolumn{1}{c}{(8)} &      
\multicolumn{1}{c}{(9)} 
}     
\startdata
NGVS1 & NGVS J120345.06+273702.0 & 180.9377500 & 27.6172222 & 0 & \nd & 3361.8 \err 6.0 & SDSS & \nd \\
NGVS2 & NGVS J120451.83+265933.1 & 181.2159583 & 26.9925278 & 0 & \nd & \nd & \nd & \nd \\
NGVS3 & NGVS J120745.73+120337.0 & 181.9405522 & 12.0602775 & 2 & \nd & \nd & \nd & \nd \\
NGVS4 & NGVS J120801.21+124856.0 & 182.0050598 & 12.8155639 & 2 & \nd & \nd & \nd & \nd \\
NGVS5 & NGVS J120804.43+131317.0 & 182.0184583 & 13.2213889 & 2 & \nd & \nd & \nd & \nd \\
NGVS6 & NGVS J120805.48+124550.3 & 182.0228320 & 12.7639790 & 0 & \nd & \nd & \nd & \nd \\
NGVS7 & NGVS J120820.90+123007.8 & 182.0870827 & 12.5021665 & 0 & \nd & 2321.7 \err 11.1 & NED & \nd \\
NGVS8 & NGVS J120821.35+121547.1 & 182.0889422 & 12.2630971 & 2 & \nd & \nd & \nd & \nd \\
\enddata 
\begin{center}\begin{minipage}{0.8\linewidth}
\tablenotetext{\dagger}{The Table lists fundamental properties for all galaxies within the NGVS survey area and classified as {\it bona fide} members of the Virgo cluster based on our analysis, as well as for the 10 galaxies that, after the fact, were found to have radial velocities beyond the 3,500 km s$^{-1}$ boundary adopted for the cluster. Columns are as follows: (1) the galaxy nickname; (2) the galaxy name following IAU conventions; (3) the J2000 Right Ascension and (4) Declination of the isophotal center of each galaxy; (5) the degree of confidence in the Virgo membership of each galaxy: Membership Class 0 are certain members; Membership Class 1 are likely members,  Membership Class 2 are possible members, Membership Class 4 are galaxies deemed to be Virgo members based on our analysis but found, after the fact, to have radial velocity beyond the 3,500 km s$^{1}$ boundary adopted for the cluster (see also NGVS-XIV); (6) the VCC number for galaxies also cataloged by Binggeli et al. (1985); (7) the radial velocity and associated error, when available; and (8) the code for the reference of the radial velocity. The references for the latter are: Keck: R. Guhathakurta et al. (in preparation); MMT: E. W. Peng et al. (in preparation); AAT: E. W. Peng et al. (in preparation); SDSS: Data Release 13 from the SDSS; NED: as listed in the NASA Extragalactic Database as of August 1, 2025; S11: \citet[][]{Strader+11}; (9) the distance modulus and associated error derived from SBF measurements by \citet{Cantiello+24}. Note that only the first 60 rows are included; the full table is available in machine readable format.}
\end{minipage}\end{center}
\end{deluxetable*}
\end{longrotatetable}
\normalsize

\small
\begin{longrotatetable}
\startlongtable
\begin{deluxetable*}{lclclcccccc}
\tablecaption{NGVS Galaxy catalog: General Properties II\tablenotemark{$\dagger$}} \label{tab:Tab6}
\tablehead{
\multicolumn{1}{c}{Nickname} &       
\multicolumn{1}{c}{Membership} &
\multicolumn{1}{c}{Morphological} &
\multicolumn{1}{c}{Nuclear} &
\multicolumn{1}{c}{Fit} &
\multicolumn{1}{c}{FWHM,$u^*$} &
\multicolumn{1}{c}{FWHM,$g$} &
\multicolumn{1}{c}{FWHM,$i$} &
\multicolumn{1}{c}{FWHM,$z$} &
\multicolumn{1}{c}{$\rho_{All}$} &
\multicolumn{1}{c}{$\rho_{Bright}$} \\
\multicolumn{1}{c}{} &
\multicolumn{1}{c}{Class} &
\multicolumn{1}{c}{Classification} &
\multicolumn{1}{c}{Code} &
\multicolumn{1}{c}{Model} &
\multicolumn{1}{c}{(arcsec)} &
\multicolumn{1}{c}{(arcsec)} &
\multicolumn{1}{c}{(arcsec)} &
\multicolumn{1}{c}{(arcsec)} &
\multicolumn{1}{c}{($\#$gal. deg$^{-2}$)} &
\multicolumn{1}{c}{($\#$gal. deg$^{-2}$)} \\
\multicolumn{1}{c}{(1)} &       
\multicolumn{1}{c}{(2)} &            
\multicolumn{1}{c}{(3)} &          
\multicolumn{1}{c}{(4)} &  
\multicolumn{1}{c}{(5)} &      
\multicolumn{1}{c}{(6)} &  
\multicolumn{1}{c}{(7)} &      
\multicolumn{1}{c}{(8)} &  
\multicolumn{1}{c}{(9)} &      
\multicolumn{1}{c}{(10)} &      
\multicolumn{1}{c}{(11)} 
}
\startdata
  NGVS1 & 0 & ES(sp$\vert$du$\vert$b:)$\mh$A(g$\vert$sp$\vert$p$\vert$f) & 0 & G/S & 0.75 & 0.87 & 0.41 & 0.93 & \nd & \nd\\
  NGVS2 & 0 & S(du$\vert$di)$\mh$A(g$\vert$d$\vert$f:) & 0 & G/S & 0.74 & 0.86 & 0.45 & 0.93 & \nd & \nd\\
  NGVS3 & 2 & E$\mh$Q & 0 & G/S & 0.80 & 0.71 & 0.51 & 0.79 & 11.16 & \phantom{1}4.32\\
  NGVS4 & 2 & E(n:)$\mh$Q & 2 & G/S+P & 0.85 & 0.74 & 0.49 & 0.73 & 13.45 & \phantom{1}5.09\\
  NGVS5 & 2 & E$\mh$Q & 0 & G/S & 0.85 & 0.73 & 0.49 & 0.73 & 13.81 & \phantom{1}5.39\\
  NGVS6 & 0 & E$\mh$Q & 0 & G/S & 0.85 & 0.74 & 0.49 & 0.73 & 14.49 & \phantom{1}4.73\\
  NGVS7 & 0 & EI(sp:$\vert$du:$\vert$b:$\vert$sh:$\vert$pm:)$\mh$I(c$\vert$p$\vert$g) & 0 & G/S & 0.80 & 0.71 & 0.51 & 0.79 & 13.11 & \phantom{1}4.65\\
  NGVS8 & 2 & E$\mh$Q & 0 & G/S & 0.80 & 0.70 & 0.51 & 0.79 & 12.52 & \phantom{1}4.88\\
\enddata
\end{deluxetable*}
\end{longrotatetable}
\normalsize

\small
\begin{longrotatetable}
\startlongtable
\begin{deluxetable*}{lcccccccccc}
\tablecaption{NGVS Galaxy catalog: Curve of Growth Photometry\tablenotemark{$\dagger$}} \label{tab:Tab7}
\tablehead{
\multicolumn{1}{c}{} &       
\multicolumn{1}{c}{} &
\multicolumn{1}{c}{} &
\multicolumn{1}{c}{} &
\multicolumn{1}{c}{} &
\multicolumn{1}{c}{} &
\multicolumn{5}{c}{$g$-band}\\
\multicolumn{1}{c}{} &       
\multicolumn{1}{c}{} &
\multicolumn{1}{c}{} &
\multicolumn{1}{c}{} &
\multicolumn{1}{c}{} &
\multicolumn{1}{c}{} &
\multicolumn{5}{c}{\rule{300pt}{0.5pt}}\\ 
\multicolumn{1}{c}{Nickname} &       
\multicolumn{1}{c}{Membership} &
\multicolumn{1}{c}{$u^*$} &
\multicolumn{1}{c}{$g$} &
\multicolumn{1}{c}{$i$} &
\multicolumn{1}{c}{$z$} &
\multicolumn{1}{c}{$R_e$} &
\multicolumn{1}{c}{$\mu_{0}$} &
\multicolumn{1}{c}{$\mu_{R_e}$} &
\multicolumn{1}{c}{$<\mu_{R_e}>$} &
\multicolumn{1}{c}{C$_{80/20}$} \\
\multicolumn{1}{c}{} &
\multicolumn{1}{c}{Class} &
\multicolumn{1}{c}{(mag)} &
\multicolumn{1}{c}{(mag)} &
\multicolumn{1}{c}{(mag)} &
\multicolumn{1}{c}{(mag)} &
\multicolumn{1}{c}{(arcsec)} &
\multicolumn{1}{c}{(mag arcsec$^{-2}$)} &
\multicolumn{1}{c}{(mag arcsec$^{-2}$)} &
\multicolumn{1}{c}{(mag arcsec$^{-2}$)} &
\multicolumn{1}{c}{} \\
\multicolumn{1}{c}{(1)} &       
\multicolumn{1}{c}{(2)} &            
\multicolumn{1}{c}{(3)} &          
\multicolumn{1}{c}{(4)} &  
\multicolumn{1}{c}{(5)} &      
\multicolumn{1}{c}{(6)} &  
\multicolumn{1}{c}{(7)} &      
\multicolumn{1}{c}{(8)} &  
\multicolumn{1}{c}{(9)} &      
\multicolumn{1}{c}{(10)} &  
\multicolumn{1}{c}{(11)} 
}
\startdata
NGVS17 & 0 & 16.26 \err 0.03 & 15.39 \err 0.01 & 14.60 \err 0.01 & 14.48 \err 0.02 & 14.39 & 20.68 & 23.72 & 23.18 & 3.1\\ 
NGVS22 & 0 & 17.05 \err 0.03 & 16.12 \err 0.01 & 15.54 \err 0.01 & 15.39 \err 0.02 & \phantom{1}4.31 & 20.19 & 22.16 & 21.29 & 4.2\\ 
NGVS26 & 0 & 16.25 \err 0.03 & 15.36 \err 0.01 & 14.72 \err 0.01 & 14.55 \err 0.02 & \phantom{1}4.20 & 19.06 & 21.57 & 20.47 & 5.5\\ 
NGVS34 & 0 & 15.12 \err 0.03 & 13.95 \err 0.01 & 13.12 \err 0.01 & 12.89 \err 0.02 & \phantom{1}5.49 & 17.34 & 20.86 & 19.65 & 7.3\\ 
NGVS35 & 0 & 16.06 \err 0.03 & 14.92 \err 0.01 & 14.10 \err 0.01 & 13.93 \err 0.02 & \phantom{1}9.05 & 20.42 & 22.45 & 21.70 & 3.9\\ 
NGVS36 & 0 & 15.11 \err 0.03 & 14.49 \err 0.01 & 14.19 \err 0.01 & 13.69 \err 0.02 & 18.75 & 21.13 & 23.71 & 22.85 & 5.4\\ 
NGVS62 & 0 & 15.20 \err 0.03 & 13.93 \err 0.01 & 13.03 \err 0.01 & 12.82 \err 0.02 & 15.99 & 19.38 & 22.36 & 21.95 & 3.3\\ 
NGVS63 & 0 & 15.04 \err 0.03 & 14.43 \err 0.01 & 13.91 \err 0.01 & 13.77 \err 0.02 & 24.15 & 21.53 & 24.13 & 23.34 & 4.0\\
\enddata 
\begin{center}\begin{minipage}{0.8\linewidth}
\tablenotetext{\dagger}{The Table lists non parametric curve of growth photometry and structural parameters for the 897 NGVS galaxies analyzed with {\sf Typhon} (see \S\ref{subsec:typhon}). This includes all galaxies in the core region, i.e. the four fields (roughly 3.7 deg$^2$) centered on M87 (NGVS+0+0, NGVS+0+1, NGVS-1+0, and NGVS-1+1), all galaxies brighter than $g\sim16$ mag (as measured in the VCC), and the occasional galaxy of interest. Note that $r$-band photometry (which exists only for the galaxies in the core region) was published in NGVS-XIV and is not duplicated here. Columns are as follows: (1) the galaxy nickname, as in Table \ref{tab:Tab5}; (2) the degree of confidence in the Virgo membership of each galaxy (see notes to Table \ref{tab:Tab5}); (3)-(6) the curve of growth total magnitudes (see \S\ref{subsec:typhon}) in the $u^*$-,$g$-,$i$-, and $z$-band; (7) the geometric effective radius, in arcseconds, measured in the $g$-band and defined as the radius enclosing half the total curve-of-growth magnitude. Note that the geometric radius is equal to the radius measured along the semimajor axis multiplied by $\sqrt{1-\epsilon}$, where $\epsilon$ is the ellipticity tabulated in Table \ref{tab:Tab12}; (8) the central surface brightness, measured at 0.187 arcsec (1 MegaCam pixel); (9) the $g$-band surface brightness at the effective radius and (10) averaged within the effective radius; (11) the $g$-band concentration, defined as the ratio of the radii including 80\% and 20\% of the total flux. All magnitudes and surface brightnesses are uncorrected for extinction. Note that only the first 60 rows are included; the full table is available in machine readable format.}
\end{minipage}\end{center}
\end{deluxetable*}
\end{longrotatetable}
\normalsize

\small
\begin{longrotatetable}
\startlongtable
\begin{deluxetable*}{lccccccccc}
\tablecaption{NGVS Galaxy catalog: Total Parametric Photometry and Structural Parameters\tablenotemark{$\dagger$}} \label{tab:Tab8}
\tablehead{
\multicolumn{1}{c}{} &       
\multicolumn{1}{c}{} &
\multicolumn{1}{c}{} &
\multicolumn{1}{c}{} &
\multicolumn{1}{c}{} &
\multicolumn{1}{c}{} &
\multicolumn{4}{c}{$g$-band}\\
\multicolumn{1}{c}{} &       
\multicolumn{1}{c}{} &
\multicolumn{1}{c}{} &
\multicolumn{1}{c}{} &
\multicolumn{1}{c}{} &
\multicolumn{1}{c}{} &
\multicolumn{4}{c}{\rule{250pt}{0.5pt}}\\ 
\multicolumn{1}{c}{Nickname} &       
\multicolumn{1}{c}{Membership} &
\multicolumn{1}{c}{$u^*$} &
\multicolumn{1}{c}{$g$} &
\multicolumn{1}{c}{$i$} &
\multicolumn{1}{c}{$z$} &
\multicolumn{1}{c}{$R_e$} &
\multicolumn{1}{c}{$n$} &
\multicolumn{1}{c}{$\mu_{R_e}$} &
\multicolumn{1}{c}{$<\mu_{R_e}>$}\\
\multicolumn{1}{c}{} &
\multicolumn{1}{c}{Class} &
\multicolumn{1}{c}{(mag)} &
\multicolumn{1}{c}{(mag)} &
\multicolumn{1}{c}{(mag)} &
\multicolumn{1}{c}{(mag)} &
\multicolumn{1}{c}{(arcsec)} &
\multicolumn{1}{c}{} &
\multicolumn{1}{c}{(mag arcsec$^{-2}$)} &
\multicolumn{1}{c}{(mag arcsec$^{-2}$)}\\
\multicolumn{1}{c}{(1)} &       
\multicolumn{1}{c}{(2)} &            
\multicolumn{1}{c}{(3)} &          
\multicolumn{1}{c}{(4)} &  
\multicolumn{1}{c}{(5)} &      
\multicolumn{1}{c}{(6)} &  
\multicolumn{1}{c}{(7)} &      
\multicolumn{1}{c}{(8)} &  
\multicolumn{1}{c}{(9)} &  
\multicolumn{1}{c}{(10)}
}
\startdata
NGVS1  & 0  & 17.52 \err 0.03  &  16.67 \err 0.01  &  16.08 \err 0.01  &  16.01 \err 0.02  & \phantom{1}6.39 \err 0.03  & 0.72 \err 0.01  & 23.22 \err 0.01  & 22.68 \err 0.01 \\
NGVS2  & 0  & 14.53 \err 0.03  &  13.63 \err 0.01  &  12.92 \err 0.01  &  12.74 \err 0.02  & 13.95 \err 0.03  & 0.67 \err 0.01  & 21.85 \err 0.01  & 21.33 \err 0.01 \\
NGVS3  & 2  & 21.81 \err 0.06  &  20.58 \err 0.04  &  19.68 \err 0.03  &  19.70 \err 0.06  & \phantom{1}2.45 \err 0.18  & 0.83 \err 0.01  & 25.12 \err 0.17  & 24.51 \err 0.17 \\
NGVS4  & 2  & 20.42 \err 0.04  &  19.43 \err 0.02  &  18.72 \err 0.02  &  18.70 \err 0.04  & \phantom{1}3.69 \err 0.05  & 0.72 \err 0.01  & 24.84 \err 0.07  & 24.26 \err 0.02 \\
NGVS5  & 2  & 21.51 \err 0.08  &  20.55 \err 0.07  &  19.81 \err 0.07  &  19.68 \err 0.10  & \phantom{1}3.63 \err 0.55  & 1.11 \err 0.02  & 26.08 \err 0.35  & 25.33 \err 0.34 \\
NGVS6  & 0  & 18.38 \err 0.03  &  17.26 \err 0.02  &  16.47 \err 0.02  &  16.45 \err 0.03  & 16.82 \err 0.29  & 0.85 \err 0.01  & 25.99 \err 0.04  & 25.37 \err 0.04 \\
NGVS7  & 0  & 18.83 \err 0.03  &  18.01 \err 0.02  &  17.49 \err 0.02  &  17.44 \err 0.03  & \phantom{1}7.84 \err 0.17  & 1.01 \err 0.01  & 25.16 \err 0.05  & 24.46 \err 0.05 \\
NGVS8  & 2  & 24.54 \err 0.55  &  22.61 \err 0.19  &  22.16 \err 0.25  &  21.42 \err 0.20  & \phantom{1}3.27 \err 1.19  & 0.34 \err 0.06  & 27.43 \err 0.88  & 27.17 \err 0.85 \\
\enddata
\end{deluxetable*}
\end{longrotatetable}

\small
\begin{longrotatetable}
\startlongtable
\begin{deluxetable*}{lcccccccc}
\tablecaption{NGVS Galaxy catalog: Parametric Photometry and Structural Parameters for the Inner (NSC) Component \tablenotemark{$\dagger$}} \label{tab:Tab9}
\tablehead{
\multicolumn{1}{c}{} &       
\multicolumn{1}{c}{} &
\multicolumn{1}{c}{} &
\multicolumn{1}{c}{} &
\multicolumn{1}{c}{} &
\multicolumn{1}{c}{} &
\multicolumn{3}{c}{$g$-band}\\
\multicolumn{1}{c}{} &       
\multicolumn{1}{c}{} &
\multicolumn{1}{c}{} &
\multicolumn{1}{c}{} &
\multicolumn{1}{c}{} &
\multicolumn{1}{c}{} &
\multicolumn{3}{c}{\rule{200pt}{0.5pt}}\\ 
\multicolumn{1}{c}{Nickname} &       
\multicolumn{1}{c}{Membership} &
\multicolumn{1}{c}{$u^*$} &
\multicolumn{1}{c}{$g$} &
\multicolumn{1}{c}{$i$} &
\multicolumn{1}{c}{$z$} &
\multicolumn{1}{c}{$R_e$} &
\multicolumn{1}{c}{$\mu_{R_e}$} &
\multicolumn{1}{c}{$<\mu_{R_e}>$}\\
\multicolumn{1}{c}{} &
\multicolumn{1}{c}{Class} &
\multicolumn{1}{c}{(mag)} &
\multicolumn{1}{c}{(mag)} &
\multicolumn{1}{c}{(mag)} &
\multicolumn{1}{c}{(mag)} &
\multicolumn{1}{c}{(arcsec)} &
\multicolumn{1}{c}{(mag arcsec$^{-2}$)} &
\multicolumn{1}{c}{(mag arcsec$^{-2}$)}\\
\multicolumn{1}{c}{(1)} &       
\multicolumn{1}{c}{(2)} &            
\multicolumn{1}{c}{(3)} &          
\multicolumn{1}{c}{(4)} &  
\multicolumn{1}{c}{(5)} &      
\multicolumn{1}{c}{(6)} &  
\multicolumn{1}{c}{(7)} &      
\multicolumn{1}{c}{(8)} &  
\multicolumn{1}{c}{(9)}
}
\startdata
NGVS4 & 2 & 24.70 \err 0.04 & 23.75 \err 0.02 & 22.95 \err 0.02 & 22.93 \err 0.05 & \nd & \nd & \nd\\
NGVS11 & 1 & 24.70 \err 0.05 & 23.50 \err 0.02 & 22.56 \err 0.01 & 22.35 \err 0.03 & \nd & \nd & \nd\\
NGVS16 & 0 & 20.82 \err 0.01 & 20.23 \err 0.01 & 19.99 \err 0.01 & 19.90 \err 0.01 & \nd & \nd & \nd\\
NGVS17 & 0 & 22.16 \err 0.01 & 20.94 \err 0.01 & 20.27 \err 0.01 & 20.09 \err 0.02 & 0.05 & 17.47 & 16.43\\
NGVS18 & 2 & 26.26 \err 0.16 & 25.34 \err 0.09 & 24.50 \err 0.06 & 24.14 \err 0.16 & \nd & \nd & \nd\\
NGVS31 & 0 & 22.75 \err 0.01 & 21.98 \err 0.01 & 21.55 \err 0.01 & 21.38 \err 0.01 & \nd & \nd & \nd\\
NGVS35 & 0 & 22.98 \err 0.01 & 22.11 \err 0.01 & 21.33 \err 0.01 & 21.30 \err 0.01 & 0.05 & 18.64 & 17.60\\
NGVS36 & 0 & 23.48 \err 0.01 & 22.56 \err 0.01 & 21.79 \err 0.01 & 21.55 \err 0.01 & 0.05 & 19.08 & 18.05\\
\enddata 
\begin{center}\begin{minipage}{0.75\linewidth}
\tablenotetext{\dagger}{For the 835 galaxies believed to host a NSC (i.e. Nuclear Code equal 1, 1,off, 2 or 2,off, see \ref{tab:Tab6}) and for which a two components fit (i.e. either T/D or G/S+P, see Table \ref{tab:Tab6}) could be performed, the Table lists the NSC parameters derived from a S\'ersic fit in the case of a {\sf Typhon}, or a PSF in the case of a {\sf GalFit}. Note that when a T/D fit was used to better describe a complex surface brightness profile, but for which the inner component does not represent a NSC, only the combined parameters are given (Table \ref{tab:Tab8}): these cases are identified by a `*' following the Nuclear Code listed in Table \ref{tab:Tab6}, column 4.  Columns are as follows: (1) the galaxy nickname, as in Table \ref{tab:Tab5}; (2) the Membership Class (see notes to Table \ref{tab:Tab5}); (3)-(6) the NSC magnitude in the $u^*$-,$g$-,$i$- and $z$-band; (7)-(9) the effective radius, surface brightness at the effective radius and averaged within an effective radius, respectively, all measured in the $g$-band. All uncertainties are as returned by the fitting codes and should be treated with caution (see \S\ref{sec:fin}), indeed we do not report errors on surface brightnesses as these are clearly significantly underestimated. Note that if a parameter was held fixed during the fits, no uncertainties are given; additionally, only magnitudes are available for a G/S+P fit, since the inner component is modeled as an unresolved PSF. For a T/D fit, with one exception, the S\'ersic index for the inner component is fixed at $n=2$ (the one exception is NGVS81, for which $n$ was fitted as 0.041 \err 0.02), and the ellipticity was fixed to 0. All magnitudes and surface brightnesses are uncorrected for extinction. Note that only the first 60 rows are included; the full table is available in machine readable format.}
\end{minipage}\end{center}
\end{deluxetable*}
\end{longrotatetable}

\small
\begin{longrotatetable}
\startlongtable
\begin{deluxetable*}{lccccccccc}
\tablecaption{NGVS Galaxy catalog: Parametric Photometry and Structural Parameters for the Outer (Main Galaxy) Component \tablenotemark{$\dagger$}} \label{tab:Tab10}
\tablehead{
\multicolumn{1}{c}{} &       
\multicolumn{1}{c}{} &
\multicolumn{1}{c}{} &
\multicolumn{1}{c}{} &
\multicolumn{1}{c}{} &
\multicolumn{1}{c}{} &
\multicolumn{4}{c}{$g$-band}\\
\multicolumn{1}{c}{} &       
\multicolumn{1}{c}{} &
\multicolumn{1}{c}{} &
\multicolumn{1}{c}{} &
\multicolumn{1}{c}{} &
\multicolumn{1}{c}{} &
\multicolumn{4}{c}{\rule{270pt}{0.5pt}}\\ 
\multicolumn{1}{c}{Nickname} &       
\multicolumn{1}{c}{Membership} &
\multicolumn{1}{c}{$u^*$} &
\multicolumn{1}{c}{$g$} &
\multicolumn{1}{c}{$i$} &
\multicolumn{1}{c}{$z$} &
\multicolumn{1}{c}{$R_e$} &
\multicolumn{1}{c}{$n$} &
\multicolumn{1}{c}{$\mu_{R_e}$} &
\multicolumn{1}{c}{$<\mu_{R_e}>$}\\
\multicolumn{1}{c}{} &
\multicolumn{1}{c}{Class} &
\multicolumn{1}{c}{(mag)} &
\multicolumn{1}{c}{(mag)} &
\multicolumn{1}{c}{(mag)} &
\multicolumn{1}{c}{(mag)} &
\multicolumn{1}{c}{(arcsec)} &
\multicolumn{1}{c}{(1)} &       
\multicolumn{1}{c}{(mag arcsec$^{-2}$)} &
\multicolumn{1}{c}{(mag arcsec$^{-2}$)}\\
\multicolumn{1}{c}{(1)} &       
\multicolumn{1}{c}{(2)} &            
\multicolumn{1}{c}{(3)} &          
\multicolumn{1}{c}{(4)} &  
\multicolumn{1}{c}{(5)} &      
\multicolumn{1}{c}{(6)} &  
\multicolumn{1}{c}{(7)} &      
\multicolumn{1}{c}{(8)} &  
\multicolumn{1}{c}{(9)}&
\multicolumn{1}{c}{(10)}
}
\startdata
NGVS4 & 2 & 20.44 \err 0.01 & 19.45 \err 0.01 & 18.75 \err 0.01 & 18.73 \err 0.01 & 3.74 \err 0.05 & 0.72 \err 0.01 & 24.86 \err 0.03 & 24.31 \err 0.03\\  
NGVS11 & 1 & 19.58 \err 0.01 & 18.38 \err 0.01 & 17.52 \err 0.01 & 17.41 \err 0.01 & 3.97 \err 0.05 & 0.99 \err 0.01 & 24.05 \err 0.03 & 23.35 \err 0.03\\  
NGVS16 & 0 & 18.55 \err 0.01 & 17.51 \err 0.01 & 16.78 \err 0.01 & 16.66 \err 0.01 & 11.97 \err 0.15\phantom{1} & 1.09 \err 0.01 & 25.63 \err 0.03 & 24.89 \err 0.03\\  
NGVS17 & 0 & 16.30 \err 0.01 & 15.37 \err 0.01 & 14.55 \err 0.01 & 14.41 \err 0.01 & 15.27 \err 0.19\phantom{1} & 1.12 \err 0.01 & 24.04 \err 0.03 & 23.28 \err 0.03\\  
NGVS18 & 2 & 21.74 \err 0.01 & 20.67 \err 0.01 & 19.90 \err 0.01 & 20.03 \err 0.02 & 3.01 \err 0.17 & 0.72 \err 0.01 & 25.59 \err 0.14 & 25.04 \err 0.13\\  
NGVS31 & 0 & 18.30 \err 0.01 & 17.26 \err 0.01 & 16.48 \err 0.01 & 16.35 \err 0.01 & 7.10 \err 0.06 & 0.98 \err 0.01 & 24.19 \err 0.02 & 23.50 \err 0.02\\  
NGVS35 & 0 & 16.11 \err 0.01 & 14.94 \err 0.01 & 14.13 \err 0.01 & 13.96 \err 0.01 & 8.94 \err 0.02 & 1.02 \err 0.01 & 22.41 \err 0.01 & 21.70 \err 0.01\\  
NGVS36 & 0 & 15.47 \err 0.01 & 14.72 \err 0.01 & 14.31 \err 0.01 & 14.04 \err 0.01 & 14.51 \err 0.15\phantom{1} & 1.02 \err 0.01 & 23.24 \err 0.03 & 22.52 \err 0.02\\
\enddata 
\begin{center}\begin{minipage}{0.75\linewidth}
\tablenotetext{\dagger}{For the 835 galaxies believed to host a NSC (i.e. Nuclear Code equal 1, 1,off, 2 or 2,off, see \ref{tab:Tab6}) and for which a two components fit (i.e. either T/D or G/S+P, see Table \ref{tab:Tab6}) could be performed, the Table lists the parameters for the outer components described as a S\'ersic law in both the case of {\sf Typhon} and {\sf GalFit}. Note that when a T/D fit was used to better describe a complex surface brightness profile, but for which the inner component does not represent a NSC, only the combined parameters are given (Table \ref{tab:Tab8}): these cases are identified by a `*' following the Nuclear Code listed in Table \ref{tab:Tab6}, column 4. Columns are as follows: (1) the galaxy nickname, as in Table \ref{tab:Tab5}; (2) the Membership Class (see notes to Table \ref{tab:Tab5}); (3)-(6) the outer (main galaxy) component magnitude in the $u^*$-,$g$-,$i$- and $z$-band. Note that for the reasons listed in \S\ref{subsec:typhon}, \S\ref{subsec:galfit}, and \S\ref{sec:fin}, these magnitudes should not be used to calculate colors, which are instead listed in Table \ref{tab:Tab11}; (7)-(10) the geometric effective radius, S\'ersic index, surface brightness at the effective radius and averaged within an effective radius, respectively, all measured in the $g$-band. All uncertainties are as returned by the fitting codes and should be treated with caution (see \S\ref{sec:fin}). Note that if a parameter was held fixed during the fits, no uncertainties are given. As a reminder, the geometric radius is equal to the radius measured along the semimajor axis multiplied by $\sqrt{1-\epsilon}$, where $\epsilon$ is the ellipticity tabulated in Table \ref{tab:Tab12}. All magnitudes and surface brightnesses are uncorrected for extinction. Note that only the first 60 rows are included; the full table is available in machine readable format.}
\end{minipage}\end{center}
\end{deluxetable*}
\end{longrotatetable}

\small
\startlongtable
\begin{deluxetable*}{lccrrrcc}
\tablecaption{NGVS Galaxy catalog: Colors and Stellar Masses\tablenotemark{$\dagger$}} \label{tab:Tab11}
\tablehead{
\multicolumn{1}{c}{Nickname} &       
\multicolumn{1}{c}{Membership} &
\multicolumn{1}{c}{$E(B-V)$} &
\multicolumn{1}{c}{$u^*-g$} &
\multicolumn{1}{c}{$g-i$} &
\multicolumn{1}{c}{$g-z$} &
\multicolumn{1}{c}{\mstar} &
\multicolumn{1}{c}{\mstar~ range}\\
\multicolumn{1}{c}{} &
\multicolumn{1}{c}{Class} &
\multicolumn{1}{c}{(mag)} &
\multicolumn{1}{c}{(mag)} &
\multicolumn{1}{c}{(mag)} &
\multicolumn{1}{c}{(mag)}  &
\multicolumn{1}{c}{(M$_{\odot}$)}  &
\multicolumn{1}{c}{(M$_{\odot}$)}  \\
\multicolumn{1}{c}{(1)} &       
\multicolumn{1}{c}{(2)} &            
\multicolumn{1}{c}{(3)} &          
\multicolumn{1}{c}{(4)} &  
\multicolumn{1}{c}{(5)} &      
\multicolumn{1}{c}{(6)} &  
\multicolumn{1}{c}{(7)} &  
\multicolumn{1}{c}{(8)}
}
\startdata
NGVS1 & 0 & 0.019 & 0.83 \err  0.03 & 0.56 \err  0.02 & 0.62 \err  0.03 & 7.34e+07 & [5.42e+07 8.47e+07]\\
NGVS2 & 0 & 0.026 & 0.88 \err  0.03 & 0.66 \err  0.02 & 0.82 \err  0.03 & 1.70e+09 & [1.29e+09 2.15e+09]\\
NGVS3 & 2 & 0.025 & 1.20 \err  0.07 & 0.86 \err  0.05 & 0.82 \err  0.07 & 4.05e+06 & [3.42e+06 4.52e+06]\\
NGVS4 & 2 & 0.027 & 0.96 \err  0.04 & 0.66 \err  0.03 & 0.66 \err  0.04 & 6.02e+06 & [4.82e+06 8.13e+06]\\
NGVS5 & 2 & 0.030 & 0.93 \err  0.10 & 0.69 \err  0.09 & 0.80 \err  0.12 & 2.89e+06 & [1.96e+06 3.71e+06]\\
NGVS6 & 0 & 0.027 & 1.09 \err  0.04 & 0.74 \err  0.03 & 0.74 \err  0.03 & 6.35e+07 & [5.51e+07 7.40e+07]\\
NGVS7 & 0 & 0.026 & 0.79 \err  0.04 & 0.47 \err  0.03 & 0.51 \err  0.04 & 1.58e+07 & [1.20e+07 1.83e+07]\\
NGVS8 & 2 & 0.026 & 1.90 \err  0.58 & 0.41 \err  0.31 & 1.14 \err  0.27 & 5.78e+05 & [3.99e+05 7.80e+05]\\
\enddata 
\begin{center}\begin{minipage}{0.9\linewidth}
\tablenotetext{\dagger}{The Table lists colors derived from parametric fits to each galaxy's surface brightness profile (see \S \ref{subsec:typhon} and \ref{subsec:galfit}) as well as stellar masses. Columns are as follows: (1) the galaxy nickname; (2) the Membership Class (see notes to Table \ref{tab:Tab5}); (3) the Galactic reddening $E(B-V)$ following Schlegel et al. (1998). The extinction coefficients adopted to correct each band are $A_{u^*}/E(B-V) = 4.594$, $A_g/E(B-V) =  3.560$, $A_i/E(B-V) =  1.813$, and $A_z/E(B-V) =  1.221$; (4)-(6): the $u^*-g$, $g-i$, and $g-z$ colors, dereddened using the reddening and extinction coefficients given above, and measured as described in  \S\ref{sec:fin}; (7)-(8) the stellar mass and the mass range corresponding to the 50th percentile of the posterior probability distribution (see \S\ref{sec:mass}). Note that only the first 60 rows are included; the full table is available in machine readable format.}
\end{minipage}\end{center}
\end{deluxetable*}
\normalsize

\newpage
\small
\startlongtable
\begin{deluxetable*}{lcccc}
\tablecaption{NGVS Galaxy catalog: $g$-band Structural Parameters\tablenotemark{$\dagger$}} \label{tab:Tab12}
\tablehead{
\multicolumn{1}{c}{Nickname} &       
\multicolumn{1}{c}{Membership} &
\multicolumn{1}{c}{$\epsilon \pm \delta \epsilon$} &
\multicolumn{1}{c}{$PA \pm \delta PA$} &
\multicolumn{1}{c}{$B4 \pm \delta B4$}  \\
\multicolumn{1}{c}{} &       
\multicolumn{1}{c}{Class} &
\multicolumn{1}{c}{} &
\multicolumn{1}{c}{(degrees)} &
\multicolumn{1}{c}{($\times100$)}  \\
\multicolumn{1}{c}{(1)} &       
\multicolumn{1}{c}{(2)} &            
\multicolumn{1}{c}{(3)} &          
\multicolumn{1}{c}{(4)} &  
\multicolumn{1}{c}{(5)}
}
\startdata
NGVS1 & 0 & 0.08 \err 0.01 &  \phantom{$-$}70.78 \err  0.33 &  \nd\\
NGVS2 & 0 & 0.53 \err 0.01 &  $-$60.53 \err  0.01 &  \nd\\
NGVS3 & 2 & 0.37 \err 0.01 &  \phantom{$-$}73.00 \err  0.75 &  \nd\\
NGVS4 & 2 & 0.26 \err 0.01 &  $-$32.33 \err  0.48 &  \nd\\
NGVS5 & 2 & 0.51 \err 0.01 &  \phantom{$-$}18.98 \err  0.59 &  \nd\\
NGVS6 & 0 & 0.10 \err 0.01 &  \phantom{$-$}17.14 \err  0.79 &  \nd\\
NGVS7 & 0 & 0.30 \err 0.01 &  \phantom{$-$}54.95 \err  0.26 &  \nd\\
NGVS8 & 2 & 0.19 \err 0.06 &  \phantom{1}$-$84.77 \err 12.21 &  \nd\\
\enddata 
\begin{center}\begin{minipage}{0.7\linewidth}
\tablenotetext{\dagger}{The Table lists structural parameters for all NGVS Virgo galaxies, based on either the {\sf Typhon} or {\sf GalFit} analysis. Columns are as follows: (1) the galaxy nickname, as in Table \ref{tab:Tab5}; (2) the Membership Class (see notes to Table \ref{tab:Tab5}); (3)-(5) the average ellipticity, position angle (measured from North to East) and  B4 coefficient, respectively.  The B4 coefficient is multiplied by 100 for ease of reading, and is only available for the galaxies analyzed with {\sf Typhon}. All parameters are measured in the $g$-band. Note that  the ellipticity and position angle in the {\sf GalFit} analysis are radially invariant; in this case the errors listed in the table are 
the formal errors returned by the code. In the {\sf Typhon} analysis, the structural parameters are derived from a non-parametric isophotal analysis and are completely independent from the parametric fit. The {\sf Typhon} parameters listed here are averaged between 1$\arcsec$ and one (geometric) effective radius (Table \ref{tab:Tab8}) and the uncertainties are the standard deviations in the mean (no uncertainties are given if the parameter was held fixed while fitting the isophotes). Note that only the first 60 rows are included; the full table is available in machine readable format.}
\end{minipage}\end{center}
\end{deluxetable*}
\normalsize

\small
\startlongtable
\begin{deluxetable*}{llcll}
\tablecaption{NGVS CFHT/MegaCam $r$-band Exposures\tablenotemark{$\dagger$}} \label{tab:rband}
\tablehead{
\multicolumn{1}{c}{Field} &       
\multicolumn{1}{c}{RA} &             
\multicolumn{1}{c}{DEc} &             
\multicolumn{1}{c}{Observing Dates} &             
\multicolumn{1}{c}{Exposure Time}\\
\multicolumn{1}{c}{} &       
\multicolumn{1}{c}{(h:m:s)} &            
\multicolumn{1}{c}{(d:m:s)} &  
\multicolumn{1}{c}{} &            
\multicolumn{1}{c}{N$\times$t(s)}\\
\multicolumn{1}{c}{(1)} &       
\multicolumn{1}{c}{(2)} &            
\multicolumn{1}{c}{(3)} &            
\multicolumn{1}{c}{(4)} &            
\multicolumn{1}{c}{(5)}
}          
\startdata
NGVS+0+0 & 12:32:12 & +12:00:34 & 2008-03-03 to 2008-03-14 & 7$\times$687\\
NGVS+0+1 & 12:32:12 & +12:56:34 & 2008-03-01 to 2008-03-02 & 7$\times$687\\
NGVS+0$-$4 & 12:32:14 & +08:12:49 & 2009-02-22 to 2012-04-13 & 4$\times$687\\
NGVS+0$-$5 & 12:32:11 & +07:14:49 & 2009-04-26 to 2013-04-19 & 6$\times$697,1$\times$623,1$\times$466,1$\times$251,1$\times$214\\
NGVS+1+0 & 12:36:06 & +11:58:49 & 2013-04-18 to 2013-04-19 & 1$\times$616,1$\times$464,1$\times$218,1$\times$160\\
NGVS+2+0 & 12:39:57 & +11:59:49 & 2013-04-18 to 2013-04-19 & 1$\times$687,1$\times$458,1$\times$380,1$\times$302\\
NGVS+2$-$1 & 12:39:55 & +11:02:49 & 2013-04-18 to 2013-04-19 & 1$\times$445,1$\times$424,1$\times$361,1$\times$314\\
NGVS+3+0 & 12:43:50 & +11:59:49 & 2012-04-13 to 2013-04-19 & 2$\times$687,1$\times$571,1$\times$457,1$\times$414,1$\times$302\\
\enddata 
\tablenotetext{\dagger}{$r$-band exposures available for the NGVS fields. Columns are as follows: (1) the name of the NGVS field, with the two numbers roughly indicating the location of the field measured in degrees East and North of M87, with NGVS+0+0 being the field that includes M87; (2) the J2000 Right Ascention and (3) Declination of the field's center; (4) the range of dates over which the exposures were acquired; (5) the number of exposures and the exposure time, in seconds, of each exposure.}
\end{deluxetable*}
\normalsize

\clearpage
\small 
\startlongtable
\setlength{\tabcolsep}{5.0pt}
\begin{deluxetable*}{l c l}
\tablecaption{Notes on Individual Galaxies\tablenotemark{$\dagger$}}
\label{tab:notes}
\tablehead{
 \colhead{Nickname} &
 \colhead{Official Name} &
 \colhead{Notes} \\
 \colhead{(NGVS)} & 
 \colhead{(NGVS J)} & 
 \colhead{}\\
 \colhead{(1)} & \colhead{(2)} & \colhead{(3)}
}
\startdata
1 & 120345.06+273702.0 &   Face-on ES galaxy with global star formation, as well as SF in spiral arms and\\
& & patches, embedded in elliptical isophotes with possible bar. Classified as certain member.\\
2 & 120451.83+265933.1 &   Inclined LTG with active SF across disk. Classified as certain member.\\
3 & 120745.73+120337.0 &   Very compact LSB ETG. Classified as possible member.\\
4 & 120801.21+124856.0 &   Compact LSB ETG with possible nucleus. Classified as possible member.\\
5 & 120804.43+131317.0 &   Very compact, elongated LSB ETG. Classified as possible member.\\
6 & 120805.48+124550.3 &   LSB ETG with round isophotes. Classified as certain member.\\
7 & 120820.90+123007.8 &   Possible EI galaxy with irregular structure and star formation. Post merger system.\\
\enddata
\tablenotetext{\dagger}{Note that only the first 60 rows are included; the full table is available in machine readable format.}
\end{deluxetable*}

\small 
\startlongtable
\setlength{\tabcolsep}{5.0pt}
\begin{deluxetable*}{l c r c c c l}
\tablecaption{Additional Structural Parameters for Core-S\'ersic Galaxies\tablenotemark{$\dagger$}} \label{tab:coresersic}
\tablehead{
 \colhead{Nickname} &
 \colhead{$I_b$} &
 \colhead{$\gamma$} &
 \colhead{$R_b$} &
 \colhead{$\alpha$} \\
 \colhead{} & 
 \colhead{(mag arcsec$^{-2}$)} & 
 \colhead{} & 
 \colhead{(arcsec)} & 
 \colhead{} \\ 
 \colhead{(1)} &
 \colhead{(2)} & 
 \colhead{(3)} & 
 \colhead{(4)} & 
 \colhead{(5)} & 
 \colhead{}
}
\startdata
NGVS401  & 16.65 \err 0.06 & $-$0.75 \err 0.01 & 0.634 \err 0.003 & 99.00 \err \nd\\
NGVS1064 & 16.57 \err 0.03 &  0.11 \err 0.01 & 1.103 \err 0.001 & 99.00 \err \nd\\
NGVS1259 & 15.75 \err 0.08 & $-$1.00 \err 0.02 & 0.225 \err 0.001 & 99.00 \err \nd\\
NGVS1926 & 16.81 \err 0.03 &  0.08 \err 0.01 & 1.632 \err 0.003 &  8.30 \err 0.12\\
NGVS2115 & 18.09 \err 0.07 &  0.36 \err 0.01 & 5.227 \err 0.009 & 23.08 \err 0.94\\
\enddata
\tablenotetext{\dagger}{Additional parameters derived for the five galaxies best fitted with a core-S\'ersic profile \citep{Graham+03}. Columns are as follows: (1) the galaxy nickname; (2) the surface brightness profile at the break radius (uncorrected for extinction), $R_b$, and associated error; (3) the slope $\gamma$ of the inner power law and associated error; (4) the break radius $R_b$ and associated error; (5) the sharpness $\alpha$ of the transition between inner power law and outer S\'ersic law, and associated error.}
\end{deluxetable*}

\end{document}